\documentclass[
reprint,
superscriptaddress,
 amsmath,amssymb,
 aps,
 rmp,
]{revtex4-1}

\usepackage{graphicx}% Include figure files
\usepackage{dcolumn}% Align table columns on decimal point
\usepackage{bm}% bold math
\usepackage{epigraph}

\usepackage{xcolor}

\begin{document}

\title{Memory formation in matter}% Force line breaks with \\

\author{Nathan C.\ Keim}%
 \email[Equal contribution. ]{nkeim@calpoly.edu}%
 \affiliation{Department of Physics, California Polytechnic State University, San Luis Obispo, CA 93407, USA}%
 \affiliation{Kavli Institute for Theoretical Physics, Santa Barbara, CA 93106, USA}%
\author{Joseph D. Paulsen}
 \email[Equal contribution. ]{jdpaulse@syr.edu}%
 \affiliation{Department of Physics and Soft and Living Matter Program, Syracuse University, Syracuse, NY 13244, USA}%
 \affiliation{Kavli Institute for Theoretical Physics, Santa Barbara, CA 93106, USA}%
\author{Zorana Zeravcic}%
 \email{zorana.zeravcic@espci.fr}%
 \affiliation{PCT, Gulliver Lab UMR 7083, ESPCI PSL Research University, 75005 Paris, France}
 \affiliation{Kavli Institute for Theoretical Physics, Santa Barbara, CA 93106, USA}%
\author{Srikanth Sastry}
 \email{sastry@jncasr.ac.in}
 \affiliation{Jawaharlal Nehru Centre for Advanced Scientific Research, Bengaluru 560074, India}
 \affiliation{Kavli Institute for Theoretical Physics, Santa Barbara, CA 93106, USA}%
\author{Sidney R. Nagel}%
 \email{srnagel@uchicago.edu}%
 \affiliation{The James Franck and Enrico Fermi Institutes and The Department of Physics, The University of Chicago, Chicago, IL 60637, USA}%
 \affiliation{Kavli Institute for Theoretical Physics, Santa Barbara, CA 93106, USA}%

\date{\today}

\begin{abstract}
Memory formation in matter is a theme of broad intellectual relevance; it sits at the interdisciplinary crossroads of physics, biology, chemistry, and computer science. 
Memory connotes the ability to encode, access, and erase signatures of past history in the state of a system. Once the system has completely relaxed to thermal equilibrium, it is no longer able to recall aspects of its evolution.  Memory of initial conditions or previous training protocols will be lost. Thus many forms of memory are intrinsically tied to far-from-equilibrium behavior and to transient response to a perturbation. 
This general behavior arises in diverse contexts in condensed matter physics and materials: phase change memory, shape memory, echoes, memory effects in glasses, return-point memory in disordered magnets, as well as related contexts in computer science. 
Yet, as opposed to the situation in biology, there is currently no common categorization and description of the memory behavior that appears to be prevalent throughout condensed-matter systems. 
Here we focus on material memories.  We will describe the basic phenomenology of a few of the known behaviors that can be understood as constituting a memory.  We hope that this will be a guide towards developing the unifying conceptual underpinnings for a broad understanding of memory effects that appear in materials. 
\end{abstract}

\maketitle

\tableofcontents

%\epigraph{There are as many forms of memory as there are ways of perceiving, and every one of them is worth mining for inspiration}{Twyla Tharp, \emph{The Creative Habit}}
%\epigraph{It's the greatest curse that's ever been inflicted on the human race: memory}{Jedediah Leland, ``Citizen Kane''}
\bigskip
\emph{``There are as many forms of memory as there are ways of perceiving, and every one of them is worth mining for inspiration''} (Twyla Tharp, The Creative Habit).

\section{Introduction} 
Memories come in many forms and strike us in odd and seemingly unpredictable ways.  We experience this daily: we have long-term memories of our childhood, we have short-term memory of where we left our overcoat, we have muscle memory of how to walk or ride a bicycle, we have memories of smell.  The list goes on and is overwhelming in its variety.  While memory is acutely present in our consciousness, it is less well recognized as an organizing principle for studying the properties and dynamics of matter.  However, once we acknowledge this possibility, we realize that there are, as in our experience of consciousness, many forms of memory that are stored in untold numbers of ways in the matter surrounding us.  Some are obvious and mundane while others require a much greater degree of sophistication and ingenuity to encode or retrieve from a material.  In contrast, memory loss seems to be less problematic and we often take it for granted. However, these aspects of memory retention and loss are intrinsically linked; neither are simple processes.  

As Twyla Tharp, the dancer and choreographer, proclaimed in the epigraph above, each form of memory should be an ``inspiration'' for asking new questions. When we apply this dictum to memory in materials, it gives us a chance to examine the nature of far-from-equilibrium behavior in a new light.  As we will show, there are a great many ways in which materials can retain an imprint of their previous history that can be read out at a later time by following protocols that often are specialized to the type of information encoded.  While undeniably we take inspiration from the idea of memory in our world of consciousness, we will focus not on such phenomena but rather on its counterpart in the material world.  Our aim in this review is to indicate the great variety of issues that can be addressed in this context.  %As we will see throughout the discussion, only a few of them are coming into focus.   

In the physical world, we find many different forms of memory formation. 
We are taught early on to store certain memories by making pencil marks on a sheet of paper. 
In a more sophisticated fashion, media such as compact discs store information as binary markings.
In one form of computer memory, digitized information is stored  in the form of magnetic bits. 
On paper or a computer, one can store unlimited amounts of information by increasing the system size.
However, some forms of memory can encode only small amounts of information. 
For example, a common feature of glassy physics is that a system can retain an imprint of what was the largest strain that was applied to it either by  compression or shear in a specified direction.  
Similarly, %in the Kovacs effect the 
a system can store the time over which it has been subjected to an applied stress.
 
As these examples suggest, memory engages us in a study that targets phenomena related to transient or far-from-equilibrium behavior. 
A system that has not yet fully relaxed to equilibrium may retain memories of its creation while one that is in equilibrium has no memory of its past; the very process of reaching equilibrium erases memory of previous training. 
In the study of evolving systems, such as in geophysics and on an even grander scale, in cosmology and astrophysics, one uses information from the local terrain or the current state of the Universe to infer previous conditions. 

Disordered, out-of-equilibrium systems are often described by a vast rugged potential or free-energy landscape.  This allows a memory to be formed by falling into a recognizable state in this terrain.  Some memories that are encoded in this way include, among others, the Hopfield paradigm of associative memory and the memories caused by oscillatory shear of a jammed system. 
Similar considerations underlie the deep neural networks now used in machine learning~\cite{Pankaj18}, a subject which is beyond the scope of this review. 

Memories may be stored in a myriad of different systems: from solids to fluids; from paper to stone; from atomic positions to spin orientations; from chemical-reaction pathways to avalanches in transition dynamics. 
Memories can be encoded in time, as in the spacing between pulses in a spin echo; in position, as in the spacing of particles in a sheared suspension; in temperature as in the rejuvenation, aging and memory of glassy systems; or in chemical bonding, as in a chemically-controlled soup of colloids with designed inter-particle interactions. 
For each of these forms of information there are specific training protocols; some systems may need only a single training pulse while others may require repeated cyclic training before a signal can be read out reliably. This last is reminiscent of training by rote that many of us have experienced in school to learn the alphabet. 
Although many materials exhibit memory of past conditions by some form of history dependence, we will focus here on examples where a readout method also exists that recovers the encoded information with some fidelity; in the examples we consider there is a protocol for recovering as well as storing a specific input.

\begin{figure*}
    \begin{center}
        \includegraphics[width=7.0in]{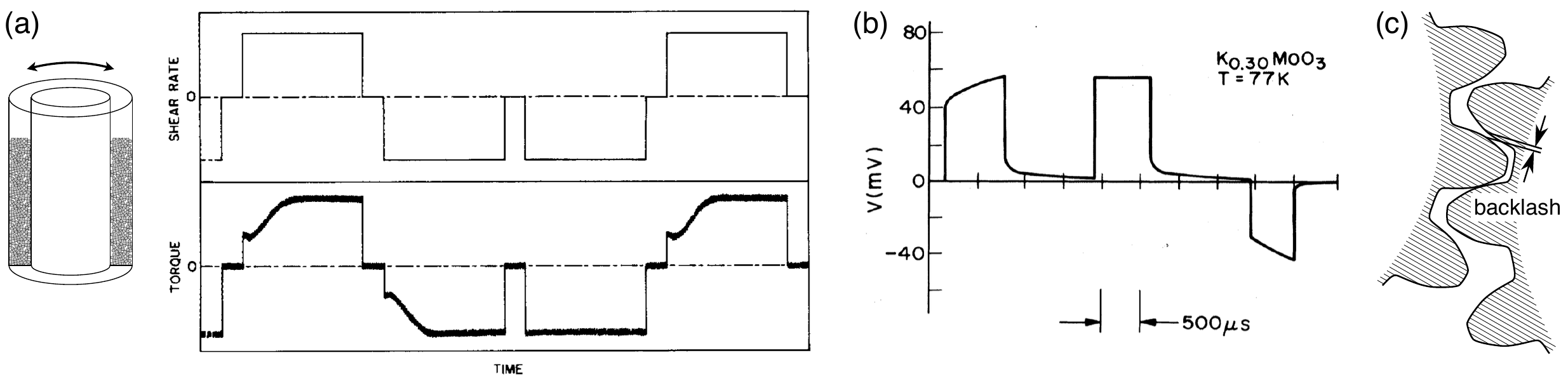}
    \end{center}
\caption{
    \textbf{Memory of a direction.} 
    \textbf{(a)} 
    Two concentric cylinders that are rotated with respect to one another may be used to measure the shear stresses in a suspension of neutrally-buoyant particles in a viscous liquid that fills the gap between the cylinders. 
    The top graph shows a series of rotations of the outer cylinder (counterclockwise, clockwise, clockwise, counterclockwise), with pauses between each rotation. 
    The bottom graph shows a schematic curve of the torque felt on the stationary inner cylinder; the presence (or absence) of a transient may be viewed as a memory of the previous shear direction. 
    %The same effect is observed in a dry granular material~\cite{Toiya04}. 
    \textbf{(b)} Pulse-sign memory in the traveling charge-density wave conductor K$_{0.30}$MoO$_3$. 
    If an applied current pulse has the same sign as the previous one, the voltage response lacks a strong transient. 
    \textbf{(c)} Backlash between two gears as a memory of a direction. 
    If the right gear is rotated clockwise, the left gear responds immediately. 
    If the right gear is rotated counterclockwise, there is a small lag. 
    %This effect can be viewed as a memory of the last direction of rotation. 
    % Adapted from: https://commons.wikimedia.org/wiki/File:Backlash.svg
    Plots in panel \textit{a} reproduced from~\citet{Gadala-Maria80} with permission; panel \textit{b} reproduced from~\citet{Fleming83} with permission; and panel \textit{c} adapted from David Richfield, Wikimedia Commons.
    \label{fig:direction}
}
\end{figure*}

Each form of memory is a stimulus for asking new questions and for examining the nature of far-from-equilibrium materials in a new light. The observation that memories can be stored in a seemingly countless number of ways raises the question of whether different kinds of memories share common principles. 
Other questions can be asked: 
What constitutes a memory? 
Are there different categories of memory and can they be enumerated? 
How many memories can be stored in a system---that is, what is the capacity of the memory storage? 
How fast can a memory be stored or retrieved? 
What is the entropy associated with a memory? 
What gives rise to plasticity, that is, the ability of a system to continue to store new memories? 
A study of these questions can be an entr\'{e}e for understanding the nature of the non-equilibrium world. 

As an invitation into this study, this article outlines a set of memory behaviors.  We describe a collection of distinct physical systems, and show how their responses may be considered as memories. 
The set of behaviors and systems described here are not meant to be exhaustive. In particular, we will not attempt to cover the vast range of technology associated with memory storage in computers, nor the fascinating array of memory effects in biological function such as in the immune system \cite{Osterholm12,Barton15}.
%such as immunological memory that is the basis of vaccines %JDP: commented out; potentially distracting \cite{Hammarlund03}
Rather, we intend to illustrate the breadth of memory phenomena in materials and, in the words of Twyla Tharp, to ``inspire'' new questions about, and new ways of classifying, material properties.
We will outline recent advances and raise open questions that may guide future work.  These will help to identify some of the issues that one must confront when trying to build a broader understanding of memories in materials. 
We hope that our perspective will help guide the beginnings of such a venture. 

\section{Simplest forms of memory: direction and magnitude}
\label{sec:simplest}

\subsection{Memory of a direction}\label{sec:Direction}
One of the simplest memory phenomenologies is when a material remembers the most recent direction in which it was driven. 
A well known application of this behavior is digital magnetic storage, where an external field puts individual magnetic regions in one of two polarities to represent a $1$ or a $0$. 
Yet, the same phenomenology occurs in other materials where it is not commonly associated with information storage, such as amorphous materials made of large particles, from suspensions of non-Brownian particles in a liquid (\emph{e.g.}, cement) to packings of dry grains. 

For example, this behavior has been studied in a Couette cell holding a sample of neutrally-buoyant hard spheres as illustrated in Fig.~\ref{fig:direction}a. 
When the outer cylinder \text{holding the sample} is rotated azimuthally, there will be a torque on the inner cylinder in the direction of shear. 
There is a transient period where the torque gradually evolves until it reaches a constant steady-state value as shown in the first portion of the response curve of Fig.~\ref{fig:direction}a~\cite{Gadala-Maria80}. 
If the rotation is stopped and then restarted in the \textit{same direction}, the torque immediately adopts the steady-state value with no transient. 
This behavior occurs because the dynamics are overdamped and inertia can be neglected completely, so the particles immediately stop moving when the shearing stops, and they immediately restart almost exactly where they left off when the shearing resumes. 
If the shearing direction is \textit{reversed}, however, the response once again undergoes a transient. This shows that the particle structure in the steady-state has become anisotropic due to the previously applied shear. 
One can thus detect the most recent shear direction by moving the inner cylinder in one direction and looking for the presence or absence of a transient in the torque response. 
Similar behavior has been seen in dry granular material~\cite{Toiya04}. 

The same gross phenomenology was discovered in the electrical response of charge-density wave conductors and has been called ``pulse-sign memory"~\cite{Gill81,Fleming83}. 
Figure~\ref{fig:direction}b shows the observed voltage in a sample of K$_{0.30}$MoO$_3$ when several current pulses were applied sequentially. 
The voltage has a transient response to the first pulse; the transient disappears for the second current pulse which is in the same direction as the first. 
When the direction of the current is reversed, the transient in the voltage reemerges. 
Thus, the response of the voltage depends on the direction of the last applied current pulse. 

In materials science, the Bauschinger effect also displays a memory of the last direction of driving~\cite{Bannantine90}. 
This effect refers to a phenomenon where the yield stress of a material decreases when the direction of working is reversed, and it occurs in polycrystalline metals and also amorphous materials~\cite{Karmakar10}. 
Although the details may certainly differ, the basic idea is the same as for a sheared suspension or granular material discussed above: shearing the material introduces anisotropy in its microstructure. 
In this case, plastic deformations encode a direction, which may be detected in subsequent measurements of the stiffness. 

The above examples deal with bulk materials that are inherently disordered, but the same phenomenology can be seen in an even simpler system: the coupling between two gears. 
Suppose the left gear in Fig.~\ref{fig:direction}c presents some resistance when it is driven either clockwise or counterclockwise by the right gear. 
If the right gear is turned in one direction, pauses, and then resumes in the same direction, resistance will immediately be felt when the driving resumes. 
However, if the rotation is reversed, there will be a small interval before contact is established with the left gear, due to the small gap between the gear teeth. 
This clearance, called ``backlash,'' is essential to prevent jamming of the gears. 
Its existence means that whenever we walk away from a simple gear box, we may leave behind a bit of information that is stored in the contacts between the gears. 

Our discussion of this simple form of memory has raised several themes that will appear again as we consider more complicated memory behaviors. 
One theme is that similar memory phenomena can occur in systems that seem very different, such as a slurry or a charge-density wave. 
Some memories also have a counterpart at the macro-scale that is material independent, as in the example of the gears given above.  We will show several more examples of such phenomena throughout the text.
These observations may prompt us to ask how deep the connections are among such systems: when does similar memory behavior imply similar underlying physics?  

There is also a distinction that can be made: in some systems memories persist for extremely long times, whereas in others the memories are constantly fading and must be continually reinstated to preserve them.
In a colloidal suspension, the memory of the previous shear direction is eventually lost as the particles diffuse, losing their positional correlations. 
In contrast, in a granular material, because the energy for any particle rearrangement is much larger than thermal energy, the memory will remain until the system is disturbed. 
In computers too, there is so-called ``volatile memory,'' which refers to a device that only retains data when provided with power, in contrast to non-volatile memory such as magnetic storage, optical discs, or punched cards, which do not change their state when left alone. 

Finally, the ability of many disordered materials to remember a direction highlights another theme: memories can have important practical consequences.
When performing rheological characterization, sample preparation must be done carefully so as to avoid influencing the measurements because of the material's history. 
This memory is one facet of the complex history-dependence displayed by emulsions, slurries, and dry grains, which can complicate their industrial handling and transport, since their effective bulk properties may depend on what was previously done to them~\cite{Kim17,Jaeger96,Lasanta18}. 

\subsection{Memory of largest input: Kaiser and Mullins effects}\label{sec:Kaiser}
We have just discussed how a material may remember one aspect of its most recent driving. 
Slightly more complex is the ability of many physical systems to retain a memory of the maximum value of any previously applied perturbation. 
For memories of this type, when the system has been trained by application of an input of a given magnitude, it shows reversible behavior as long as the input is kept below that initial training magnitude; in this range, varying the input leads to reproducible behavior.  However, if that input value is exceeded, the system evolves to a new state so that the system now displays a new response curve.  This curve is itself reproducible as long as the input does not exceed the new maximum value.

One example, which is particularly easy to visualize, is a very thin crumpled sheet confined by a piston of mass, $m$, under gravity in a vertical cylindrical tube~\cite{Matan02}.  As the weight compresses the sheet, the height of the piston, $h$, decreases logarithmically in time. The training of the system is accomplished using a mass $m=M_1$. Because it is not feasible to wait until all relaxation has stopped, the training is done for a fixed initialization time. 
Once the training is finished, the mass $m=M_1$ is removed, and the height $h(m)$ is measured for different values of mass $m$, where for each measurement the fixed waiting time is much smaller than the initialization time. 

\begin{figure}[t]
    \begin{center}
        \includegraphics[width=3.4in]{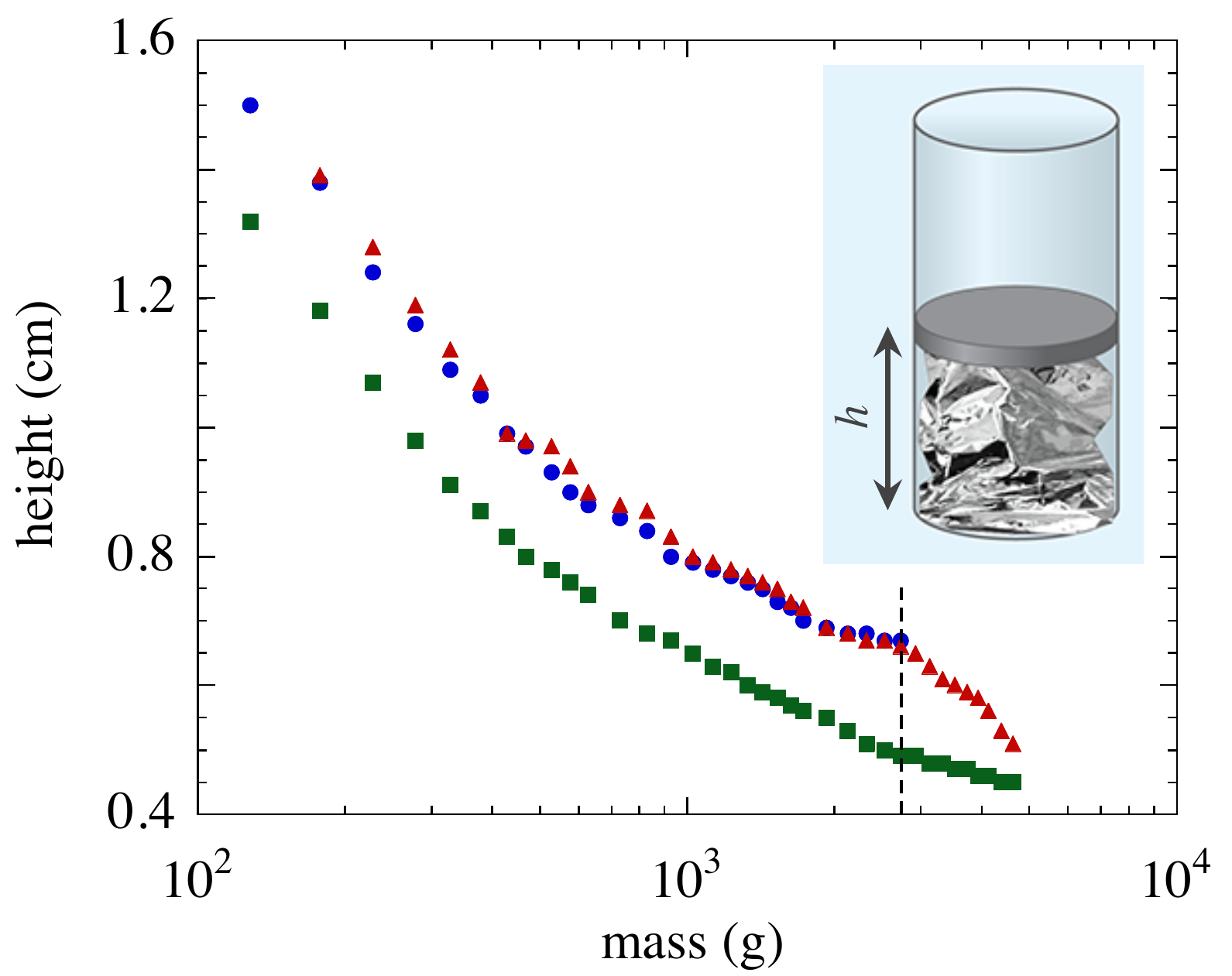}
    \end{center}
\caption{
    \textbf{Kaiser effect.} 
    Height of a piston compressing a mylar sheet following the protocol described in the text. The inset shows the mylar sheet compressed within a cylindrical tube by a piston of mass $m$.  The piston height, $h(m)$ is reproducible (shown by blue circles for diminishing mass and by red triangles for increasing mass) as long as the maximum mass does not exceed the previously applied maximum value of $m=M_1=$ $2.6$~kg indicated by the dashed line. When a larger mass (red triangles above $2.6$~kg) is placed on the piston, the curve changes; green square data points show the reproducible curve after training with $m=M_2=5$~kg. Plot adapted from~\citet{Matan02} with permission; inset: APS/Carin Cain. \label{fig:Kaiser} 
}
\end{figure}

As long as $m < M_1$, the measurements are reproducible as shown for a Mylar sheet in Fig.~\ref{fig:Kaiser}; the same values of $h(m)$ are obtained both on increasing (red triangles) and decreasing (blue disks) $m$.  However, once the training mass has been exceeded so that the mass on the piston is $M_2 > M_1$, the situation changes; the curve $h(m)$ is extended to higher values of $m$ and appears to have a different dependence above $M_1$ than it did below that value.  This is shown by the red triangles above $m=2.6$~kg in the figure.  Now, when the height is measured again starting at low values of $m$, $h(m)$ no longer follows the original curve but drops to a lower value.  

If the crumpled sheet is now trained at $M_2$ (either by allowing it to sit under this mass for a long time or by cycling the mass multiple times up to $m=M_2$), then a new reproducible curve is found for $h(m<M_2)$ as shown by the green squares in the figure.  The crumpled sheet has a memory of the largest weight to which it has been subjected.  When the system is pushed into a new regime into which it had never been previously exposed, it changes irreversibly.

The case of the crumpled sheet is not the earliest example of this rather ubiquitous effect.  It has been observed in many other systems and goes by different names depending on the material and the measurement. 
The Kaiser Effect was originally observed in the acoustic emission of a metal under strain~\cite{Kaiser50}; the acoustic emission of the sample vanishes if the applied stress is smaller than the previously applied maximum value.  The material thus retains a memory of the largest strain to which it was subjected.  Similar behavior is seen in other materials such as rock~\cite{Kurita79} where acoustic emission is a harbinger of material failure. Another close analogy is the reversible and irreversible compaction of soil, where reading out the memory of maximum load (``overconsolidation'') can be crucial in predicting how a new building will settle \cite{Budhu10}. 

\begin{figure}
    \begin{center}
        \includegraphics[width=3.0in]{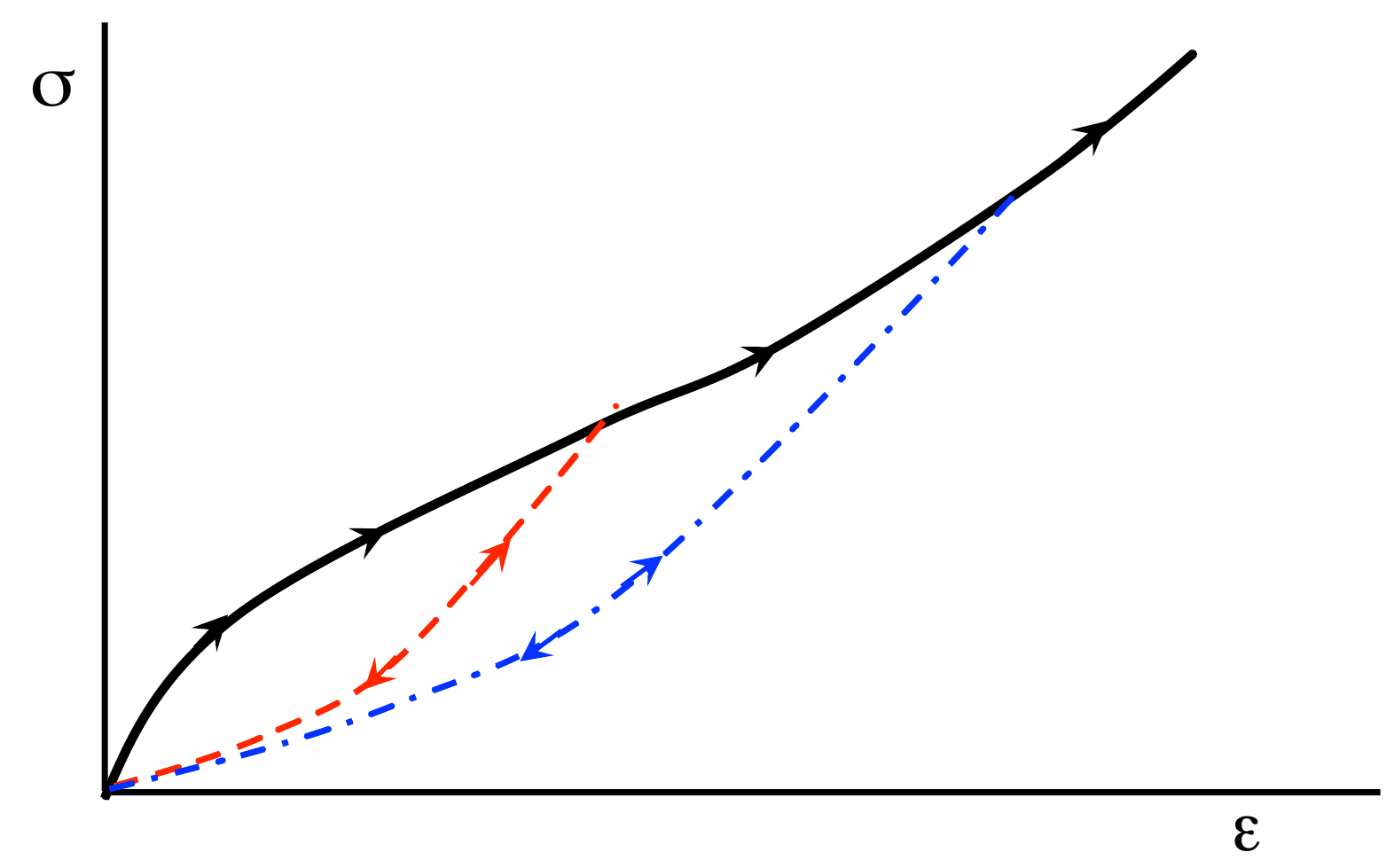}
    \end{center}
\caption{
    \textbf{Mullins effect.} 
    Schematic stress-strain curve for the Mullins effect in rubber adapted from~\citet{Cantournet09}. 
    The pristine loading curve (black full line) is smooth.  
    After partial loading, when the stress, $\sigma$, is removed the response changes and the unloading curves (red dashed and blue dash-dot curves) do not follow the original loading. 
    On reloading, the response is reversible up to where the previous largest stress had been applied. 
    At that point there is a kink where the curve rejoins the original pristine loading curve.
    \label{fig:Mullins}
}
\end{figure}

Another example of this type of behavior is the Mullins Effect, which occurs in rubber after it has been stretched~\cite{Mullins48,Diani09}.  A schematic stress-strain curve is shown in Fig.~\ref{fig:Mullins}. The black (full) line shows the pristine loading curve that occurs on the first application of stress.  When the stress, $\sigma$, is removed, the curve does not retrace the %original 
pristine loading curve but drops more rapidly as shown by the red (dashed) curve. This new response %(red dashed curve) 
is reversible on reloading up to the point where originally the stress was removed.  At that point there is a kink where it rejoins the pristine loading curve.  When the stress is increased further, the strain increases as in the original pristine behavior. This unloading/loading procedure can be repeated at different values of the stress as shown by the %red (dashed) and the 
blue (dash-dot) curve.  The Mullins effect, as in the Kaiser effect and in the crumpled sheet, %has
demonstrates a memory of the largest stress that had been applied.

\subsection{Memory of a duration: Kovacs effect} 
A different type of memory where a single input value is remembered is the Kovacs effect~\cite{Kovacs63,Mossa2004d,Bouchbinder2010}.  Originally observed in polymer glasses, the time-dependent evolution of a glassy system is observed to depend sensitively on its thermal history. 
In the conventional protocol, a sample
is cooled and allowed to relax 
for some duration at a low temperature, and then warmed up to a higher temperature. 
The subsequent evolution of the sample, observed in quantities such as volume, can be non-monotonic, exhibiting a peak at a time that depends on the duration spent at the lower temperature \cite{Volkert89,Bertin03,Cugliandolo04}. %, with a form that depends on the time spent at the lower temperature.
Although the relationship between the waiting time and the peak time is not always emphasized, one may view the response as a memory of the duration of the aging at the low temperature.

This aspect of the Kovacs effect has a simple phenomenology (\emph{i.e.}, a single value is remembered), but the mechanism is more subtle than the memories presented in sections~\ref{sec:Direction} and ~\ref{sec:Kaiser}.  
Because it also has some features in common with echoes, we wait until Sec.~\ref{sec:Kovacs} to describe the Kovacs effect in more detail.

\section{Hysteresis and return-point memory}
\label{sec:rpm}

In Sec.~\ref{sec:simplest}, we considered a system that remembers the most recent direction of driving, and we modeled it as being in one of two states. We can build on this simple kind of hysteresis by considering a system that responds to a scalar field $H$---perhaps from an electrical current in a magnetic coil---and can be in a ``$+1$'' or ``$-1$'' state. 
The two values $H^+$ and $H^-$ are properties of the system that specify when it switches states. 
Conventionally, $H^+ > H^-$ for a dissipative system; in this case the system is always in the $+1$ state when $H > H^+$, and always in the $-1$ state when $H < H^-$, but in between, the state depends on the driving history. 
(We assume the driving is quasistatic, meaning $H$ is varied much more slowly than the system's response.)
These systems are basic elements of hysteresis, sometimes termed ``hysterons.''

One conceptually simple and important application of such hysterons is in digital memory storage.
If $H^+ > 0$ and $H^- < 0$, the hysteron retains its state even after the field is removed and in the presence of noise, thereby storing a single bit of information indefinitely. 
In the case of magnetism the state sets the direction of a magnetic field made by the hysteron itself; a nearby probe can read the bit.
This models a building block of digital magnetic memory---the hard disks and tapes that currently store most of the world's data \cite{Jiles16}.

\subsection{Single return-point memories}
\label{sec:rpmsingle}

\begin{figure}
    \begin{center}
        \includegraphics[width=2.75in]{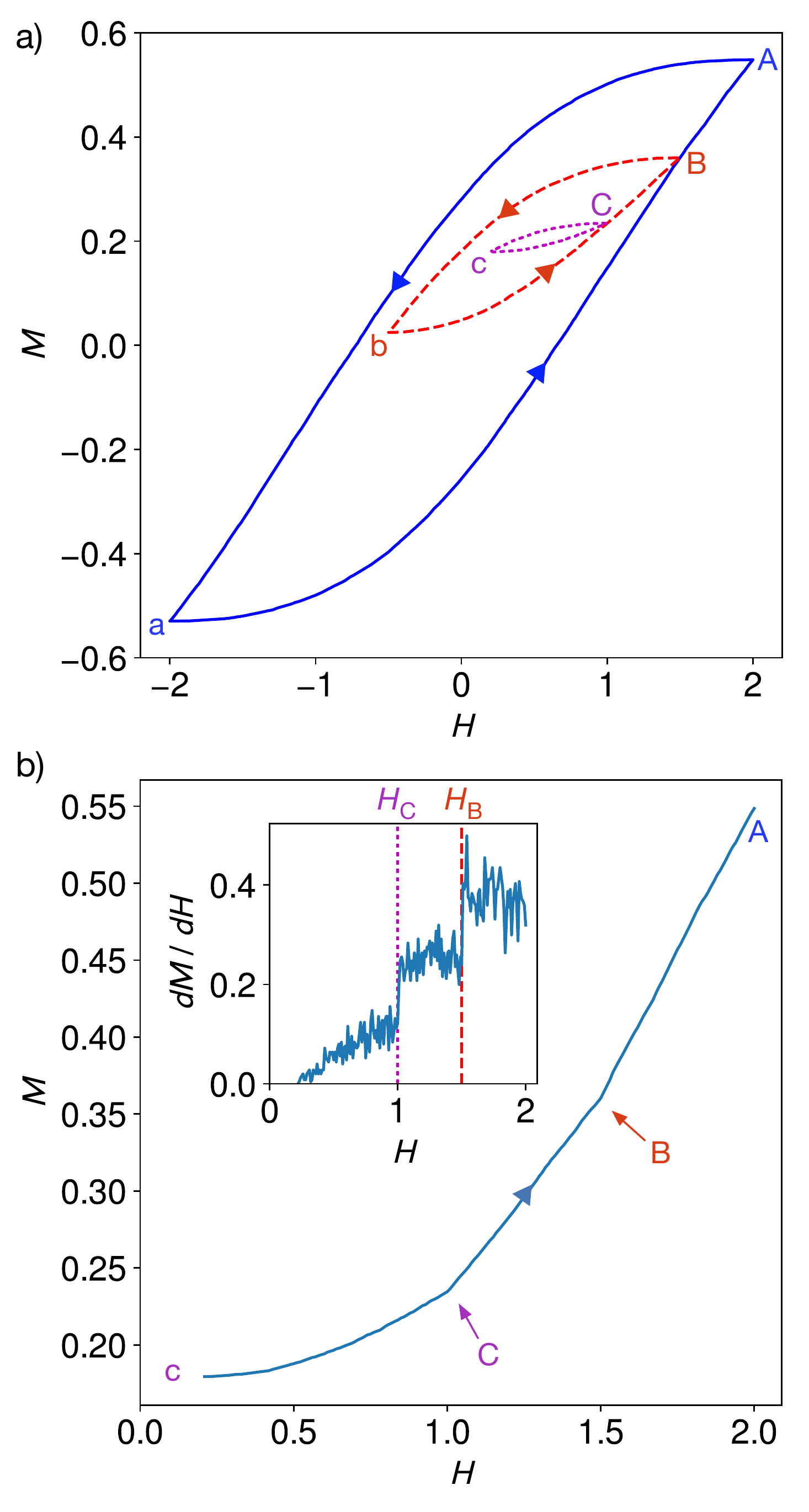}
    \end{center}
\caption{
\textbf{Return-point memory.}
    \textbf{(a)} Magnetization of a simulated ferromagnet model \cite{Preisach35} as the applied magnetic field $H$ is varied. The sequence $a \to B \to b$ creates a memory at $H = H_B$. As long as $H_a \le H \le H_B$, returning to $H_B$ will restore the system to the same state $B$ every time, regardless of intervening events (such as the excursion to $b$). The sequence $b \to C \to c$ creates a new sub-loop and encodes a second memory at $H = H_C$. 
    \textbf{(b)} Detail of trajectory from panel (a), showing signatures of the two nested memories as $H$ is swept from $H_c$ to $H_A$.  
Each time a sub-loop is exited (points $C$ and $B$), its history is erased and the slope of the curve changes. \textit{Inset:} Slope $dM/dH$ during readout. The jumps at $H_C$ and $H_B$ indicate memories. For this figure $5\times 10^4$ hysterons were simulated, with a Gaussian distribution of $H^+_i$ and $H^-_i$ with standard deviation 1.8, and selected so that $H^+_i > H^-_i$.
    \label{fig:rpm}
}
\end{figure}

When they make up a larger system, these hysterons can give rise to a rich behavior called 
return-point memory, 
which describes the system's ability to recall a previous state when $H$ is returned to a previous value \cite{Barker83,Sethna93}. 
To illustrate this behavior we use the model of a ferromagnet described by \citet{Preisach35}, in which each hysteron represents a magnetic domain that is coupled only to the applied magnetic field $H$, and there is a distribution of $H^+$ and $H^-$ to represent the material's disorder. 
Figure~\ref{fig:rpm}a shows return-point memory in the model's magnetization $M$ (the average state of the hysterons) versus $H$, where $H$ was varied in the directions shown by the arrows.
The switchbacks and loops on the plot reveal the hysteresis of the system's many components.
As we follow the evolution along some trajectory, such as the one through the labeled states $a \to B \to b \to B$, we 
can define an interval of time that starts when we first reach state $B$, and lasts as long as $H_a \le H \le H_B$, where $H_a$ and $H_B$ are the magnetic field for states $a$ and $B$, respectively.
Return-point memory means that during this interval, returning to $H = H_B$ will always restore the system to state $B$. By contrast, when 
$H < H_B$, the state depends on the history of $H$ since $H_B$ was last visited. Returning to $H_B$ erases this history.

There is an important side effect of the erasure of history at $H = H_B$: as we continue past $H_B$ toward $H_A$ there is a transition from behavior that depended on recent history (\emph{i.e.}\ the excursion to $b$) to behavior that does not. Here, we exit the sub-loop delimited by states $(b, B)$ and rejoin the older outer loop delimited by states $(a, A)$. There is a subtle change in slope at point $B$ where we first switched from increasing $H$ to decreasing it. %; it indicates that we stored a memory at $H_B$ when we first began the sub-loop, by switching from increasing $H$ to decreasing it. 
Analogous to the Mullins effect, discussed in Sec.~\ref{sec:simplest} and illustrated in Fig.~\ref{fig:Mullins}, we cannot see this signature of the memory without overwriting it.

\subsection{Multiple memories through nested hysteresis loops}
\label{sec:rpms}
Simple memories of one or two quantities, like the Mullins and Kaiser effects, seem exceptional given the disorder and many degrees of freedom within most non-equilibrium matter. 
Return-point memory gives us our first example of multiple memories.
We can see this by applying the definition of return-point memory recursively. In Fig.~\ref{fig:rpm}a, at point $C$ we again change from increasing $H$ to increasing it. This begins a time interval in which $H$ is bounded by $H_b$ and $H_C$. 
We go on to traverse a sub-loop delimited by $(c, C)$ that is nested inside $(b, B)$. 
This nested structure means that if we wish to encode a given set of $H$ values, there is only one ordering of the values that works \citep{Middleton92,Sethna93}.
To read out the memories, we sweep $H$ continuously toward $H_A$ and look for changes in the slope $dM/dH$. Figure~\ref{fig:rpm}b shows a close-up of the trajectory we would follow, and its derivative, which reveal the memories of our reversals at $H_C$ and $H_B$. 
(Note that we could have instead chosen to sweep the field in the other direction toward $H_a$ and observed memories of our reversals at $H_c$ and $H_b$.)
While we are accustomed to the readout of magnetic memory by observing only the present value of the local magnetization (as in audio tape or a computer hard disk), this method lets us recover the
value
of the applied field that formed each memory in the first place~\cite{Perkovic97}.

\subsection{Generality}

Many types of matter can be modeled as collections of subsystems that are individually hysteretic. It should therefore come as no surprise that return-point memory is observed in a 
wide range of systems beyond ferromagnets, from spin ice 
\cite{Deutsch04} and high-temperature superconductors \cite{Panagopoulos06}, to adsorption of gases on surfaces \citep{Emmett47}, to solids with shape memory \cite{Ortin91} (not to be confused with the shape-memory effect itself, discussed in Sec.~\ref{sec:shape}).
Interactions within these systems can give rise to cooperative and even critical phenomena such as avalanches, so that the actual dynamics are usually dramatically different from the Preisach model we have just examined~\cite{Sethna01}.
Nonetheless, \citet{Sethna93} proved that these systems will have return-point memory as long as the interactions are ferromagnetic (\emph{i.e.}\ a hysteron flipping to the +1 state encourages others to do so). 
Even when this condition is violated, return-point memory can still hold approximately or under certain additional conditions~\cite{Deutsch03, Deutsch04, Hovorka08, Gilbert15}; \citet{Mungan18} have begun work toward a general framework for describing these behaviors. We will revisit these possibilities in Sec.~\ref{sec:jammed}.

We can also demonstrate the generality of return-point memory by finding it in a perhaps unexpected context: the backlash between gears from Sec.~\ref{sec:simplest} (Fig.~\ref{fig:direction}c). Consider a long train of $N$ gears, each one driving the next, with a large amount of backlash between each pair. We can turn only the first gear, and we neglect inertia so that the $n$th gear moves only when turned by gear $n-1$.
After a long period of forward rotation, we turn the first gear backward, but only enough to overcome the backlash among the first $n$ gears. Those $n$ gears begin to reverse, but gear $n+1$ is left disengaged by some amount --- a fraction of the total available backlash. 
Finally, the first gear is once again rotated forward. As it reaches the position where it was originally reversed, the entire system returns to its previous state, fulfilling return-point memory. 

Just as in the other systems we can encode multiple memories by repeatedly reversing the direction of the first gear. Each time we reverse, we rotate by a smaller amount so that we progressively manipulate fewer and fewer gears. For each reversal, a pair of gears is left disengaged by some fractional amount we choose---we can place a distinct memory in each of the $N-1$ couplings. This same principle is used in a single-dial combination lock to store multiple values from the history of a single input. We can then read out half the memories without disassembling the gearbox using the same method as in Fig.~\ref{fig:rpm}b: we turn the first gear unidirectionally and note the positions at which the torque abruptly increases as each gear is engaged in sequence---a sensation familiar to anyone who has reset a combination lock dial after using it.

\section{Memories from cyclic driving}\label{sec:cyclic}

The memories discussed so far may be written by applying a deformation or changing a field just once. But repeated \emph{cyclic} driving is also ubiquitous: buildings and bridges are repeatedly loaded and unloaded, temperatures change between day and night, and we practice a skill repeatedly in the hope of learning it. These forms of driving may create memories.

Driving that lasts for multiple cycles may also be used to store multiple values, by varying its parameters (\emph{e.g.}\ strain amplitude) from cycle to cycle.
In the previous section we encountered one way that a system can remember multiple values of a single variable. %, but it is not unique.
By considering the case of cyclic driving, we will be able to talk about these various behaviors using similar language. 

\begin{figure}
    \begin{center}
        \includegraphics[width=3.3in]{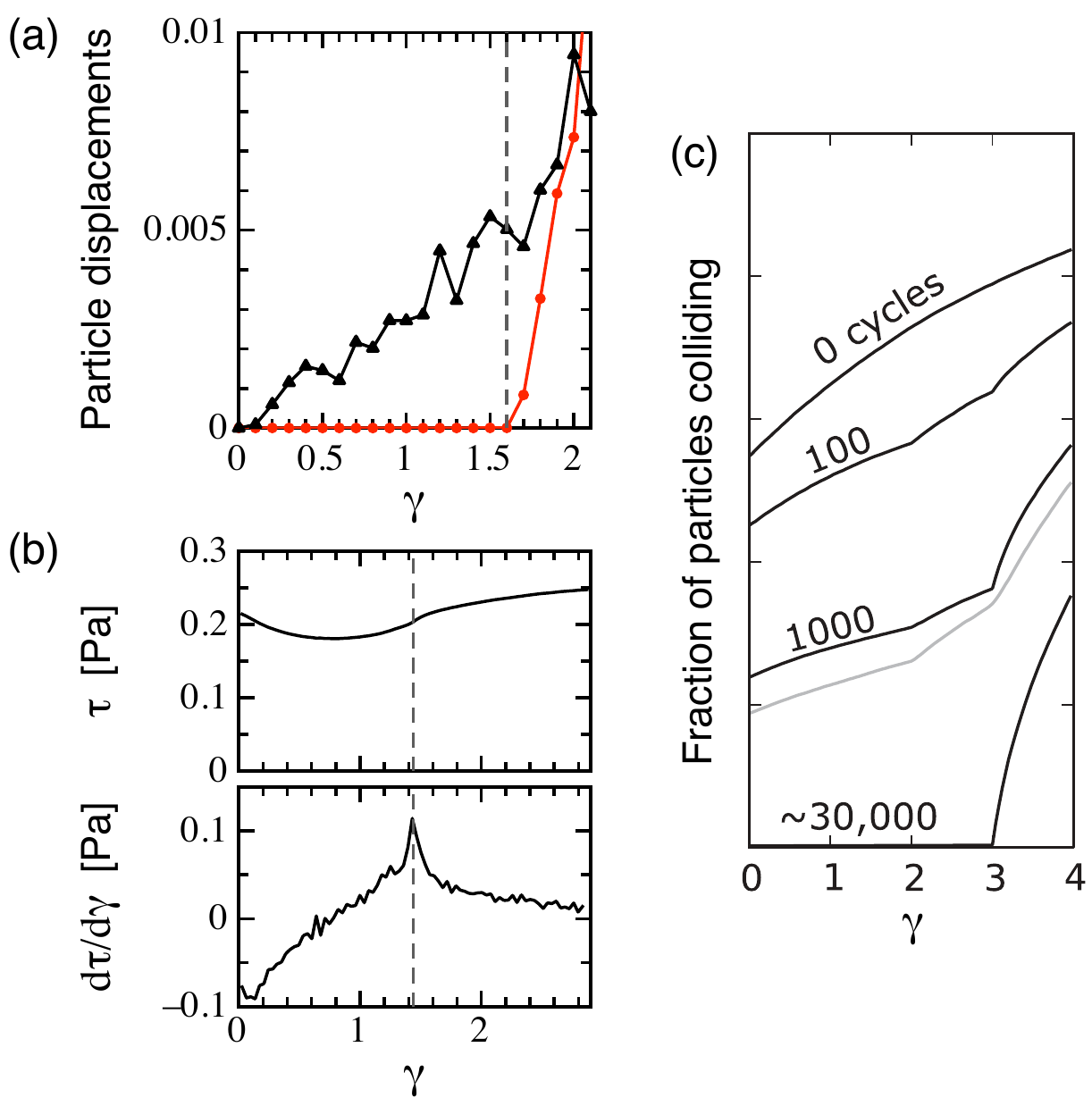}
    \end{center}
\caption{
\textbf{Memories from cyclic driving.}
    \textbf{(a)} To read out memory in a sheared suspension, cycles are applied with successively larger strain amplitude (horizontal axis). After each cycle, each particle's position along the direction of shear is compared with its position at the start of the cycle (vertical axis). An untrained system (black triangles) shows no memory, while a system trained for many cycles with amplitude $\gamma_0 = 1.6$ (red circles) shows a memory.
    \textbf{(b)} After 10 cycles with $\gamma_0 = 1.44$, measuring shear stress $\tau$ vs.\ strain $\gamma$, and its derivative $d\tau/d\gamma$, shows a clear memory in the mechanical response.  
    \textbf{(c)} Multiple transient memories: In a simulation of the sheared suspension, memories are read out by monitoring collisions as strain amplitude is increased, analogous to panel \textit{a}. Curves are labeled with the number of training cycles applied. Training with amplitudes 2.0 and 3.0 is evident after just 100 cycles, but eventually a steady state is reached with only one memory. The shaded curve shows the result after many cycles when noise is added. 
    Panels \textit{a} and \textit{b} adapted from \citet{Paulsen14} with permission; panel \textit{c} adapted from \citet{Keim11} with permission.
    }
    \label{fig:mtm}
\end{figure}

%    \comm{=== John Sharpe (Cal Poly) ===
%    Typo in multiple transient memory figure: Strain amplitude in experiment is 1.44, not 0.44.}

\subsection{Memory of an amplitude}
When some systems are driven repeatedly, \textit{e.g.}, by shear, electrical pulses or temperature, they eventually reach a steady state in which the system is left virtually unchanged by further repetitions. This behavior is astonishingly common among non-equilibrium systems, including granular matter~\citep{Toiya04,Mueggenburg05,Ren13}; crystalline~\citep{Laurson12} and amorphous solids made of colloids \citep{Haw98, Petekidis02}, bubbles \cite{Lundberg08}, or molecules \cite{Packard10};
colloidal suspensions \citep{Ackerson88,Corte08} and gels \citep{Lee08}; liquid crystals \citep{Sircar10}; vortices in superconductors \citep{Mangan08}; charge-density wave conductors \cite{Fleming86,Brown86}; and even crumpled sheets of plastic \cite{Lahini17}.
Often there is also an amplitude past which the system cannot reach a steady state.

It is natural to suspect that the steady state contains a memory of the driving that formed it over many cycles. An early and important example is a charge-density wave conductor \cite{Thorne96}, which when subjected to many identical voltage pulses, comes to ``anticipate'' the end of each pulse with a rush of current \cite{Fleming86,Brown86,Coppersmith87b}. 

An accessible recently-studied example is a suspension of particles in a liquid when inertia and Brownian motion are negligible~\citep{Corte08,Keim11,Paulsen14}. 
For pure Stokes flow, cyclically shearing such a suspension back and forth will return each particle exactly to its starting point. However, pairs of particles that come too close to each other during shearing may touch and change their trajectories irreversibly~\citep{Pine05,Popova07,Corte08,Pham15}. 
Over many cycles of shearing between strain $\gamma = 0$ and a constant amplitude $\gamma_0$, the particles may move into new positions where they no longer disturb each other. When viewed stroboscopically (once per cycle), the system stops changing. But this steady state persists only if strain stays between the extrema  encountered so far (0 and $\gamma_0$). To read out the memory of $\gamma_0$, we begin with a cycle of smaller strain amplitude, which does not change the system, and then apply cycles with larger and larger amplitudes until a change is first observed as shown in Fig.~\ref{fig:mtm}a \citep{Keim11,Paulsen14}. Just past $\gamma_0$, many pairs of particles that have been ``swept out'' of the interval $[0, \gamma_0]$ promptly come into contact~\cite{Keim13b}. Experiments by \citet{Paulsen14} also show an immediate, sharp change in mechanical response at $\gamma_0$ as shown in Fig.~\ref{fig:mtm}b, similar to how the Mullins effect is read out.

Memories formed over many cycles of driving are sometimes termed ``self-organized''~\citep{Coppersmith87a,Tang87,Coppersmith87b,Coppersmith97}.
This refers to the somewhat efficient way that the system evolves its many degrees of freedom to conform to the driving. If a charge-density wave conductor, suspension, or granular packing simply explored new states at random each time it was driven, it might never find a steady state. Instead, the evolution is regulated and directed toward the ``goal'' of conforming to the driving. For example, in suspensions, driving does not change the entire system uniformly, but instead disrupts only the regions where particle positions are inconsistent with a steady state~\citep{Corte08,Keim11}, and leaves other regions untouched. Because the driving regulates evolution, rather than just activating it, such systems seem predisposed to form specific memories.

\subsection{Multiple transient memories}

When the driving is varied from one cycle to the next, some of these systems are known to retain memories of multiple values, but only \emph{before} the transient self-organization has finished. This behavior, called \emph{multiple transient memories}, was first seen in charge-density waves when the duration of the pulses was varied \citep{Coppersmith97}. The principles are easier to illustrate for non-Brownian suspensions. If successive cycles of shear begin at strain $\gamma = 0$ and alternate between amplitudes $\gamma_1$ and $\gamma_2 > \gamma_1$, particle collisions will be reduced for strains in the interval from $0$ to $\gamma_1$ more rapidly than from $\gamma_1$ to $\gamma_2$ --- the system will respond differently in each interval. This has been observed in simulations and experiments~\citep{Keim11,Paulsen14}, as illustrated in Fig.~\ref{fig:mtm}c. In principle, for large enough systems, we can store arbitrarily many memories in this way. There is no restriction on the order in which we apply these amplitudes; this is in contrast to the case of return-point memories. However, once self-organization is completed for $[0, \gamma_2]$, the system's response is uniform (no collisions) in this interval. There is no way to read or write the memory of $\gamma_1$, and only the memories at 0 and $\gamma_2$ remain. 
Thus at long times and without noise, systems with multiple transient memories recover the Mullins- or Kaiser-effect memory of the extrema of driving. 

Remarkably, this long-term memory loss can be avoided. \emph{Adding} noise to the charge-density wave and suspension systems~\cite{Povinelli99,Paulsen14}
has the effect of prolonging the transient indefinitely, preserving the ability to retain multiple memories and allowing the memory content to evolve as inputs change.  This is a concrete example of ``memory plasticity'' whereby a system has the ability to  continue storing new memories.  Non-Brownian suspensions also have a critical strain amplitude above which complete self-organization is impossible~\cite{Corte08}; driving the system just above this amplitude has the same memory-enhancing effect~\cite{Keim13b}.

\begin{figure}
    \begin{center}
        \includegraphics[width=3.4in]{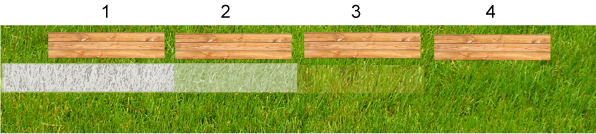}
    \end{center}
    \caption{\textbf{Multiple transient memories in a row of benches on a lawn}. The schematic shows the park viewed from above; the park entrance is at the left, and lighter colors indicate worn grass. Even though there is only one path, the grass heights encode the memory that benches 2 and 3 were visited. If this pattern of visits continues without the grass regrowing, eventually only the largest memory (bench 3) will remain. See~\citet{Paulsen18}.
    \label{fig:benches}
}
\end{figure} 

The examples of charge-density waves and non-Brownian suspensions show that two very different systems, one electronic and one fluid, can show this same type of memory.  It must therefore be a canonical form of encoding inputs. 
We describe a third physical example (albeit at a macroscopic scale) that drives this point home.  Consider the lawn of a small park, with several benches arranged in a row as shown in Fig.~\ref{fig:benches}. Visitors can only enter the park at the left. Some weeks after the park opens, there is a narrow path worn into the grass. 
The heights of the grass along this path encode which benches were visited, as illustrated by the shading in the figure.  If the grass does not regrow, eventually the path to the farthest bench that receives visitors will be worn bare, erasing the memory of all the other benches' popularity. This is similar to the charge-density waves, or the non-Brownian suspensions without noise.  However, if the grass does regrow at an appreciable rate, the system can reach a steady state that encodes the full set of memories --- equivalent to the effect of noise in the other systems~\cite{Paulsen18}.  Thus all three physical examples show the same form of memory.

\subsection{Cyclic memory in jammed and glassy systems}
\label{sec:jammed}

Multiple transient memories in a suspension were associated with particles moving to positions where they no longer made any contact with their neighbors during a shear cycle. 
As the particle density is increased, the system may no longer be able to find such positions. Thus, one might expect that a steady state with memory of a shear-cycle amplitude could no longer be formed. However, even in this higher-density regime, one can still observe memory formation with a surprisingly similar phenomenology to that of the sheared suspensions, but with crucial differences.

At sufficiently high density, the system undergoes a dramatic transition: it jams and becomes rigid \cite{Liu10,VanHecke10,Cubuk17}. In the jammed state, contacts endure throughout the oscillation cycle, except for sporadic and abrupt contact changes; the system traverses a rugged potential-energy landscape and visits different distinct energy minima.
Yet under cyclic shear, athermal jammed systems can  still reach a steady state in which subsequent cycles leave the system unchanged \citep{Hebraud97,Petekidis02}.
As in dilute suspensions the steady state encodes a memory, and a suitable readout protocol can recover the strain amplitude \citep{Fiocco14}. But the character of this memory is different---in dilute suspensions, the steady-state quasi-static motion was fully reversible, with particles following the same paths in forward and reverse directions in each cycle; in jammed systems the motion is periodic but \emph{not} reversible, so that the particles trace different paths in the forward and return parts of the cycle. Thus, each particle traverses a \emph{loop} so that, at the end of a cycle, it returns to the identical position it had at the start of that cycle \cite{Slotterback12,Keim14,Nagamanasa14}.  
In some cases periodic states are found where it takes \emph{multiple cycles} to return to a previous configuration~\cite{Regev13,Royer15,Lavrentovich17,Mungan19}. 

\begin{figure}
    \begin{center}
        \includegraphics[width=2.5in]{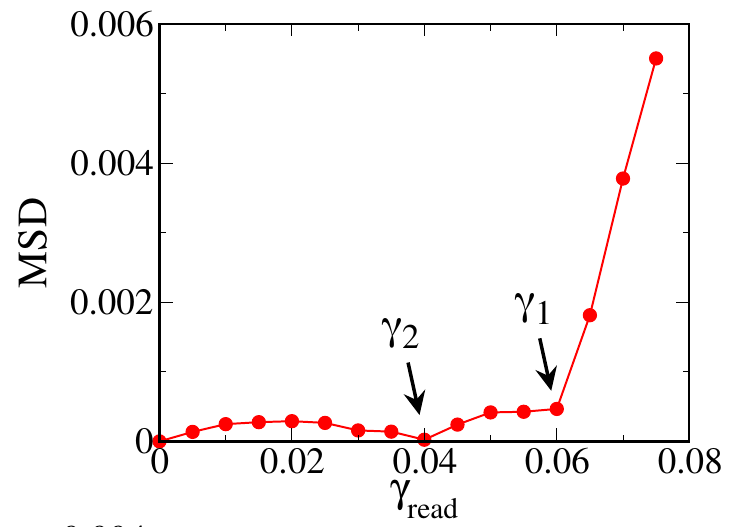}
    \end{center}
    \caption{
    \textbf{Multiple memories in a jammed solid.}
    Readout of memories of two strain amplitudes in a simulation of a jammed, amorphous solid, adapted from \citet{Adhikari18} with permission.
The system was trained by alternating between strain amplitudes $\gamma_1=0.06$ and $\gamma_2=0.04$, for 30 repetitions of the pattern. Readout involves applying cycles of increasing strain amplitude $\gamma_\text{read}$, starting from zero, and comparing the state of the system after each cycle to the state after training. Here mean-squared displacement (MSD) of particles is used to measure differences. The system returns to the same state when $\gamma_\text{read}=\gamma_2$, and shows a sharp change in behavior past $\gamma_\text{read}=\gamma_1$.   
    \label{fig:jammed-readout}
}
\end{figure} 

Simulations of these systems \cite{Fiocco14,Adhikari18} and  experiments \cite{Sood18, Keim18} show that multiple strain amplitudes can be remembered from past shearing, as illustrated in Fig.~\ref{fig:jammed-readout}. But this memory seems distinctly different from the multiple transient memories in the dilute case. These systems never lose their capacity for multiple memories, but the order in which memories were encoded is important --- traits reminiscent of return-point memory (Sec.~\ref{sec:rpms}). Indeed, microscopic observations and further modeling show that the steady-state behavior is approximately consistent with return-point memory \cite{Keim18}.

Systems at densities just \emph{below} jamming can show similar behavior \cite{Schreck13,Adhikari18}, but that behavior appears to be much less stable at finite temperature and strain rate than in jammed systems.
Continuing a theme of this review, some superficially very different systems also share this behavior; similar results were found in simulations of a system of magnetic spins \cite{Fiocco14} and in an abstract model of driven state transitions \cite{Fiocco15}. 
These examples, and a form of spin ice recently studied in experiments \cite{Gilbert15}, seem to represent an extension of return-point memory in which a system's disorder must be ``trained'' for multiple cycles before it exhibits the return-point behavior~\cite{Mungan18}.

We note that in systems with ``native'' return-point memory that need no training, the disorder is quenched: the hysterons, their interactions, and their coupling to an external field cannot ordinarily change. In contrast, these properties are generally not as stable in 
systems that require extended training to learn one or more inputs;
there must be some mechanism by which quenched disorder emerges~\cite{Perez-Reche16}.
Finally, the systems we have just discussed are dominated by disorder, but the presence of limit cycles, avalanches, and other nonequilibrium phenomena in crystals hints that 
those solids may also harbor rich cyclic memory behaviors~\cite{Laurson12,Sethna17,Perez-Reche16}.

\begin{figure*}
    \begin{center}
        \includegraphics[width=5.5in]{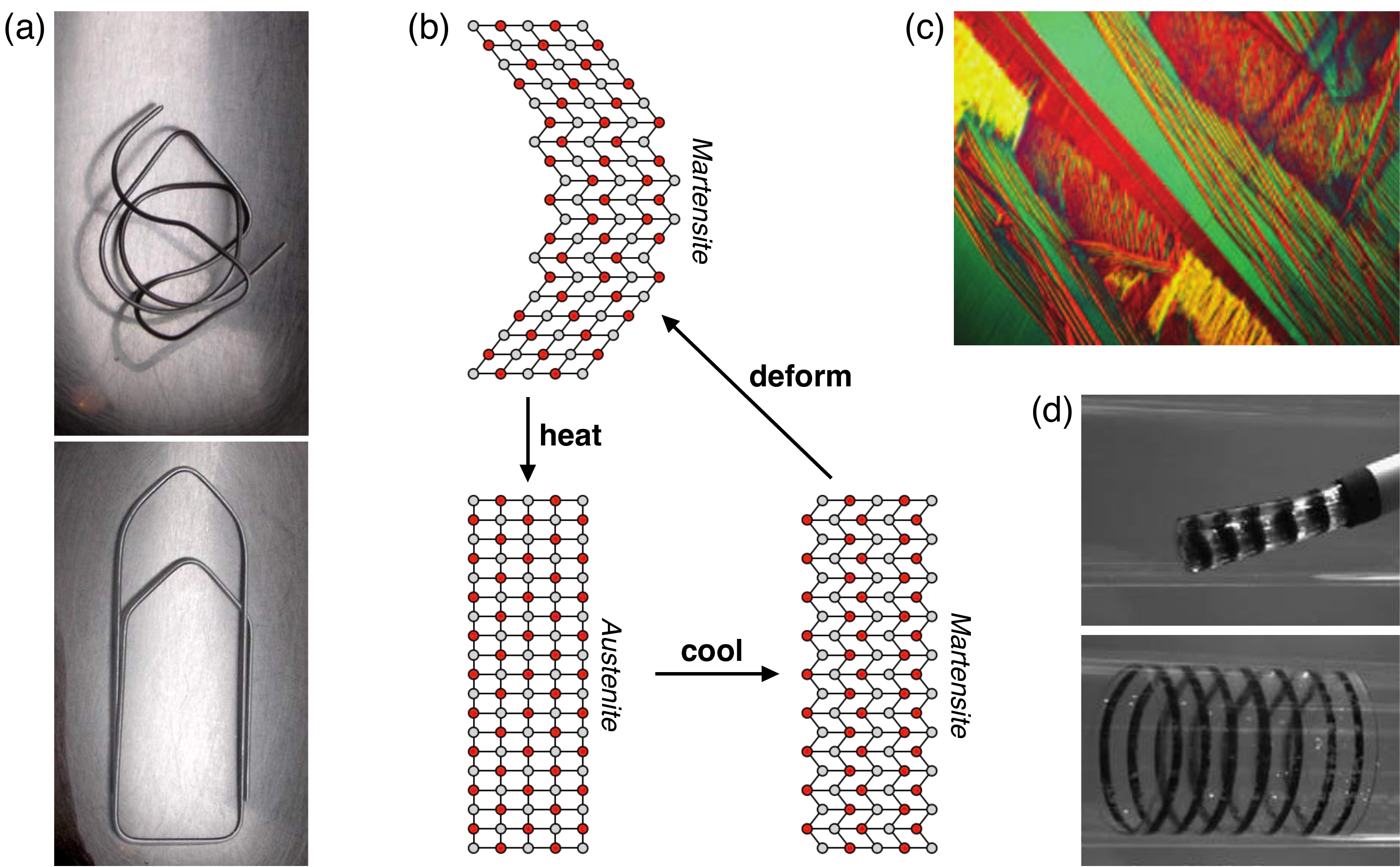}
    \end{center}
\caption{
    \textbf{Shape memory.} 
    \textbf{(a)} When this bent titanium-nickel wire (top) is submerged in hot water, it spontaneously reconfigures into a paperclip (bottom). 
    \textbf{(b)} Simplified description of shape-memory alloys. 
    Bottom row: When the sample is cooled, it undergoes a phase transition to the martensite phase \textit{without} changing its macroscopic shape. 
    The material may then be deformed to relatively large strains without changing the topology of the bond network (top). 
    Heating up the sample recovers the original shape by restoring the cubic lattice (bottom left). 
    \textbf{(c)} Optical microscopy image of the shape-memory alloy Au$_{30}$Cu$_{25}$Zn$_{45}$. % between thermal cycles. 
    Colors (different shades) indicate different orientations of the martensitic phase. 
    \textbf{(d)} A medical stent made from a shape-memory polymer, shown expanding in a 22 mm glass tube. 
    Black rings were drawn for visualization. 
    Panel \textit{a} adapted from Peter Maerki, Wikimedia Commons; panel \textit{c} adapted from~\citet{Song13} with permission; and panel \textit{d} adapted from~\citet{Yakacki07} with permission. 
    \label{fig:shape}
}
\end{figure*}

\section{Shape memory} \label{sec:shape}

Another dramatic instance of a material remembering its past is the phenomenon of ``shape memory.'' 
Figure~\ref{fig:shape}a shows a nickel-titanium wire in a messy crumpled configuration. 
Note that the wire is in mechanical equilibrium in this shape, and if kept at constant ambient conditions, it would stay in this shape indefinitely. 
Remarkably, when the wire is heated, it spontaneously reconfigures into a paperclip. 
Evidently, this shape was somehow programmed into the material at a previous time.

At the heart of the effect is a phase transformation~\cite{Bhattacharya03}. 
For sufficiently high temperatures the thermodynamically-stable phase is a simple cubic lattice called austenite, whereas at lower temperatures the material transforms to a lower-symmetry martensite phase. 
Crucially, when the paperclip is cooled down, it can transition from austenite to martensite \textit{without significantly changing its macroscopic shape}, by forming the martensite phase in alternating orientations (i.e., a ``twinned martensite''). 
This solid-to-solid phase transition from one crystal structure to another is drawn schematically in Fig.~\ref{fig:shape}b. 
Then, when stress is applied to the material, the orientation of any martensite region may flip to its mirror-image version (\emph{i.e.}, its twin), allowing large strains in the material while keeping the bond network unchanged. 
Because the different twinned microstructures are all mechanically stable, the material will hold a new shape when the external stress is released. 
Thus, despite the different macroscopic appearance of the entire wire, the atoms in the two macroscopic configurations in Fig.~\ref{fig:shape}a have approximately the same network of atomic bonds---this is the key to the recovery of a pre-programmed shape. 
When the wire is heated, the multitude of microscopic shear deformations are removed as the microstructure returns to the austenite phase, and the original macroscopic shape is recovered. 

Shape memory alloys may also be reprogrammed to learn a different shape. 
This is done by forming the austenite phase when the rod is in a new configuration (typically by clamping the sample tightly, raising it to a much higher temperature, and cooling rapidly). 
In this way, the bond network in the austenite phase is re-wired to prefer a different shape. 
Thus, these materials may learn new memories while forgetting old ones. 

This simplified picture gets at the essence of the ``one-way effect,'' but it ignores other rich aspects of real shape memory alloys. 
For instance, the samples generally require multiple thermal or mechanical training cycles to learn a particular shape, in order to coax crystal defects and grain boundaries into desirable configurations. 
Figure~\ref{fig:shape}c shows one example of a complex arrangement of the martensitic domains, from~\citet{Song13}. 
Surprisingly, these materials also show signs of critical behavior, even though the relevant phase transition is first order~\cite{Gallardo10}. 
This criticality appears to be a self-organized behavior that arises in the steady-state of cyclic driving~\cite{Perez-Reche07,Perez-Reche16}.
Other surprising phenomena that we do not address here are the ability to train two shapes into a single sample (the ``two-way effect"), as well as superelasticity, in which the austenite-martensite transition occurs at ambient
temperature---here the sample is driven into the martensite phase by an external stress, and the material returns to the austenite phase when the stress is removed, thereby allowing unusually large reversible deformations. 
Recent efforts try to understand some of these behaviors 
by focusing on the constraints that arise from fitting together separate phases of a material \cite{James19}. 

How does shape memory compare to the other memories we have considered? 
In this case, the memory is stored in the network topology of the constituent atoms, namely, the set of bonds between the atoms. 
If the material is not deformed too far, the bonds remain intact, which allows the atoms to return to their original positions upon heating. 
In this sense, the behavior is not so different than the simple ``memory'' of an elastic solid---the familiar and commonplace phenomenon that an elastically deformed solid will return to its rest state when loading is removed. 
What makes shape-memory materials different from typical elastic solids is a mechanism for temporarily arresting a deformed configuration. 
This is achieved by mixing together different symmetry-broken domains of the martensite phase to assume a different macroscopic shape, as we described in Fig.~\ref{fig:shape}b. 
As a result, the deformed configuration is mechanically stable, yet the material has retained its original network of bonds. 

A similar shape memory effect can occur in polymers~\cite{Lendlein02} but by a different microscopic mechanism. 
Here the effect relies on polymer chains that can be tuned from being flexible to being rigid as a function of temperature, typically by vitrification or crystallization~\cite{Mather09}. 
Starting with a suitable polymer sample, a deformation is applied and held while the sample is cooled below a transition temperature where the polymer chains become rigid. 
As a result, the sample holds the deformed shape when it is released. 
Raising the temperature returns the polymer chains to their flexible state, and the sample recovers its original macroscopic shape. 
This shape recovery is primarily entropic in nature---the original macroscopic shape is preferred because it allows the largest number of microscopic polymer conformations. 
As in the case of shape memory alloys, the topology of the bond network is maintained throughout the entire process. 

Due in part to their relatively low cost, shape-memory polymers are finding use in a wide range of applications. 
Figure~\ref{fig:shape}d shows a medical stent that is activated by body temperature to expand inside a blood vessel. 
In addition to medicine~\cite{Mano08}, shape-memory polymers are starting to be used in textiles~\cite{Chan-Vili07}, structural repairs such as self-peeling adhesives~\cite{Xie08}, and self-deployable structures for space applications~\cite{Sokolowski07}. 
Alternative triggering mechanisms such as light~\cite{Lendlein05} further expand the range of possible uses. 

While more expensive, shape-memory alloys can withstand much larger loads than polymers, and they are used as sensors and actuators in a variety of aerospace, automotive, and biomedical settings~\cite{Lagoudas08}. 
There are also important engineering challenges for improving these materials, such as delaying their eventual failure due to many cycles of actuation~\cite{Chluba15}.

\section{Aging and rejuvenation} \label{sec:aging}

\begin{figure*}
    \begin{center}
        \includegraphics[width=7in]{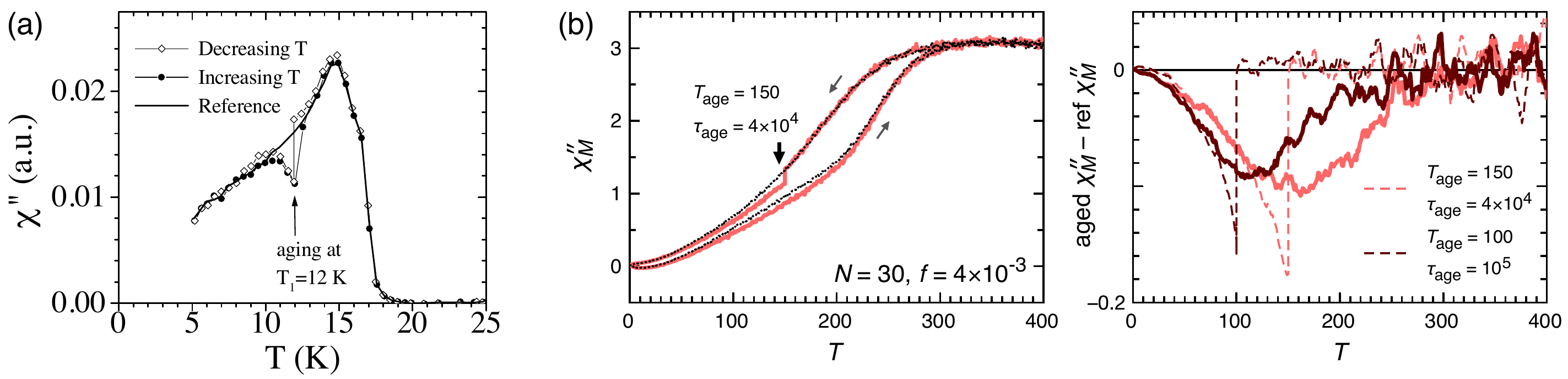}
    \end{center}
\caption{
    \textbf{Aging and rejuvenation.} 
    \textbf{(a)} Relaxation of the spin-glass CdCr$_{1.7}$In$_{0.3}$S$_4$, measured by the imaginary part of the magnetic susceptibility. 
    The thick curve is obtained when cooling from $25$ K down to $5$ K at a constant rate. 
    Solid symbols: If the system is held at an intermediate temperature, here $12$ K, the susceptibility slowly drops (``aging"), but upon further cooling it returns to the curve it was previously following (``rejuvenation"). 
    Remarkably, the same curve is traced out upon reheating at a constant rate (open symbols). 
    \textbf{(b)} Left: Similar behavior is observed in a simple number-sorting algorithm, where nearest neighbors are swapped with a probability given by a Boltzmann factor.  
    Right: Subtracting the reference curve (obtained for a constant cooling rate) makes the trend more clear. 
    Panel \textit{a} adapted from \citet{Jonason98} with permission, and panel \textit{b} reproduced from \citet{Zou10} with permission. 
    \label{fig:aging}
}
\end{figure*}

A very different type of memory can form in spin glasses, which are magnetic materials with quenched disorder, arising for instance due to a mixture of ferromagnetic and antiferromagnetic bonds. 
One way of probing these materials is by their response to an applied magnetic field via the magnetic susceptibility, $\chi$. 
This quantity may be measured continually to monitor the material as a function of temperature and time. 
Figure~\ref{fig:aging}a shows the out-of-phase part of the ac magnetic susceptibility, $\chi''$, in the spin glass CdCr$_{1.7}$In$_{0.3}$S$_4$, measured at low frequency and low applied field~\cite{Jonason98}. 
The thick curve shows the measurements starting from $25$ K and going down to $5$ K at a constant cooling rate of $0.1$ K per minute. 
If the experimenter repeats the same cooling protocol but holds the sample at some intermediate temperature for an interval of time (e.g., at $12$ K as shown in the figure), the out-of-phase susceptibility gradually drops to lower and lower values. 
This phenomenon is called ``aging,'' wherein the sample relaxes through a series of states with progressively lower energy. 
(The dynamics of aging is in general very slow and may proceed logarithmically in time, as in the example of the piston compressing a crumpled mylar sheet in Sec.~\ref{sec:Kaiser}. 
Thus, the lowest value of $\chi''$ that may be measured at any given temperature is set by the amount of time available to the experimenter---the lower open symbol at $12$ K shows the final value obtained after 7 hours of aging, which represents the lowest-energy state the system was able to find.) 
Remarkably, as the temperature is lowered again, $\chi''$ climbs up to where it ``should have been''---the system seems to have forgotten everything about the aging. 
Evidently, finding a lower-energy state at $12$ K had little effect on the properties of the system at $10$ K. 

This behavior, called ``rejuvenation,'' is certainly surprising, but something even more remarkable happens as the sample is subsequently heated up. 
The solid symbols show that upon reheating at a constant rate, $\chi''$ dips at $12$ K---there is a memory of the temperature where it was aged! 
After that point in the experiment, all of these features are erased; it is only on increasing the temperature that one can retrace what happened on the way down. 

Further experiments show that multiple memories may be stored simultaneously within a single sample~\cite{Jonason98,Jonason00,Bouchaud01,Vincent2006b}. 
Here the sample is aged at multiple temperatures during the cool-down. 
When the temperature is increased, a dip is observed at each temperature where the system was aged. 

The origin of this memory behavior is not yet fully understood. 
In principle, numerical simulations of spin-glass models could provide valuable insight into the relaxation processes and the relevant length-scales involved. 
Unfortunately, reproducing the phenomenon on a computer has proven difficult, and several studies have reported different interpretations of the situation \cite{Komori00,Picco01,Berthier02,Takayama02}. 
In light of this, Maiorano \textit{et al.}~focused on identifying the basic phenomenology of finite-dimensional Edwards-Anderson spin-glass models \cite{Maiorano05}. 
They found only simple cumulative aging, which is incompatible with rejuvenation over the timescales they studied. 
In the same year, Jim\'enez 
\textit{et al.}~reported deviations from cumulative aging in these models \cite{Jimenez05}. 
Whether rejuvenation and memory might be recovered at longer timescales, or whether further physical effects must be added to the Edwards-Anderson model to reproduce the experimental phenomenology, remains an open question.

If this memory behavior only occurred in spin glasses, while very interesting it might be just an isolated phenomenon. 
However, one can also observe this behavior in molecular glasses \cite{Yardimci03} and polymers \cite{Bellon02,Fukao05}. 
Remarkably, the same effect occurs in a simple model where a list of numbers is sorted in a thermally-activated manner \cite{Zou10}. 
Suppose you are given the numbers 1 through 5 in a random order, and you want to put them into increasing order. 
One way would be to pick a random nearest-neighbor pair and swap them with a probability given by a Boltzmann factor where temperature is replaced by an effective temperature and the energy is a function of the difference between adjacent pairs of numbers.  In addition there is an energy term that couples the system to an external field that favors sequences of numbers in ascending (or descending) order.
This is actually a terrible algorithm in terms of how many steps are required, but it has the flavor of a physical annealing process. 
Interestingly, this sorting algorithm turns out to show glassy behavior. 
To draw an analogy with the spin-glass experiments above, one may define a susceptibility in this system as its linear response with respect to variations in the external field. 
That susceptibility, $\chi''_M$, displays a logarithmically slow relaxation to the fully-sorted state. 
Even more than this,
if you take just 30 numbers and do a similar cooling protocol, you can observe aging, rejuvenation, and memory, as shown in Fig.~\ref{fig:aging}b. 
The relevant features are somewhat subtle in the raw curves, but they become clearly apparent when the reference curve is subtracted, as shown in the second panel of Fig.~\ref{fig:aging}b. 
This result is not trivial at all---it is not easy to think about how this arises in thermally-activated list sorting. 
What we can say is that this looks like a generic kind of memory formation. 

An eventual explanation of this memory behavior must address how the memory is stored and survives so that it can be read out at a later time. 
One promising view is that aging coarsens the system, organizing it over larger and larger lengthscales. 
This makes the memory robust to any subsequent evolution over shorter times, which would correspond to smaller lengthscales. 
This physical picture has been exploited by a recent simulation algorithm called ``patchwork dynamics,'' which accesses the non-equilibrium behavior of spin glasses over a broad range of timescales by directly equilibrating the model glass on successively larger lengthscales~\cite{Thomas08, Yang17}.

\section{Memory through path reversal: echoes}
A shout across a mountain valley often results in an acoustic echo as the sound is returned to its source a few moments later.  The sound waves reflecting from the far side of the valley follow in reverse the path along which they propagated in the first half of their journey.  This is as close as one can get to time reversal---the velocities of the wave packets are reversed and they follow the identical path both to and from the valley's far side.  The sound waves that return to their source contain a memory of what was shouted including the timing between different syllables. 

\begin{figure*}
    \begin{center}
        \includegraphics[width=6.75in]{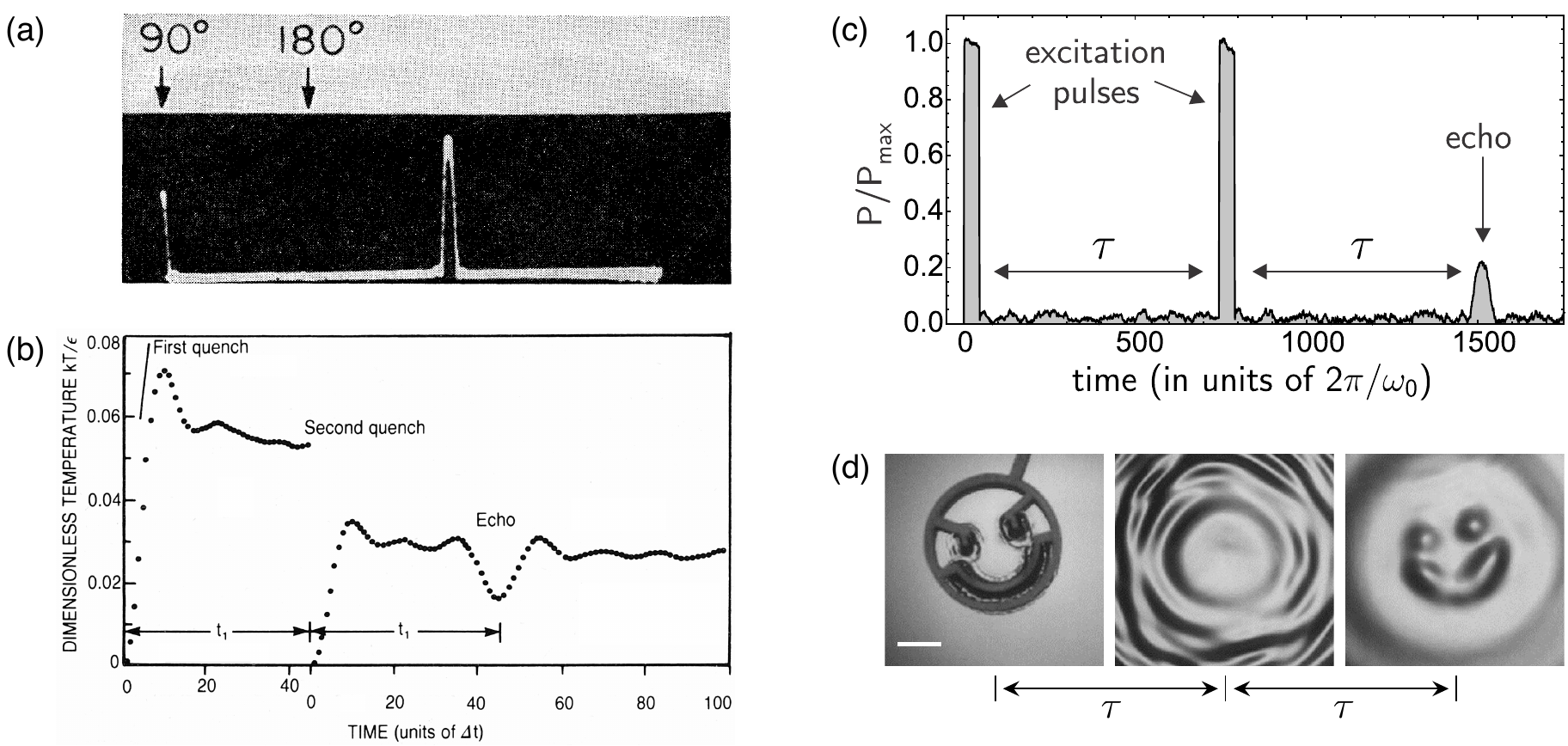}
    \end{center}
\caption{
    \textbf{Echo phenomena.} 
    \textbf{(a)} Spin echo produced with the pulse sequence described in the text. 
    The timing of the two input pulses are indicated above a photograph of the oscilloscope trace; the subsequent peak is the echo. 
    Note that there is no signal coinciding with the ``$\pi$'' pulse, as this simply flips the (dephased) spins at that instant. 
    \textbf{(b)} Quench echo in a simulated Lennard-Jones glass of 500 particles. 
    The two quenches consist of setting the kinetic energy of the particles equal to zero instantaneously. 
    Because they have potential energy, they will start to move again after each quench. 
    Remarkably, this leads to an echo at a later time. 
    \textbf{(c)} Echo in a jammed amorphous solid due to anharmonic modes.
    The signal is an average of 10,000 independent simulations, each composed of 1,000 soft spheres in three dimensions. 
    \textbf{(d)} Echo in a system of capillary waves. A dish of water is disturbed by a face-shaped template, creating surface waves that radiate outwards. The entire dish is then accelerated downwards, causing the existing ripples to act as sources, sending a new set of ``time-reversed'' waves back to the site of the initial disturbance. The face reappears as the waves reconvene. 
    Panel \textit{a} reproduced from \citet{Carr54} with permission; panel \textit{b} reproduced from \citet{Nagel83} with permission; panel \textit{c} reproduced from \citet{Burton16} with permission; and panel \textit{d} adapted from \citet{Bacot16} with permission.
    \label{fig:echoes}
}
\end{figure*}

\subsection{Spin echoes}
Perhaps due to our familiarity with this phenomenon, such acoustic echoes are captivating but they no longer challenge our intuition. 
However, they have counterparts in various material systems which are not at all intuitive; they are subtle, challenging to understand, and ultimately very surprising. 
A well-known example is the case of the spin echo first demonstrated by Edwin \citet{Hahn50} and developed by \citet{Carr54}---one of their measurements is shown in Fig.~\ref{fig:echoes}a. 
Such echoes are now an integral component of how MRI (Magnetic Resonance Imaging) produces a three-dimensional representation of water- or oil-containing samples. 
As with the more familiar acoustic echo, these spin echoes likewise preserve a memory---they encode the precise timing between discrete radio-frequency pulses applied to an ensemble of spins. 

In the spin echo, a large static magnetic field $H_0\hat{z}$ produces a small polarization of the spins in the sample; slightly more of the spins point in the $+\hat{z}$ than in the $-\hat{z}$ direction. 
Each spin precesses about the $\hat{z}$ axis at its Larmor frequency $\omega$, determined both by the applied static field $H_0\hat{z}$, and the small magnetic inhomogeneities in the sample. 
When a radio-frequency (RF) pulse tuned to the average $\langle \omega_0 \rangle$ is applied in a direction perpendicular to $H_0$ the spins rotate away from $+\hat{z}$ axis. 
The trick with the spin echo is that the RF pulse can be applied for only a short time so that when it is turned off, the spins instantaneously point in an arbitrary direction---in the simplest case for this explanation, a ``$\pi/2$ pulse'' rotates spins into the $x$-$y$ plane perpendicular to $\hat{z}$, where they again precess about the static field $H_0\hat{z}$.  
Because of the local inhomogeneities of the magnetic field in the sample, the Larmor frequency varies from one spin to another so some spins will precess farther. 
This causes the spins to dephase so that they will eventually fan out and point in all directions in the $x$-$y$ plane.  

After the spins have been allowed to dephase for a time $\tau$, another RF pulse of twice the duration (called a $\pi$ pulse) is applied, flipping the spins in their orientation in the $x$-$y$ plane. 
The spins will once again precess about the field with the same Larmor frequency that they had originally.  
However, the fast and slow spins have been switched. 
The fast spins (that had rotated farther during the period between the pulses) will now be in a position where they lag behind those that precess more slowly. 
Now, because the fast spins rotate more rapidly, after waiting an additional time $\tau$ they will catch up with the slow ones that were in a lower field.  
This realignment can be read out in a pulse. 
Such an echo is a memory of the time delay, $\tau$, between the two RF pulses. 
Multiple RF pulses, with different durations, can also be applied; these protocols have helped to make nuclear magnetic resonance a useful tool for measuring and unscrambling many different spin interactions in solids.

The spin echo is only one of many related effects. 
Similar behavior is associated with other ensembles of two-state systems. 
These include photon echoes~\cite{Kurnit64} as well as phonon echoes~\cite{Golding76}, which have been interpreted as due to quantum mechanical tunneling between two nearly degenerate configurations in low-temperature glasses. 

\textit{Other echoes---}
Other types of echoes do not require a two-state system in order to return an imposed signal. 
An example is the temperature quench echo shown in Fig.~\ref{fig:echoes}b, which only requires a set of classical normal modes that share energy between kinetic and potential energy contributions~\cite{Nagel83}. 
Another echo is the anharmonic echo that appears in a set of anharmonic modes (Fig.~\ref{fig:echoes}c). 
This requires the frequency of a mode to vary with its amplitude~\cite{Gould65,Kegel65,Hill65,Burton16}.

In all these cases, the memory is encoded in the coherence of the spins (or oscillators). 
The initial pulse sets up a reference time when all the phases are in synchrony with one another. 
The system then evolves according to its dynamics and apparently de-phases so that the phase of each oscillator appears to be unrelated to that of all the others. 
A snapshot at any instant of time would not appear to have any special order within the sample. 
However, the second pulse allows all the phases to coalesce at a later time. 

Capillary waves present another arena where an echo may be produced. 
Raindrops striking a puddle or pond create surface waves that are easily observed with the naked eye. 
Starting from an initially-localized wave front, the different wavelength components radiate outwards at different speeds, so the total energy is spread over a growing area and the waves become harder to see. 
If you were to watch a video of this scene in reverse, it would look awry: the waves would instead travel inwards towards a single point and arrive there in concert, focusing all their energy at that location. 
Remarkably, recent experiments have shown that this reversal of dynamics may be achieved in real time \cite{Bacot16}. 
By imposing a sudden downward acceleration to a bath of water at some time $\tau$ after an initial surface perturbation, the existing ripples act as sources that create two waves -- one continuing onward plus a ``time-reversed'' wave that propagates in the opposite direction.
As in other echo phenomena, the refocusing occurs a time $\tau$ after the second pulse (in this case the downward acceleration of the bath). 
Moreover, this protocol even reproduces the shape of the initial perturbation, as shown in Fig.~\ref{fig:echoes}d.

In the study of all these echo phenomena, the crucial ingredient for creating a memory is the ability to retrace the dynamics in the reverse order from how the initial system evolved. 
That is, it relies on a form of time reversal in which the spins (or waves) are manipulated to be in a configuration so that they effectively retrace their previous dynamics. 
(For anharmonic echoes, the effect is attributed to phase conjugation \cite{korpel81}, which is closely related to time reversal.)
This is one distinctive way in which matter can retain a memory of the inputs.  
 
\subsection{Apparent time reversal in viscous fluids}
Such behavior is not only relegated to the case of an echo. 
It is well known that a liquid of sufficiently high viscosity (\emph{i.e.}, low Reynolds number) is reversible if the boundaries are distorted and then returned to their original positions along precisely the reverse sequence of motions as they were originally deformed. 
This phenomenon is showcased in a video-recorded demonstration by G.~I.\ Taylor for the National Committee for Fluid Mechanics Films~\cite{Taylor85}. 
He begins with a synopsis: ``Low Reynolds number flows are reversible when the direction of motion of the boundaries which gave rise to the flow is reversed. This may lead to some surprising situations, which might almost make one believe that the fluid has a memory of its own." 

In the demonstration, a blob of dye is introduced into a viscous liquid occupying a narrow gap between two cylindrical walls. 
When the inner cylinder is rotated, the fluid is sheared and the spot of dye smears out into a sheet. 
After the first shear, it appears as if the fluid is completely mixed and that there is no way of regaining the original conformation. 
But there is a subtle memory in the fluid that ``remembers" where it came from. 
When the shear is then performed in the opposite direction, the spot reappears. 

This memory, like the spin echo, is surprising but is based on the idea of time (or more precisely, path) reversal. 
If the system can be manipulated so that after a first set of deformations, the dynamics can be made to repeat itself but in the opposite direction then there is a reversal of dynamics (if not in the time itself).

\subsection{The Kovacs effect}\label{sec:Kovacs}

Another situation in which this type of behavior shows up is in a form of relaxation phenomenon. 
Consider a system that has a set of modes by which relaxation can take place. 
Each of these modes has a specified relaxation time (or spectrum of times). The origin of these relaxation times is a detail of the system that is often unclear, though in some cases they are simply the relaxations of a material's subsystems individually \cite{Chakraverty05}.
Independent of where the system is started, it will approach an equilibrium state via these relaxation modes.
The ones with short time constants will equilibrate first and the ones with longer time constants will equilibrate more slowly. 
It is important to specify that the time constants themselves do not change due to a perturbation of the system. 

\begin{figure}
    \begin{center}
        \includegraphics[width=2.75in]{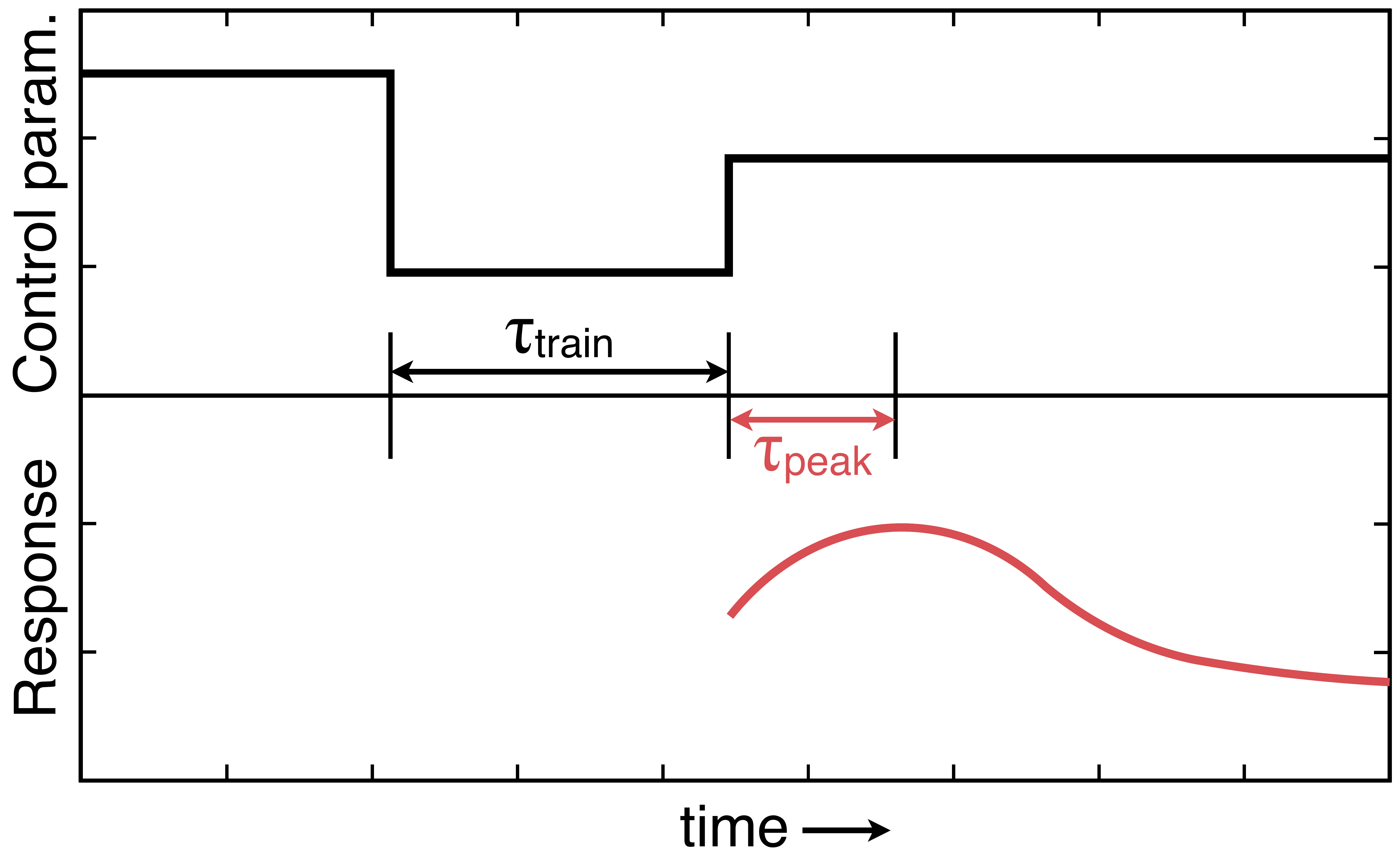}
    \end{center}
    \caption{
    \textbf{Kovacs effect.}
    Top curve shows the general experimental protocol, whereby a perturbation is applied to a sample for a duration $\tau_\text{train}$, and then the size of the perturbation is reduced and held constant at a value closer to the initial one. Bottom curve shows the response, which peaks at a time $\tau_\text{peak}$ that is proportional to $\tau_\text{train}$, indicating a memory of $\tau_\text{train}$.  Experiments show such a memory of waiting time in a crumpled mylar sheet \cite{Lahini17}, where the control parameter is the volume controlled by a piston, and the response is the force exerted by the sheet on the piston.  (See the image inset in Fig.~\ref{fig:Kaiser}.)
    \label{fig:crumpled-kovacs}
}
\end{figure} 

%JDP: Some of the text seemed repetitive -- I commented out a few sentences and added some text to orient the reader with the figure.
If the system is perturbed so that it relaxes toward a new state, all the modes will relax. 
However, if the system is only held in this new position for a waiting time $\tau_\text{train}$ before it is returned closer to its starting configuration, then only the modes that have relaxation times smaller than $\tau_\text{train}$ will contribute appreciably to that relaxation. 
The modes with longer relaxation times will not have had a chance to evolve much  \cite{Chakraverty05}.
%When the system is returned after time $\tau_\text{train}$ to a condition closer to where it started, the modes which had previously relaxed in the forward direction must now relax in the opposite direction. 
%The modes that did not have a chance to relax due to the initial perturbation will still relax---but slowly---towards this new equilibrium. 
%This gives rise to a non-monotonic behavior. 
The memory may be seen in the behavior of the system \textit{after} the set of two perturbations has been applied (shown by the response curve in the bottom of Fig.~\ref{fig:crumpled-kovacs}). 
In particular, the initial relaxation appears to be in the reverse direction of the first perturbation, but then at a later time the system responds as if the intermediate perturbation had never happened.
This non-monotonic behavior contains a memory of the time the system was forced to wait after the initial perturbation.

This effect is named after Andr\'e J. Kovacs who discovered it in amorphous polymers, as noted earlier~\cite{Kovacs63, Kovacs79}. 
An analogous example of this behavior is in the physics of crumpled sheets~\cite{Lahini17}. We note that the behavior of a crumpled thin sheet was also used above in Section~\ref{sec:Kaiser} to illustrate the rudiments of another form of memory, the Kaiser effect. 
In these experiments, a sheet of plastic is prepared by crumpling it many times, then confined in a cylinder. It is compressed into a smaller volume for a time $\tau_\text{train}$ and then allowed to expand into an intermediate volume. Once the sheet expands it begins to exert a force on the container.
This force continues to grow for a time comparable to $\tau_\text{train}$, but then starts to decay as the slower modes become involved---modes slow enough to effectively ignore the time spent at the smallest volume. 
The elapsed time until the peak 
force, $\tau_\text{peak}$, is proportional to $\tau_\text{train}$ and is a memory of the waiting time. 
A similar behavior has been observed in measurements of the total area of microscopic contact between two solid objects, where a large normal force is applied for a duration $\tau_\text{train}$ and then reduced to a smaller value, thus demonstrating that frictional interfaces exhibit a similar type of memory of their loading history~\cite{Dillavou18}. 

The model proposed for the Kovacs effect in compressed crumpled sheets relied on having a set of relaxation pathways that could retrace their paths once the driving compressive force is released \cite{Amir12}. 
This way of looking at the problem suggests that echo phenomena and the Kovacs effect (seen in a myriad of different examples) have a common underlying origin in the reversal of path or relaxation dynamics.

\textit{Relation to aging and rejuvenation---}
Once thought of in this context, the rejuvenation, aging, and memory experiments in spin glasses might also be included as a form of path reversal.  In that much more complicated case, when the spin glass is quenched and held for some time at a temperature $T_0$, it evolves most rapidly on short length scales and more and more sluggishly as the system tries to reach its ground state on longer scales. In the model by Middleton~\cite{Thomas08,Yang17} for how spin glasses age under a temperature quench, a small change in temperature creates a drastic readjustment of all the inter-spin interactions. Each time the temperature is changed, the system must start to search once again for an adequate ground state. In that model, the evolution can be thought of as a series of relaxation events that have a time scale that is tied to the spatial length over which the relaxation has taken place. When the system is re-heated to the temperature where it was first partially aged, the spin configuration again goes through a sequence of relaxations that occur first on the smallest lengths (and therefore fastest timescales) and proceed to larger and slower relaxation modes. Through the states of these slower modes, the spin glass remembers that it was aged at a certain temperature as it passes through that temperature in the heating cycle. This model has a flavor of the physics that was required for the Kovacs effect in the crumpled sheet. In both cases, precisely how the ingredients for memory might arise from the collective dynamics of the system remains an open question.

\section{Associative memory}

A memory that we all have firsthand experience with is \textit{associative memory}. 
At one point or another you probably had a hard time remembering the name of a person you knew or a place you once visited. 
Yet if you can conjure up some specific detail associated with it---a scene or an event or even a smell or taste---suddenly the name comes back to you. 
What makes this particular kind of memory special is that you can use a partial or approximate version of the memory to recall much more.

\begin{figure}[b]
    \begin{center}
        \includegraphics[width=2.9in]{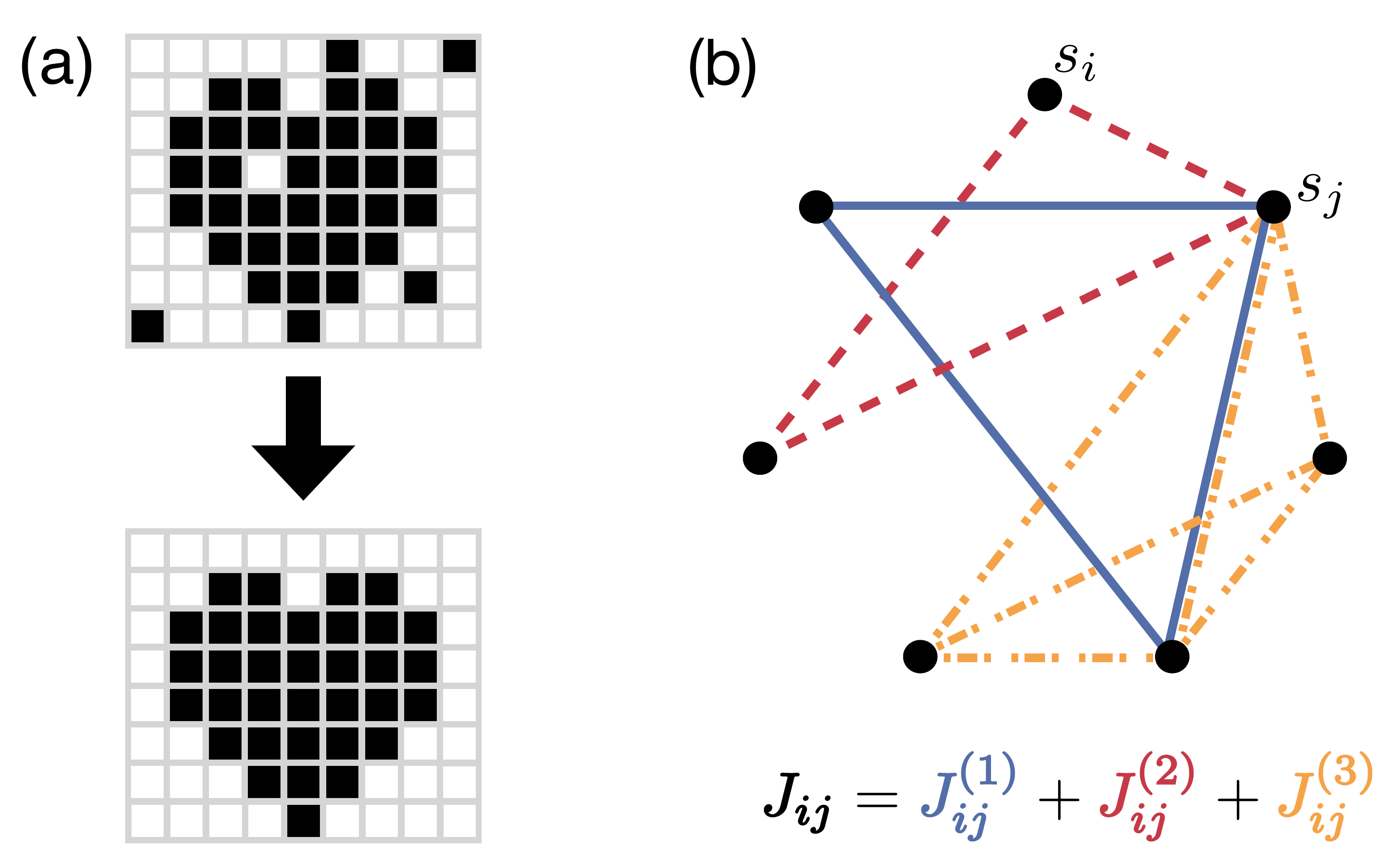}
    \end{center}
\caption{
    \textbf{Associative memory.} 
    \textbf{(a)} Partial or corrupted information may be used to retrieve a memory. 
    \textbf{(b)} Schematic of Hopfield model, showing three sets of connections (solid, dashed, and dash-dotted edges) defined on the same set of neurons (vertices). 
    %For simplicity, only the edges joining neurons in the ``firing" state for each memory are shown. 
    For simplicity, only the bonds between the ``on'' neurons $(s_i=1)$ are drawn. 
    Under the right conditions, the single set of weighted connections $J_{ij} = J^{(1)}_{ij} + J^{(2)}_{ij} + J^{(3)}_{ij}$ may be used to store the three individual memories. 
    \label{fig:Hopfield}
}
\end{figure}

This experience is by its very nature subjective, and there may be many important biological factors that affect this processing of information in our brains. 
Nonetheless, the physics community was able to formulate model systems which also exhibit the phenomena of associative memories. Some of this work has contributed to the study of artificial neural networks with associative memory --- systems that have growing applications in technology and scientific research \cite{Pankaj18}. 

\subsection{Hopfield neural networks}

Consider a network of identical nodes, each node denoting a neuron $i$, and each connection $ij$ being assigned a weight $J_{ij}$.
Each neuron may be ``on'' $(s_i=1)$ or ``off'' $(s_i=-1)$.
The connection weights govern the evolution of the system: starting from some initial state (e.g., the top image in Fig.~\ref{fig:Hopfield}a), the neurons are updated by turning $s_i$ ``on'' if $\sum_j J_{ij} s_j \geq 0$ and ``off'' otherwise. 
A memory is one particular state of the system, $\vec{s}_\text{mem}=\{s_{1}^\text{mem},s_{2}^\text{mem}\ldots\}$, defined by our choice of which neurons are ``on,'' as represented for example by the pixels in Fig.~\ref{fig:Hopfield}a (lower panel).
There is an appealing and intuitive idea that you store a memory by making certain connections stronger. 
Following this notion, we set the weights $J_{ij} = 1$ if $s_i^\text{mem} = s_j^\text{mem}$ (strengthening the connection between neurons which are both ``on'' or ``off'' in the memory) and $J_{ij} = -1$ if $s_i^\text{mem} \neq s_j^\text{mem}$. 
The model with such weights reproduces the desired associative memory behavior: if you are in a state $\vec{s}$ that is merely \textit{close} to the memory $\vec{s}_\text{mem}$, then the evolution will lead to the state $\vec{s}=\vec{s}_\text{mem}$ which is a stable fixed point unchanged by further evolution. In that sense, the stored memory is successfully retrieved by the system.

Although this behavior might be unsurprising, consider trying to store three different memories in the same network in such a way that any of them can be retrieved.
Suppose the memory of your friend would be stored in a set of bond strengths, $J^{(1)}_{ij}$, the memory of your boss is given by another set, $J^{(2)}_{ij}$, and the memory of your pet fish is yet another, $J^{(3)}_{ij}$ (Fig.~\ref{fig:Hopfield}b). 
A straightforward approach would be to take the sum of the weights for each individual memory:
\begin{equation}
J_{ij} = J^{(1)}_{ij} + J^{(2)}_{ij} + J^{(3)}_{ij}.
\label{eq:hopfield-sum}
\end{equation}
\noindent You may expect the system to form three stable fixed points; one for each of the memories. 
But trouble is right around the corner: Spurious stable fixed points also arise. Some are mixtures of the desired states (\emph{e.g.}, a combination of your boss and your pet fish) and others look completely unrelated. So while you set out to write a few memories in the system, you have unwittingly written many more ``false" memories---things that you can now remember but have never experienced!
John Hopfield and others showed in the 1980s that although this is true, the idea can still work. 
Below some threshold number of stored memories, each desired memory 
is indeed a stable fixed point, while all spurious fixed points have a smaller basin of attraction, \emph{i.e}., only very nearby states converge to them~\cite{Hopfield82,Amit1985,Hertz:1991}.
Thus, below the threshold (which is of the same order as the number of nodes) multiple memories may be simultaneously stored across many nodes without any unintended consequences.

Many other aspects of the Hopfield model have been studied vigorously.
As an illustration, the straightforward learning rule in Eq.~\eqref{eq:hopfield-sum}, called Hebbian learning, may be replaced by various others; further, different dynamics have been studied by allowing simultaneous or sequential updates of neurons; and there are extensive studies of time-dependences, stochasticity, and memory correlations~\cite{Hertz:1991}. 

Although the network model started as an abstraction of neuronal behavior, its mathematical form is reminiscent of magnetic systems in condensed matter physics, starting from the observation that each node could describe a spin state, $s_i=1$ or $-1$. We refer the reader to the excellent textbook by~\citet{Hertz:1991} for an introduction to the study of neural networks using the theory of magnetic systems and the tools of statistical physics. This relationship points to the possibility of observing associative memory in a host of settings, due to the pervasiveness of physical systems that may be modeled as coupled spins.

\subsection{Toward models of biological memory}

The abstract mathematical model of a neural network introduced by Hopfield \cite{Hopfield82} 
exhibits some remarkable features, such as a large capacity and stability of stored memories, which we also observe in very complex biological systems such as the human brain. However, this 
performance of the abstract model comes from a biologically and physically unsound assumption --- that the values of $J_{ij}$ may grow unboundedly as new memories are stored \cite{amit_1989,amitfusi,FusiAbbott}.
Early work~\cite{Parisi86} showed that by limiting the magnitude of $J_{ij}$, under certain conditions 
new memories will erase old ones as they are stored in the $J_{ij}$. 
This behavior evokes a first-in-first-out type of memory buffer in computing, where the storage of fixed capacity is emptied by first removing the oldest item in it.
Intense research continues decades later in search of toy models which capture more realistically the associative memory performance of biological neural networks; we point the reader to reviews by \citet{amit_1989} and \citet{Fusi2017}.

One key insight from such models
is the recognition of an intrinsic balance in a biological neural network: 
either new memories are easily accepted (plasticity of synapses), or old memories are retained for a long time (stability of synapses) \cite{Fusi2017}. 
The Hopfield model could maximize both as it had the freedom to store unbounded information in its $J_{ij}$ values. To optimize the memory performance of more realistic models, researchers have taken inspiration from the complexity of components in a biological neural network by considering multiple coupled hidden degrees of freedom for each synapse. These hidden variables can evolve on different timescales, and with some tuning of their mutual couplings, can control both the stability and plasticity of the modeled synapse. As a result, 
such models can 
reach an impressive capacity of $\sim$$N/\log(N)$ memories in a network of $N$ neurons, with a relatively slow power-law decay of old memories \cite{BennaFusi2015}. As opposed to the Hopfield model, these models with dynamical coupled hidden variables are explicitly studied in a steady state, so their associative memory behavior 
is an out-of-equilibrium phenomenon.

\bigskip

\subsection{Associative memory through self-assembly}

Associative memory may also arise in the seemingly disparate context of self-assembly. 
The process of self-assembly entails building blocks, driven only by thermal or other random fluctuations, that aggregate into a larger objects. Mutual interactions between different constituents may be the only guide during this otherwise mindless process of spontaneous assembly. The building blocks can come from a wide variety of natural or artificial nano- and meso-scale objects. Likewise the design for their mutual interactions can vary immensely in their range, anisotropy, specificity and physical mechanism. The emergent assembly may be quite distinct and intricate --- a specific crystal, macromolecule, biological cell component, or meta-material.

\begin{figure}
\begin{center}
\includegraphics[width=3.2in]{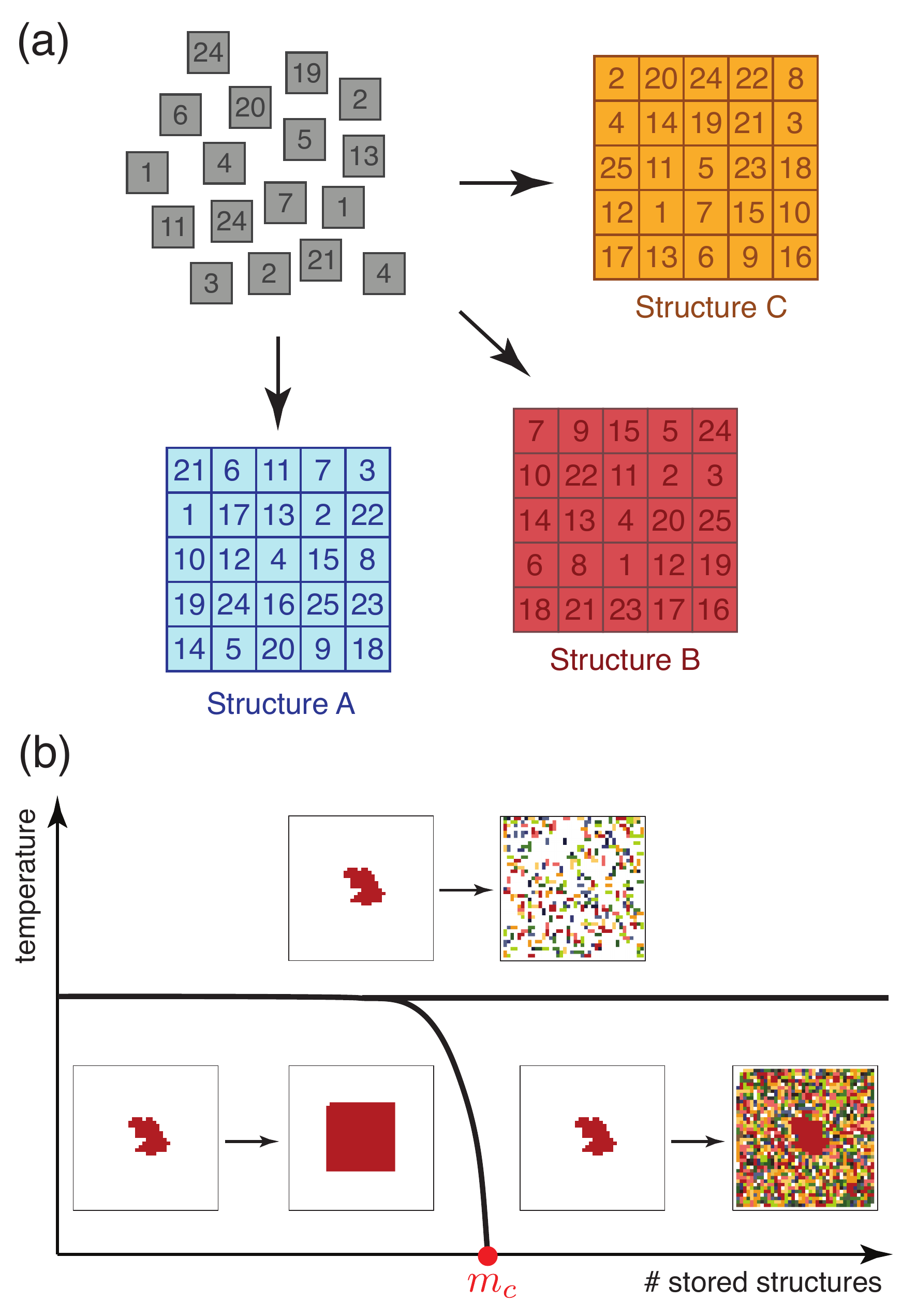}
\end{center}
\caption{\textbf{Self-assembling memories.}
\textbf{(a)} Square building blocks may bind with up to 4 nearest neighbors 
on a lattice. Three different assembled structures (A,B,C) represent three memories. The memories are simultaneously stored by defining attractive interactions between blocks which are neighbors in any of the structures.
\textbf{(b)} Simplified diagram of self-assembly regimes in simulations. Three behaviors (separated by heavy lines) are distinguished by the fate of an introduced seed for a desired structure. Building blocks are pixels in a given snapshot of the system (bordered by thin square), each pixel being colored (shaded) according to the structure whose part it is forming in that snapshot. Below a critical number of stored structures, $m_c$, and at low enough temperatures (bottom-left regime), a seed for the B structure (irregular shape in initial snapshot on the left) leads to its self-assembly, \emph{i.e.}, the desired B memory is successfully retrieved. Above storage capacity the same seed leads to an erroneous structure (bottom-right regime. When the thermal energy sufficiently exceeds the binding energy, all assembly is prevented. 
Panel \textit{b} adapted from~\citet{Murugan15}.
\label{fig:selfassembly}
}
\end{figure}

While the behavior of particles attracting or repelling each other while moving around may seem far removed from the artificial neural network or a synapse network of a brain, one may intuit the phenomena of associative memory here as well. We suggest that the desired assembled structure may be viewed as a memory that is stored in the building blocks through the design of their mutual interactions. Correspondingly, when the structure is self-assembled due to a willful trigger (say, by providing a small substructure to act as a nucleation seed), one may claim that the memory has been retrieved. 

A simple associative memory for assembly then contains a mixture of independent batches of building blocks, with each batch designed to assemble into a particular structure when the appropriate trigger is applied. This construction is however quite wasteful. In contrast, biology is much more economical and reuses the same building blocks (\textit{e.g.}, individual proteins) in many different cellular assemblies~\cite{Kuhner09}. We therefore want to consider a potent mixture of building blocks, which uses most of its blocks to assemble any of several different stored structures.  The choice of which assembly will be constructed is determined by the trigger. 

Some of this phenomenology was captured in a simple physical model of self-assembly~\cite{Murugan15}. This model considers many types of particles existing on a square lattice, while the ``memories'' are specific large contiguous arrangements of the particles as shown in Fig.~\ref{fig:selfassembly}a. To encode any given structure in the mutual interactions, one introduces a strong bonding between any pair of particles that appear as a nearest-neighbor pair in the structure; particles that are not neighbors have no attractive interactions. To encode multiple structures, the bond strengths are simply added, in the spirit of the Hebbian rule in neural networks given by Eq.~\eqref{eq:hopfield-sum}. A particle may thus bind with any one of several different particles, each of which corresponds to a desired neighbor in one of the multiple stored target structures. 

One might imagine that this simple set of rules would inevitably cause errors in the assembly of a chosen structure from among the encoded memories, analogous to the part-fish-part-boss chimera that we sought to avoid in the Hopfield model. However, it was shown that because each particle has multiple neighbors (four in the case of the square lattice), its bonds have enough specificity that this confusion can be averted.

Consider for a moment the trivial construction where each encoded structure uses a completely distinct set of blocks --- in this case only $m = N/M$ different structures can be stored, where $N$ is the number of block types and $M$ is the size of the structures. Using instead the Hebbian-like scheme, one can safely encode up to $m_c \sim \sqrt{N}(N / M)$ structures on a square lattice. 
The different memories can be retrieved by using various triggers, such as introducing a seed, or increasing certain concentrations or bond strengths so that a seed forms spontaneously. A diagram showing the various possible outcomes of such an assembly model is shown in Fig.~\ref{fig:selfassembly}b. 
Some possible connections between the self-assembly system and variants of the neural network model have been carefully elaborated~\cite{Murugan15,Zhong17}.

The relatively simple theoretical model of self-assembly described here suggests that one can design materials that change states and transform at will, while using a limited pool of constituents~\cite{Zeravcic17}. Recent experimental advances suggest that such materials will soon be achieved in the laboratory. In particular, DNA is proving a most promising ingredient as it has been used both as a building block and as a mediator of interactions that are highly designable~\cite{Rogers16,Ong2017,Zhang2018}.

\section{Memory of initial conditions in dynamics}

\begin{figure}
\begin{center}
\includegraphics[width=3.3in]{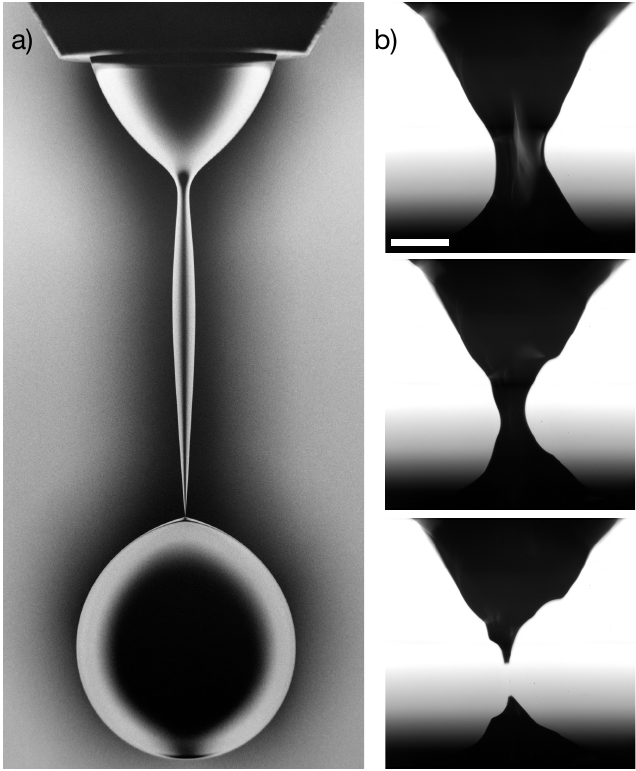}
\end{center}
\caption{\textbf{Forgetting and remembering initial conditions.}
\textbf{(a)} A drop of glycerol falls and pinches off from a faucet submerged in silicone oil. Near the top of the spherical drop, the dynamics become singular and have no dependence on the drop's initial conditions. Nozzle diameter is 0.48~cm. 
\textbf{(b)} Detail of a bubble pinching off from an underwater nozzle located just below the images. The black region is the rising air bubble and the surrounding white and gray is the water.  The pinchoff is seen in the bottom frame where the black regions become disconnected.
The dynamics become singular, but remember initial conditions strongly---in this case, the bubble's release from an asymmetrically-shaped nozzle. Scale bar is 0.5~mm and the images are 122~$\mu$s apart. 
Panel \textit{a} reproduced from~\citet{Cohen99} with permission, and panel \textit{b} reproduced from~\citet{Keim06} with permission.
\label{fig:singularity} }
\end{figure}

We conclude our survey of different forms of memory formation with a discussion of the memory of initial conditions that occur in experiments on dynamical systems.  In Fig.~\ref{fig:singularity} we show two situations where a fluid fissions into two parts.  On the left is an image of a viscous drop of glycerol breaking apart inside a bath of silicone oil of nearly equal viscosity.  The three images on the right show the rupture of an air bubble rising from an underwater nozzle.  The viscous liquid appears pristine --- symmetrical and smooth as the drop disconnects --- while the air bubble looks jagged as if it were being torn apart.  As we will indicate below, these examples demonstrate dynamics in which initial conditions, which are irrelevant for some systems, can in other situations influence the physical outcomes; they indicate a difference in terms of what initial conditions have been remembered by the fluids.
 
These are examples of a finite-time singularity in which a fluid mass (viscous liquid or air) detaches from a surface~\cite{Keller83,Goldstein93,Eggers94,Shi94,Eggers97}.  First a fluid neck stretches from the drop to the remaining fluid clinging to the surface.  The neck radius becomes progressively thinner until at some point it shrinks to atomic dimensions at which point the liquid has separated into two parts. The Laplace pressure between the inside and the outside of the drop, $P$, is proportional to the local curvature of the fluid, $\kappa$, and to the interfacial tension of the liquid, $\sigma$, that is, $P = \kappa \sigma$.  As the neck radius shrinks, $\kappa$ diverges so that there is a singularity in the pressure at the point of disconnection.  At this point, other physical quantities, such as fluid velocities or viscous stresses, may diverge as well.
 
This finite-time singularity is determined by a balance of terms in the Navier-Stokes equations governing the fluid flow.  Depending on the properties of the internal and external fluids (i.e., the fluid making up the drop and the fluid into which the drop is falling), there will be different similarity solutions for the singularity.  For the case of droplet snapoff, these situations fall into a discrete set of ``universality classes''; the singularities within a given class share behavioral signatures with other breakup events in the same class~\cite{Cohen01}. The nature of the singularity depends on a few physical dimensions of the apparatus and on the properties of the fluids, which for simple Newtonian fluids are the inner and outer viscosities, the inner and outer densities, and the interfacial tension.
 
The universality classes themselves can be divided up into two qualitatively distinct kinds. 
In the first kind, the singularity does not depend on the initial conditions but only depends on the liquid properties.  This is the situation for the breakup in Fig.~\ref{fig:singularity}a.  Because the process spans several decades in length- and time-scales, we do not expect initial conditions to influence the dynamics just before disconnection.  The shape of the nozzle, for example, does not influence the behavior about the point of breakup. This expectation of universality is borne out in pinch-offs with a wide variety of inner and outer fluids \cite{Eggers97, Cohen01}, making these non-equilibrium systems notable for their tendency to forget their initial conditions.
 
The second kind of singularity is extremely sensitive to the initial conditions as in the example of the air bubble in Fig.~\ref{fig:singularity}b.  The jagged and asymmetrical shape of the connecting neck of air depends sensitively on how the bubble was released from the nozzle.  The initial shape of the neck sets the amplitudes of a set of modes that proceed to oscillate with constant amplitudes and diverging frequencies while the average size of the neck shrinks \cite{Schmidt09}. Near the singularity, the relative amplitude of the oscillations with respect to the neck radius diverges. The uncertainty in a measurement of the \emph{amplitudes} grows but never overwhelms the signal, permitting the memory to be read out.  By contrast, the initial \emph{phase} of each mode is preserved but because its frequency diverges, the uncertainty in its measurements quickly exceeds $\pm \pi$.  Thus effectively, only half the initial conditions are remembered~\cite{Schmidt09}.  This example illustrates an idea we have used implicitly throughout this review: it is important not only to store a memory but to be able to read it out as well.

The two classes of breakup behavior presented here differ in their memory of initial conditions.  The falling liquid drop is insensitive to its evolution history and has lost the memory of its initial conditions. 
In contrast, the rising air bubble is highly sensitive to the forces present during its lifetime and remembers a great deal about how it was initiated.  These examples present a clear dichotomy of how in some cases it is nature while in other cases (that may naively appear similar) it is nurture that determines the fate.  Memory and its loss still surprise.

The type of memory that we have discussed in this section deals with the \emph{dynamics} of the system, such as remembering initial conditions. The basic forms of this memory are the conservation laws that we usually take for granted.  Remembering initial conditions may appear to have a somewhat different character than some of the memories mentioned in previous sections, in which the input is embedded in a system's steady-state configuration and can be extracted at an arbitrary later time.  However, there are also similarities with some of those more conventional material memories.  In echoes, the memory also appears in the dynamics as the spins return to their original phases relative to one another; likewise in the Kovacs effect, the memory appears only once in the relaxation dynamics of the material. Note also that in the present case of the dynamics of the disconnecting air bubble, the memory is not as fleeting as it first seems. While the memory of the amplitude is \emph{directly} observable only in the brief moments before the singularity, aspects of the dynamics can still be reconstructed because of the lasting effect that the initial conditions have on the size distribution and trajectories of the satellite bubbles~\cite{Keim11b}.

\section{Conclusions}

Focusing on memory storage and recall across the entire range of material properties has not been a widespread approach for studying matter. 
However, such a study provides a common language for physicists to discuss some of the most fascinating aspects of history-dependence, dynamics, and information storage in far-from-equilibrium systems. 
As Twyla Tharp suggests, each new memory can provide a novel way to perceive material behavior and thus may lead to an alternate framework for organizing our knowledge about matter in general. 
This prompts us to seek behaviors, patterns, and similarities that may have been overlooked. 
It can also lead to new experimental and theoretical endeavors because it compels us to ask questions of a somewhat unusual --- and therefore novel --- nature. 
This perspective may lead to new insights and even the creation of new forms of matter with unique functionality dictated by the memories stored in their creation.

One lesson from a comparative study of memory is that we should not assume that a memory behavior is unique, because we can often find another system---sometimes with no superficial similarities---that obeys the same rules for storage and recall. Thus, among the many memory behaviors we have encountered we perceive a smaller number of generic types and principles. The preceding sections have provided some 
examples of how different memories are related. 
Here we summarize some of the most relevant threads that connect or delineate them, and that motivate a broader perspective on memory formation.

\subsection{Some common threads and distinctions}

\textit{Hysteresis---}
We have tried to draw a few conclusions about ways of uniting different phenomena under some common principles. In Sec.~\ref{sec:simplest} we noted several systems that remember the direction or sign of the input that was applied most recently. This  simple principle underlies most of the world's digital memory and is the building block of the much richer return-point behavior (Sec.~\ref{sec:rpm}). 

\textit{Maximal input---} 
In Sec.\ \ref{sec:simplest}, we grouped together another set of memories that shared the somewhat obvious property that they all remembered a maximal value of an applied input.  This form of memory is rather common and has been found in many materials and with many different probes.  

\textit{Oscillatory training over multiple cycles---} 
Less obvious are the memories encoded in physical systems subjected to oscillatory perturbations.  While these seem to have rather different forms, some of these phenomena appear to be related.  A good example is the observation of multiple transient memories in charge-density-wave systems, in sheared non-Brownian suspensions, and in paths worn by visitors in grass.  These memories have the same overall phenomenology despite one of them arising in a hard-condensed matter system, another in a fluid, and the third being a macroscopic human-scale phenomenon.  All three initially learn many inputs and then forget all but one (or two) of them; however if noise is present they can remember all inputs indefinitely. 

\textit{Reversal of dynamics---} 
More ambitious and less obvious was our attempt to group echo phenomena together with examples of rejuvenation and aging in glasses and with the Kovacs effect.  The point of commonality was taken to be that they each manipulate their respective system as a way of ``reversing paths.''  This allows them to retrace their dynamical evolution to the instant when the phases were initially set to be coherent.  We also suggested that the Kovacs effect may likewise be conceived as allowing a system's relaxation dynamics to be reversed in time.

\bigskip

Besides these common phenomenologies, we can discern common physical principles among some memory-forming systems:

\textit{Memory as a marginal state or an energy minimum---}
In the example of non-Brownian dilute suspensions, a memory is encoded in a steady state with reversible particle motions, which is reached by shearing with a constant training amplitude. Yet one reads out that memory not by observing the steady state, but by observing the onset of irreversible motion when the training amplitude is exceeded slightly (Fig.~\ref{fig:mtm}a). Thus the steady state encodes a memory by being \emph{marginally stable} at the training amplitude. More generally, this memory is where many quiescent portions of the system promptly begin changing and participating in the dynamics (\emph{e.g.}, as measured by the shear stress in Fig.~\ref{fig:mtm}b). We can see this pattern in the single- and multiple-memory behaviors of dilute suspensions, but also in the Mullins and Kaiser effects, return-point memory, and charge-density waves. Most of these memories are indicated by a change in the first derivative of an observable such as particle displacements, magnetization, or stress. We can contrast these examples with the Hopfield model, in which a memory corresponds to the \emph{bottom} of a potential energy well.

\textit{Role of quenched or unquenched disorder---} 
Disorder can give a system the ability to distinguish among, and remember, many possible input values. For example, a system with return-point memory that can be modeled with many \emph{identical} hysterons (\textit{i.e.}, having identical hysteresis loops) would store no more information than a single hysteron. Instead, because the transition thresholds of the hysterons are distributed broadly (typically due to defects), 
a sufficiently large system can make fine distinctions among values of the applied field. Thus, in some of the systems we have described, disorder plays a beneficial role in providing a large ensemble of subsystems that together allow a nearly continuous range of input values to be stored. 
This principle also applies to the case of the charge-density wave that exhibits multiple transient memories, where the charge density evolves to find a metastable state in the presence of disordered pinning sites \cite{Coppersmith87b}.

In both these examples the disorder cannot evolve (\emph{i.e.}, it is quenched), 
and is encoded in the system's Hamiltonian. However in other systems 
the disorder itself can evolve. To see this,
one need only think about aging and rejuvenation in molecular glasses, % for consistency
exemplified by the thermal sorting algorithm with that behavior~\cite{Zou10}. There, every part of the system can change. The disorder is the medium in which to encode memories, and yet in small systems it can be annealed away entirely.

\textit{Training by following a path---} 
In considering these examples there is another aspect that appears to lead to different behaviors. In the case of the multiple transient memories in suspensions, the \textit{path} of the dynamics in phase space is remembered; it is not only the marginal final configuration that is trained. We observe the same memory of a path in charge-density wave conductors and jammed solids, and the notion of a path is even more explicit in the model of park benches on a lawn. In these examples training must typically be repetitive as the system learns the path. 
This is distinct from memories in the Hopfield model or self-assembly where the memory is stored (at least in some rules) by varying the interaction strengths between pairs of spins or particles. In those associative memories, a well-chosen learning rule deterministically sets the interactions to store a memory, without the need for any dynamics in the training.

\subsection{Making materials functional}

That a material carries a memory of its formation suggests that one may use that memory to create some specific functionality based on the stored information. Memory could inform the design and processing of new materials with a range of applications.

\textit{Amplitude variation---} 
Oscillatory manipulation of a solid is common and is, indeed, the essential ingredient of most AC (linear and non-linear) susceptibility measurements.  By focusing on the particular aspect of \textit{memory}, 
one is motivated to concentrate 
not only on how the response depends on the amplitude of the oscillation, but on how it depends on changes to the amplitude during the experiment --- or, in more practical terms, during the material's preparation and use. In systems such as dilute suspensions and jammed solids, amplitude variation is the basis for reading out memories---and for recognizing that memories are present in the first place.

\textit{Learned mechanical properties---} In systems where memory storage is in response to an external stress, as in the examples of dilute suspensions or jammed solids, memory can have mechanical consequences. Memories formed by cyclic training in these systems show that one can tailor a material's responses at specific amplitudes or values of strain (Fig.~\ref{fig:mtm}b). An example of how such a mechanism may be used to reduce viscous dissipation in continuously-flowing 
suspensions or grains may be found in the works of \citet{Lin16,Ness2018}. 
Rigid disordered materials can become auxetic (\textit{i.e.,} have a negative Poisson's ratio), or even develop a coordinated response at one location due to a stress applied elsewhere, when certain bonds are cut~\cite{Goodrich15,Hexner18,Hexner18PRE,Rocks17,Yan17,Rocks19}. A material that is aged under an imposed stress can remember its history and respond by altering the bonds that are stressed the most.  This, as in the case of cutting bonds, can lead to novel elastic response~\cite{Pashine2019}. 
Mechanical memory of this sort may also be one way that living matter can adapt to its environment \cite{Bieling:2016cg,Majumdar:2018kc}.

\textit{Memory capacity---}
Another type of question 
that leads to novel investigations is 
the memory capacity in a system.  For example, the pulse-duration memory in charge-density wave solids or the approach to the absorbing-state transition in non-Brownian suspensions, have shown an even richer phenomenology when, in addition to addressing the creation of a single memory, one asks how many different memories can be encoded in these systems. In the case of self-assembly, the goal of encoding a greater variety of structures with fewer types of particles 
motivates further development of theoretical design principles and experimental systems. 
Disorder and diversity 
are especially relevant to the design of multiple-memory systems because they can make various parts or coarse-grained scales of the system susceptible to different inputs. 

\bigskip

Throughout this paper we have raised a number of questions about memory that enrich our study of the physics of matter.  We have indicated a few, but certainly not all, of the ways in which memories are similar or dissimilar to one another and have suggested that addressing such questions can lead not only to interesting questions about the variety and robustness of memory formation but also to useful applications, once the science of material memory is better understood.  

In her novel \emph{Mansfield Park}, Jane Austin's protagonist claimed: ``If any one faculty of our nature may be called more wonderful than the rest, I do think it is memory. \dots  our powers of recollecting and of forgetting do seem peculiarly past finding out.'' 
While in our lived experience this may indeed be true, in the world of materials, the mysteries of memory formation have been yielding to acute scientific examination. 
Thus the ideas of memory formation, and the mechanisms that nature uses to implant information of different kinds into matter, can be articulated into sets of coherent principles that seem to bind them together. 
However, much work still needs to be done to delineate the distinctions between different types and provide a common framework for understanding the interconnection between them.  We hope that such questions strike a chord within the broader research community and call attention to this novel and exciting branch of inquiry.

\section{Acknowledgements}  

This article was inspired by a program at the Kavli Institute for Theoretical Physics in Santa Barbara, CA that was held during the Winter of 2018.  We are grateful to Greg Huber at the KITP who helped shepherd the program in its early stages.  Two of us (SRN and SS) would like to thank the other co-organizers of this program: Susan Coppersmith and Alan Middleton.  We had wonderful discussions with the attendees of this program.  In particular we would like to thank Itai Cohen, Leticia Cugliandolo, Karin Dahmen, Karen Daniels, Chandan Dasgupta, Greg Huber, Mehran Kardar, Yoav Lahini, Craig Maloney, Enzo Marinari, Muhittin Mungan, Arvind Murugan, Lev Truskinovsky and Tom Witten for very useful and inspiring feedback on many of the ideas presented here. 
We are grateful to the KITP for its hospitality during this program, supported in part by NSF PHY-1748958.  
In addition this work was supported by the NSF MRSEC Program DMR-1420709, NSF DMR-1404841 and DOE DE-FG02-03ER46088 and the Simons Foundation for the collaboration ``Cracking the Glass Problem'' award \#348125 at the University of Chicago (SRN). 
JDP acknowledges the Donors of the American Chemical Society Petroleum Research Fund for partial support of this work. 
NCK acknowledges support from NSF DMR-1708870. SS acknowledges support through the J C Bose Fellowship, SERB, DST, India. 

\bibliography{references}

%merlin.mbs apsrmp4-1.bst 2010-07-25 4.21a (PWD, AO, DPC) hacked
%Control: author (11) reversed first initials
\begin{thebibliography}{171}%
\makeatletter
\providecommand \@ifxundefined [1]{%
 \@ifx{#1\undefined}
}%
\providecommand \@ifnum [1]{%
 \ifnum #1\expandafter \@firstoftwo
 \else \expandafter \@secondoftwo
 \fi
}%
\providecommand \@ifx [1]{%
 \ifx #1\expandafter \@firstoftwo
 \else \expandafter \@secondoftwo
 \fi
}%
\providecommand \natexlab [1]{#1}%
\providecommand \enquote  [1]{``#1''}%
\providecommand \bibnamefont  [1]{#1}%
\providecommand \bibfnamefont [1]{#1}%
\providecommand \citenamefont [1]{#1}%
\providecommand \href@noop [0]{\@secondoftwo}%
\providecommand \href [0]{\begingroup \@sanitize@url \@href}%
\providecommand \@href[1]{\@@startlink{#1}\@@href}%
\providecommand \@@href[1]{\endgroup#1\@@endlink}%
\providecommand \@sanitize@url [0]{\catcode `\\12\catcode `\$12\catcode
  `\&12\catcode `\#12\catcode `\^12\catcode `\_12\catcode `\%12\relax}%
\providecommand \@@startlink[1]{}%
\providecommand \@@endlink[0]{}%
\providecommand \url  [0]{\begingroup\@sanitize@url \@url }%
\providecommand \@url [1]{\endgroup\@href {#1}{\urlprefix }}%
\providecommand \urlprefix  [0]{URL }%
\providecommand \Eprint [0]{\href }%
\providecommand \doibase [0]{http://dx.doi.org/}%
\providecommand \selectlanguage [0]{\@gobble}%
\providecommand \bibinfo  [0]{\@secondoftwo}%
\providecommand \bibfield  [0]{\@secondoftwo}%
\providecommand \translation [1]{[#1]}%
\providecommand \BibitemOpen [0]{}%
\providecommand \bibitemStop [0]{}%
\providecommand \bibitemNoStop [0]{.\EOS\space}%
\providecommand \EOS [0]{\spacefactor3000\relax}%
\providecommand \BibitemShut  [1]{\csname bibitem#1\endcsname}%
\let\auto@bib@innerbib\@empty
%</preamble>
\bibitem [{\citenamefont {Ackerson}\ and\ \citenamefont
  {Pusey}(1988)}]{Ackerson88}%
  \BibitemOpen
  \bibfield  {author} {\bibinfo {author} {\bibnamefont {Ackerson},
  \bibfnamefont {B.}}, \ and\ \bibinfo {author} {\bibfnamefont
  {P.}~\bibnamefont {Pusey}}} (\bibinfo {year} {1988}),\ \href@noop {}
  {\bibfield  {journal} {\bibinfo  {journal} {Phys. Rev. Lett.}\ }\textbf
  {\bibinfo {volume} {61}}~(\bibinfo {number} {8}),\ \bibinfo {pages}
  {1033}}\BibitemShut {NoStop}%
\bibitem [{\citenamefont {Adhikari}\ and\ \citenamefont
  {Sastry}(2018)}]{Adhikari18}%
  \BibitemOpen
  \bibfield  {author} {\bibinfo {author} {\bibnamefont {Adhikari},
  \bibfnamefont {M.}}, \ and\ \bibinfo {author} {\bibfnamefont
  {S.}~\bibnamefont {Sastry}}} (\bibinfo {year} {2018}),\ \href@noop {}
  {\bibfield  {journal} {\bibinfo  {journal} {Eur. Phys. J. E}\ }\textbf
  {\bibinfo {volume} {41}},\ \bibinfo {pages} {045504}}\BibitemShut {NoStop}%
\bibitem [{\citenamefont {Amir}\ \emph {et~al.}(2012)\citenamefont {Amir},
  \citenamefont {Oreg},\ and\ \citenamefont {Imry}}]{Amir12}%
  \BibitemOpen
  \bibfield  {author} {\bibinfo {author} {\bibnamefont {Amir}, \bibfnamefont
  {A.}}, \bibinfo {author} {\bibfnamefont {Y.}~\bibnamefont {Oreg}}, \ and\
  \bibinfo {author} {\bibfnamefont {Y.}~\bibnamefont {Imry}}} (\bibinfo {year}
  {2012}),\ \href@noop {} {\bibfield  {journal} {\bibinfo  {journal} {Proc.
  Natl. Acad. Sci.}\ }\textbf {\bibinfo {volume} {109}}~(\bibinfo {number}
  {6}),\ \bibinfo {pages} {1850}}\BibitemShut {NoStop}%
\bibitem [{\citenamefont {Amit}(1989)}]{amit_1989}%
  \BibitemOpen
  \bibfield  {author} {\bibinfo {author} {\bibnamefont {Amit}, \bibfnamefont
  {D.~J.}}} (\bibinfo {year} {1989}),\ \href {\doibase
  10.1017/CBO9780511623257} {\emph {\bibinfo {title} {Modeling Brain Function:
  The World of Attractor Neural Networks}}}\ (\bibinfo  {publisher} {Cambridge
  University Press})\BibitemShut {NoStop}%
\bibitem [{\citenamefont {Amit}\ and\ \citenamefont {Fusi}(1994)}]{amitfusi}%
  \BibitemOpen
  \bibfield  {author} {\bibinfo {author} {\bibnamefont {Amit}, \bibfnamefont
  {D.~J.}}, \ and\ \bibinfo {author} {\bibfnamefont {S.}~\bibnamefont {Fusi}}}
  (\bibinfo {year} {1994}),\ \href {\doibase 10.1162/neco.1994.6.5.957}
  {\bibfield  {journal} {\bibinfo  {journal} {Neural Computation}\ }\textbf
  {\bibinfo {volume} {6}}~(\bibinfo {number} {5}),\ \bibinfo {pages}
  {957}}\BibitemShut {NoStop}%
\bibitem [{\citenamefont {Amit}\ \emph {et~al.}(1985)\citenamefont {Amit},
  \citenamefont {Gutfreund},\ and\ \citenamefont {Sompolinsky}}]{Amit1985}%
  \BibitemOpen
  \bibfield  {author} {\bibinfo {author} {\bibnamefont {Amit}, \bibfnamefont
  {D.~J.}}, \bibinfo {author} {\bibfnamefont {H.}~\bibnamefont {Gutfreund}}, \
  and\ \bibinfo {author} {\bibfnamefont {H.}~\bibnamefont {Sompolinsky}}}
  (\bibinfo {year} {1985}),\ \href {\doibase 10.1103/PhysRevLett.55.1530}
  {\bibfield  {journal} {\bibinfo  {journal} {Phys. Rev. Lett.}\ }\textbf
  {\bibinfo {volume} {55}}~(\bibinfo {number} {14}),\ \bibinfo {pages}
  {1530}}\BibitemShut {NoStop}%
\bibitem [{\citenamefont {Bacot}\ \emph {et~al.}(2016)\citenamefont {Bacot},
  \citenamefont {Labousse}, \citenamefont {Eddi}, \citenamefont {Fink},\ and\
  \citenamefont {Fort}}]{Bacot16}%
  \BibitemOpen
  \bibfield  {author} {\bibinfo {author} {\bibnamefont {Bacot}, \bibfnamefont
  {V.}}, \bibinfo {author} {\bibfnamefont {M.}~\bibnamefont {Labousse}},
  \bibinfo {author} {\bibfnamefont {A.}~\bibnamefont {Eddi}}, \bibinfo {author}
  {\bibfnamefont {M.}~\bibnamefont {Fink}}, \ and\ \bibinfo {author}
  {\bibfnamefont {E.}~\bibnamefont {Fort}}} (\bibinfo {year} {2016}),\
  \href@noop {} {\bibfield  {journal} {\bibinfo  {journal} {Nature Physics}\
  }\textbf {\bibinfo {volume} {12}}~(\bibinfo {number} {10}),\ \bibinfo {pages}
  {972}}\BibitemShut {NoStop}%
\bibitem [{\citenamefont {Bannantine}\ \emph {et~al.}(1990)\citenamefont
  {Bannantine}, \citenamefont {Comer},\ and\ \citenamefont
  {Handrock}}]{Bannantine90}%
  \BibitemOpen
  \bibfield  {author} {\bibinfo {author} {\bibnamefont {Bannantine},
  \bibfnamefont {J.~A.}}, \bibinfo {author} {\bibfnamefont {J.~J.}\
  \bibnamefont {Comer}}, \ and\ \bibinfo {author} {\bibfnamefont {J.~L.}\
  \bibnamefont {Handrock}}} (\bibinfo {year} {1990}),\ \href@noop {} {\emph
  {\bibinfo {title} {Fundamentals of Metal Fatigue Analysis}}},\ Vol.~\bibinfo
  {volume} {90}\ (\bibinfo  {publisher} {Prentice hall Englewood Cliffs,
  NJ})\BibitemShut {NoStop}%
\bibitem [{\citenamefont {Barker}\ \emph {et~al.}(1983)\citenamefont {Barker},
  \citenamefont {Schreiber}, \citenamefont {Huth},\ and\ \citenamefont
  {Everett}}]{Barker83}%
  \BibitemOpen
  \bibfield  {author} {\bibinfo {author} {\bibnamefont {Barker}, \bibfnamefont
  {J.}}, \bibinfo {author} {\bibfnamefont {D.}~\bibnamefont {Schreiber}},
  \bibinfo {author} {\bibfnamefont {B.}~\bibnamefont {Huth}}, \ and\ \bibinfo
  {author} {\bibfnamefont {D.~H.}\ \bibnamefont {Everett}}} (\bibinfo {year}
  {1983}),\ \href@noop {} {\bibfield  {journal} {\bibinfo  {journal} {Proc. R.
  Soc. Lond. A}\ }\textbf {\bibinfo {volume} {386}},\ \bibinfo {pages}
  {251}}\BibitemShut {NoStop}%
\bibitem [{\citenamefont {Barton}\ \emph {et~al.}(2015)\citenamefont {Barton},
  \citenamefont {Kardar},\ and\ \citenamefont {Chakraborty}}]{Barton15}%
  \BibitemOpen
  \bibfield  {author} {\bibinfo {author} {\bibnamefont {Barton}, \bibfnamefont
  {J.~P.}}, \bibinfo {author} {\bibfnamefont {M.}~\bibnamefont {Kardar}}, \
  and\ \bibinfo {author} {\bibfnamefont {A.~K.}\ \bibnamefont {Chakraborty}}}
  (\bibinfo {year} {2015}),\ \href@noop {} {\bibfield  {journal} {\bibinfo
  {journal} {Proceedings of the National Academy of Sciences}\ }\textbf
  {\bibinfo {volume} {112}}~(\bibinfo {number} {7}),\ \bibinfo {pages}
  {1965}}\BibitemShut {NoStop}%
\bibitem [{\citenamefont {Bellon}\ \emph {et~al.}(2002)\citenamefont {Bellon},
  \citenamefont {Ciliberto},\ and\ \citenamefont {Laroche}}]{Bellon02}%
  \BibitemOpen
  \bibfield  {author} {\bibinfo {author} {\bibnamefont {Bellon}, \bibfnamefont
  {L.}}, \bibinfo {author} {\bibfnamefont {S.}~\bibnamefont {Ciliberto}}, \
  and\ \bibinfo {author} {\bibfnamefont {C.}~\bibnamefont {Laroche}}} (\bibinfo
  {year} {2002}),\ \href@noop {} {\bibfield  {journal} {\bibinfo  {journal}
  {The European Physical Journal B-Condensed Matter and Complex Systems}\
  }\textbf {\bibinfo {volume} {25}}~(\bibinfo {number} {2}),\ \bibinfo {pages}
  {223}}\BibitemShut {NoStop}%
\bibitem [{\citenamefont {Benna}\ and\ \citenamefont
  {Fusi}(2015)}]{BennaFusi2015}%
  \BibitemOpen
  \bibfield  {author} {\bibinfo {author} {\bibnamefont {Benna}, \bibfnamefont
  {M.~K.}}, \ and\ \bibinfo {author} {\bibfnamefont {S.}~\bibnamefont {Fusi}}}
  (\bibinfo {year} {2015}),\ \href@noop {} {\bibfield  {journal} {\bibinfo
  {journal} {arXiv.org}\ }}\Eprint {http://arxiv.org/abs/1507.07580}
  {1507.07580} \BibitemShut {NoStop}%
\bibitem [{\citenamefont {Berthier}\ and\ \citenamefont
  {Bouchaud}(2002)}]{Berthier02}%
  \BibitemOpen
  \bibfield  {author} {\bibinfo {author} {\bibnamefont {Berthier},
  \bibfnamefont {L.}}, \ and\ \bibinfo {author} {\bibfnamefont {J.-P.}\
  \bibnamefont {Bouchaud}}} (\bibinfo {year} {2002}),\ \href@noop {} {\bibfield
   {journal} {\bibinfo  {journal} {Physical Review B}\ }\textbf {\bibinfo
  {volume} {66}}~(\bibinfo {number} {5}),\ \bibinfo {pages}
  {054404}}\BibitemShut {NoStop}%
\bibitem [{\citenamefont {Bertin}\ \emph {et~al.}(2003)\citenamefont {Bertin},
  \citenamefont {Bouchaud}, \citenamefont {Drouffe},\ and\ \citenamefont
  {Godreche}}]{Bertin03}%
  \BibitemOpen
  \bibfield  {author} {\bibinfo {author} {\bibnamefont {Bertin}, \bibfnamefont
  {E.~M.}}, \bibinfo {author} {\bibfnamefont {J.}~\bibnamefont {Bouchaud}},
  \bibinfo {author} {\bibfnamefont {J.}~\bibnamefont {Drouffe}}, \ and\
  \bibinfo {author} {\bibfnamefont {C.}~\bibnamefont {Godreche}}} (\bibinfo
  {year} {2003}),\ \href@noop {} {\bibfield  {journal} {\bibinfo  {journal}
  {Journal of physics A: mathematical and general}\ }\textbf {\bibinfo {volume}
  {36}}~(\bibinfo {number} {43}),\ \bibinfo {pages} {10701}}\BibitemShut
  {NoStop}%
\bibitem [{\citenamefont {Bhattacharya}(2003)}]{Bhattacharya03}%
  \BibitemOpen
  \bibfield  {author} {\bibinfo {author} {\bibnamefont {Bhattacharya},
  \bibfnamefont {K.}}} (\bibinfo {year} {2003}),\ \href@noop {} {\emph
  {\bibinfo {title} {Microstructure of martensite: why it forms and how it
  gives rise to the shape-memory effect}}},\ Vol.~\bibinfo {volume} {2}\
  (\bibinfo  {publisher} {Oxford University Press})\BibitemShut {NoStop}%
\bibitem [{\citenamefont {Bieling}\ \emph {et~al.}(2016)\citenamefont
  {Bieling}, \citenamefont {Li}, \citenamefont {Weichsel}, \citenamefont
  {McGorty}, \citenamefont {Jreij}, \citenamefont {Huang}, \citenamefont
  {Fletcher},\ and\ \citenamefont {Mullins}}]{Bieling:2016cg}%
  \BibitemOpen
  \bibfield  {author} {\bibinfo {author} {\bibnamefont {Bieling}, \bibfnamefont
  {P.}}, \bibinfo {author} {\bibfnamefont {T.-D.}\ \bibnamefont {Li}}, \bibinfo
  {author} {\bibfnamefont {J.}~\bibnamefont {Weichsel}}, \bibinfo {author}
  {\bibfnamefont {R.}~\bibnamefont {McGorty}}, \bibinfo {author} {\bibfnamefont
  {P.}~\bibnamefont {Jreij}}, \bibinfo {author} {\bibfnamefont
  {B.}~\bibnamefont {Huang}}, \bibinfo {author} {\bibfnamefont {D.~A.}\
  \bibnamefont {Fletcher}}, \ and\ \bibinfo {author} {\bibfnamefont {R.~D.}\
  \bibnamefont {Mullins}}} (\bibinfo {year} {2016}),\ \href@noop {} {\bibfield
  {journal} {\bibinfo  {journal} {Cell}\ }\textbf {\bibinfo {volume}
  {164}}~(\bibinfo {number} {1-2}),\ \bibinfo {pages} {115}}\BibitemShut
  {NoStop}%
\bibitem [{\citenamefont {Bouchaud}\ \emph {et~al.}(2001)\citenamefont
  {Bouchaud}, \citenamefont {Dupuis}, \citenamefont {Hammann},\ and\
  \citenamefont {Vincent}}]{Bouchaud01}%
  \BibitemOpen
  \bibfield  {author} {\bibinfo {author} {\bibnamefont {Bouchaud},
  \bibfnamefont {J.-P.}}, \bibinfo {author} {\bibfnamefont {V.}~\bibnamefont
  {Dupuis}}, \bibinfo {author} {\bibfnamefont {J.}~\bibnamefont {Hammann}}, \
  and\ \bibinfo {author} {\bibfnamefont {E.}~\bibnamefont {Vincent}}} (\bibinfo
  {year} {2001}),\ \href@noop {} {\bibfield  {journal} {\bibinfo  {journal}
  {Physical Review B}\ }\textbf {\bibinfo {volume} {65}}~(\bibinfo {number}
  {2}),\ \bibinfo {pages} {024439}}\BibitemShut {NoStop}%
\bibitem [{\citenamefont {Bouchbinder}\ and\ \citenamefont
  {Langer}(2010)}]{Bouchbinder2010}%
  \BibitemOpen
  \bibfield  {author} {\bibinfo {author} {\bibnamefont {Bouchbinder},
  \bibfnamefont {E.}}, \ and\ \bibinfo {author} {\bibfnamefont {J.~S.}\
  \bibnamefont {Langer}}} (\bibinfo {year} {2010}),\ \href {\doibase
  10.1039/c001388a} {\bibfield  {journal} {\bibinfo  {journal} {Soft Matter}\
  }\textbf {\bibinfo {volume} {6}}~(\bibinfo {number} {13}),\ \bibinfo {pages}
  {3065}}\BibitemShut {NoStop}%
\bibitem [{\citenamefont {Brown}\ \emph {et~al.}(1986)\citenamefont {Brown},
  \citenamefont {Gr{\"u}ner},\ and\ \citenamefont {Mih{\'a}ly}}]{Brown86}%
  \BibitemOpen
  \bibfield  {author} {\bibinfo {author} {\bibnamefont {Brown}, \bibfnamefont
  {S.~E.}}, \bibinfo {author} {\bibfnamefont {G.}~\bibnamefont {Gr{\"u}ner}}, \
  and\ \bibinfo {author} {\bibfnamefont {L.}~\bibnamefont {Mih{\'a}ly}}}
  (\bibinfo {year} {1986}),\ \href@noop {} {\bibfield  {journal} {\bibinfo
  {journal} {Solid State Comm.}\ }\textbf {\bibinfo {volume} {57}},\ \bibinfo
  {pages} {165}}\BibitemShut {NoStop}%
\bibitem [{\citenamefont {Budhu}(2010)}]{Budhu10}%
  \BibitemOpen
  \bibfield  {author} {\bibinfo {author} {\bibnamefont {Budhu}, \bibfnamefont
  {M.}}} (\bibinfo {year} {2010}),\ \href@noop {} {\emph {\bibinfo {title}
  {Soil mechanics and foundations}}}\ (\bibinfo  {publisher}
  {Wiley})\BibitemShut {NoStop}%
\bibitem [{\citenamefont {Burton}\ and\ \citenamefont
  {Nagel}(2016)}]{Burton16}%
  \BibitemOpen
  \bibfield  {author} {\bibinfo {author} {\bibnamefont {Burton}, \bibfnamefont
  {J.~C.}}, \ and\ \bibinfo {author} {\bibfnamefont {S.~R.}\ \bibnamefont
  {Nagel}}} (\bibinfo {year} {2016}),\ \href@noop {} {\bibfield  {journal}
  {\bibinfo  {journal} {Physical Review E}\ }\textbf {\bibinfo {volume}
  {93}}~(\bibinfo {number} {3}),\ \bibinfo {pages} {032905}}\BibitemShut
  {NoStop}%
\bibitem [{\citenamefont {Cantournet}\ \emph {et~al.}(2009)\citenamefont
  {Cantournet}, \citenamefont {Desmorat},\ and\ \citenamefont
  {Besson}}]{Cantournet09}%
  \BibitemOpen
  \bibfield  {author} {\bibinfo {author} {\bibnamefont {Cantournet},
  \bibfnamefont {S.}}, \bibinfo {author} {\bibfnamefont {R.}~\bibnamefont
  {Desmorat}}, \ and\ \bibinfo {author} {\bibfnamefont {J.}~\bibnamefont
  {Besson}}} (\bibinfo {year} {2009}),\ \href@noop {} {\bibfield  {journal}
  {\bibinfo  {journal} {International Journal of Solids and Structures}\
  }\textbf {\bibinfo {volume} {46}}~(\bibinfo {number} {11-12}),\ \bibinfo
  {pages} {2255}}\BibitemShut {NoStop}%
\bibitem [{\citenamefont {Carr}\ and\ \citenamefont {Purcell}(1954)}]{Carr54}%
  \BibitemOpen
  \bibfield  {author} {\bibinfo {author} {\bibnamefont {Carr}, \bibfnamefont
  {H.~Y.}}, \ and\ \bibinfo {author} {\bibfnamefont {E.~M.}\ \bibnamefont
  {Purcell}}} (\bibinfo {year} {1954}),\ \href@noop {} {\bibfield  {journal}
  {\bibinfo  {journal} {Physical Review}\ }\textbf {\bibinfo {volume}
  {94}}~(\bibinfo {number} {3}),\ \bibinfo {pages} {630}}\BibitemShut {NoStop}%
\bibitem [{\citenamefont {Chakraverty}\ \emph {et~al.}(2005)\citenamefont
  {Chakraverty}, \citenamefont {Bandyopadhyay}, \citenamefont {Chatterjee},
  \citenamefont {Dattagupta}, \citenamefont {Frydman}, \citenamefont
  {Sengupta},\ and\ \citenamefont {Sreeram}}]{Chakraverty05}%
  \BibitemOpen
  \bibfield  {author} {\bibinfo {author} {\bibnamefont {Chakraverty},
  \bibfnamefont {S.}}, \bibinfo {author} {\bibfnamefont {M.}~\bibnamefont
  {Bandyopadhyay}}, \bibinfo {author} {\bibfnamefont {S.}~\bibnamefont
  {Chatterjee}}, \bibinfo {author} {\bibfnamefont {S.}~\bibnamefont
  {Dattagupta}}, \bibinfo {author} {\bibfnamefont {A.}~\bibnamefont {Frydman}},
  \bibinfo {author} {\bibfnamefont {S.}~\bibnamefont {Sengupta}}, \ and\
  \bibinfo {author} {\bibfnamefont {P.~A.}\ \bibnamefont {Sreeram}}} (\bibinfo
  {year} {2005}),\ \href@noop {} {\bibfield  {journal} {\bibinfo  {journal}
  {Phys. Rev. B}\ }\textbf {\bibinfo {volume} {71}}~(\bibinfo {number} {5}),\
  \bibinfo {pages} {1205}}\BibitemShut {NoStop}%
\bibitem [{\citenamefont {Chan~Vili}(2007)}]{Chan-Vili07}%
  \BibitemOpen
  \bibfield  {author} {\bibinfo {author} {\bibnamefont {Chan~Vili},
  \bibfnamefont {Y.~Y.}}} (\bibinfo {year} {2007}),\ \href@noop {} {\bibfield
  {journal} {\bibinfo  {journal} {Textile Research Journal}\ }\textbf {\bibinfo
  {volume} {77}}~(\bibinfo {number} {5}),\ \bibinfo {pages} {290}}\BibitemShut
  {NoStop}%
\bibitem [{\citenamefont {Chluba}\ \emph {et~al.}(2015)\citenamefont {Chluba},
  \citenamefont {Ge}, \citenamefont {de~Miranda}, \citenamefont {Strobel},
  \citenamefont {Kienle}, \citenamefont {Quandt},\ and\ \citenamefont
  {Wuttig}}]{Chluba15}%
  \BibitemOpen
  \bibfield  {author} {\bibinfo {author} {\bibnamefont {Chluba}, \bibfnamefont
  {C.}}, \bibinfo {author} {\bibfnamefont {W.}~\bibnamefont {Ge}}, \bibinfo
  {author} {\bibfnamefont {R.~L.}\ \bibnamefont {de~Miranda}}, \bibinfo
  {author} {\bibfnamefont {J.}~\bibnamefont {Strobel}}, \bibinfo {author}
  {\bibfnamefont {L.}~\bibnamefont {Kienle}}, \bibinfo {author} {\bibfnamefont
  {E.}~\bibnamefont {Quandt}}, \ and\ \bibinfo {author} {\bibfnamefont
  {M.}~\bibnamefont {Wuttig}}} (\bibinfo {year} {2015}),\ \href@noop {}
  {\bibfield  {journal} {\bibinfo  {journal} {Science}\ }\textbf {\bibinfo
  {volume} {348}}~(\bibinfo {number} {6238}),\ \bibinfo {pages}
  {1004}}\BibitemShut {NoStop}%
\bibitem [{\citenamefont {Cohen}\ \emph {et~al.}(1999)\citenamefont {Cohen},
  \citenamefont {Brenner}, \citenamefont {Eggers},\ and\ \citenamefont
  {Nagel}}]{Cohen99}%
  \BibitemOpen
  \bibfield  {author} {\bibinfo {author} {\bibnamefont {Cohen}, \bibfnamefont
  {I.}}, \bibinfo {author} {\bibfnamefont {M.~P.}\ \bibnamefont {Brenner}},
  \bibinfo {author} {\bibfnamefont {J.}~\bibnamefont {Eggers}}, \ and\ \bibinfo
  {author} {\bibfnamefont {S.~R.}\ \bibnamefont {Nagel}}} (\bibinfo {year}
  {1999}),\ \href@noop {} {\bibfield  {journal} {\bibinfo  {journal} {Phys.
  Rev. Lett.}\ }\textbf {\bibinfo {volume} {83}}~(\bibinfo {number} {6}),\
  \bibinfo {pages} {1147}}\BibitemShut {NoStop}%
\bibitem [{\citenamefont {Cohen}\ and\ \citenamefont {Nagel}(2001)}]{Cohen01}%
  \BibitemOpen
  \bibfield  {author} {\bibinfo {author} {\bibnamefont {Cohen}, \bibfnamefont
  {I.}}, \ and\ \bibinfo {author} {\bibfnamefont {S.~R.}\ \bibnamefont
  {Nagel}}} (\bibinfo {year} {2001}),\ \href@noop {} {\bibfield  {journal}
  {\bibinfo  {journal} {Phys. Fluids}\ }\textbf {\bibinfo {volume}
  {13}}~(\bibinfo {number} {12}),\ \bibinfo {pages} {3533}}\BibitemShut
  {NoStop}%
\bibitem [{\citenamefont {Coppersmith}(1987)}]{Coppersmith87a}%
  \BibitemOpen
  \bibfield  {author} {\bibinfo {author} {\bibnamefont {Coppersmith},
  \bibfnamefont {S.~N.}}} (\bibinfo {year} {1987}),\ \href@noop {} {\bibfield
  {journal} {\bibinfo  {journal} {Phys. Rev. A}\ }\textbf {\bibinfo {volume}
  {36}},\ \bibinfo {pages} {3375}}\BibitemShut {NoStop}%
\bibitem [{\citenamefont {Coppersmith}\ \emph {et~al.}(1997)\citenamefont
  {Coppersmith}, \citenamefont {Jones}, \citenamefont {Kadanoff}, \citenamefont
  {Levine}, \citenamefont {McCarten}, \citenamefont {Nagel}, \citenamefont
  {Venkataramani},\ and\ \citenamefont {Wu}}]{Coppersmith97}%
  \BibitemOpen
  \bibfield  {author} {\bibinfo {author} {\bibnamefont {Coppersmith},
  \bibfnamefont {S.~N.}}, \bibinfo {author} {\bibfnamefont {T.~C.}\
  \bibnamefont {Jones}}, \bibinfo {author} {\bibfnamefont {L.~P.}\ \bibnamefont
  {Kadanoff}}, \bibinfo {author} {\bibfnamefont {A.}~\bibnamefont {Levine}},
  \bibinfo {author} {\bibfnamefont {J.~P.}\ \bibnamefont {McCarten}}, \bibinfo
  {author} {\bibfnamefont {S.~R.}\ \bibnamefont {Nagel}}, \bibinfo {author}
  {\bibfnamefont {S.~C.}\ \bibnamefont {Venkataramani}}, \ and\ \bibinfo
  {author} {\bibfnamefont {X.}~\bibnamefont {Wu}}} (\bibinfo {year} {1997}),\
  \href@noop {} {\bibfield  {journal} {\bibinfo  {journal} {Phys. Rev. Lett.}\
  }\textbf {\bibinfo {volume} {78}}~(\bibinfo {number} {21}),\ \bibinfo {pages}
  {3983}}\BibitemShut {NoStop}%
\bibitem [{\citenamefont {Coppersmith}\ and\ \citenamefont
  {Littlewood}(1987)}]{Coppersmith87b}%
  \BibitemOpen
  \bibfield  {author} {\bibinfo {author} {\bibnamefont {Coppersmith},
  \bibfnamefont {S.~N.}}, \ and\ \bibinfo {author} {\bibfnamefont {P.~B.}\
  \bibnamefont {Littlewood}}} (\bibinfo {year} {1987}),\ \href@noop {}
  {\bibfield  {journal} {\bibinfo  {journal} {Phys. Rev. B}\ }\textbf {\bibinfo
  {volume} {36}},\ \bibinfo {pages} {311}}\BibitemShut {NoStop}%
\bibitem [{\citenamefont {Cort{\'e}}\ \emph {et~al.}(2008)\citenamefont
  {Cort{\'e}}, \citenamefont {Chaikin}, \citenamefont {Gollub},\ and\
  \citenamefont {Pine}}]{Corte08}%
  \BibitemOpen
  \bibfield  {author} {\bibinfo {author} {\bibnamefont {Cort{\'e}},
  \bibfnamefont {L.}}, \bibinfo {author} {\bibfnamefont {P.~M.}\ \bibnamefont
  {Chaikin}}, \bibinfo {author} {\bibfnamefont {J.~P.}\ \bibnamefont {Gollub}},
  \ and\ \bibinfo {author} {\bibfnamefont {D.~J.}\ \bibnamefont {Pine}}}
  (\bibinfo {year} {2008}),\ \href@noop {} {\bibfield  {journal} {\bibinfo
  {journal} {Nat. Phys.}\ }\textbf {\bibinfo {volume} {4}},\ \bibinfo {pages}
  {420}}\BibitemShut {NoStop}%
\bibitem [{\citenamefont {Cubuk}\ \emph {et~al.}(2017)\citenamefont {Cubuk},
  \citenamefont {Ivancic}, \citenamefont {Schoenholz}, \citenamefont
  {Strickland}, \citenamefont {Basu}, \citenamefont {Davidson}, \citenamefont
  {Fontaine}, \citenamefont {Hor}, \citenamefont {Huang}, \citenamefont
  {Jiang}, \citenamefont {Keim}, \citenamefont {Koshigan}, \citenamefont
  {Lefever}, \citenamefont {Liu}, \citenamefont {Ma}, \citenamefont
  {Magagnosc}, \citenamefont {Morrow}, \citenamefont {Ortiz}, \citenamefont
  {Rieser}, \citenamefont {Shavit}, \citenamefont {Still}, \citenamefont {Xu},
  \citenamefont {Zhang}, \citenamefont {Nordstrom}, \citenamefont {Arratia},
  \citenamefont {Carpick}, \citenamefont {Durian}, \citenamefont {Fakhraai},
  \citenamefont {Jerolmack}, \citenamefont {Lee}, \citenamefont {Li},
  \citenamefont {Riggleman}, \citenamefont {Turner}, \citenamefont {Yodh},
  \citenamefont {Gianola},\ and\ \citenamefont {Liu}}]{Cubuk17}%
  \BibitemOpen
  \bibfield  {author} {\bibinfo {author} {\bibnamefont {Cubuk}, \bibfnamefont
  {E.~D.}}, \bibinfo {author} {\bibfnamefont {R.~J.~S.}\ \bibnamefont
  {Ivancic}}, \bibinfo {author} {\bibfnamefont {S.~S.}\ \bibnamefont
  {Schoenholz}}, \bibinfo {author} {\bibfnamefont {D.~J.}\ \bibnamefont
  {Strickland}}, \bibinfo {author} {\bibfnamefont {A.}~\bibnamefont {Basu}},
  \bibinfo {author} {\bibfnamefont {Z.~S.}\ \bibnamefont {Davidson}}, \bibinfo
  {author} {\bibfnamefont {J.}~\bibnamefont {Fontaine}}, \bibinfo {author}
  {\bibfnamefont {J.~L.}\ \bibnamefont {Hor}}, \bibinfo {author} {\bibfnamefont
  {Y.~R.}\ \bibnamefont {Huang}}, \bibinfo {author} {\bibfnamefont
  {Y.}~\bibnamefont {Jiang}}, \bibinfo {author} {\bibfnamefont {N.~C.}\
  \bibnamefont {Keim}}, \bibinfo {author} {\bibfnamefont {K.~D.}\ \bibnamefont
  {Koshigan}}, \bibinfo {author} {\bibfnamefont {J.~A.}\ \bibnamefont
  {Lefever}}, \bibinfo {author} {\bibfnamefont {T.}~\bibnamefont {Liu}},
  \bibinfo {author} {\bibfnamefont {X.~G.}\ \bibnamefont {Ma}}, \bibinfo
  {author} {\bibfnamefont {D.~J.}\ \bibnamefont {Magagnosc}}, \bibinfo {author}
  {\bibfnamefont {E.}~\bibnamefont {Morrow}}, \bibinfo {author} {\bibfnamefont
  {C.~P.}\ \bibnamefont {Ortiz}}, \bibinfo {author} {\bibfnamefont {J.~M.}\
  \bibnamefont {Rieser}}, \bibinfo {author} {\bibfnamefont {A.}~\bibnamefont
  {Shavit}}, \bibinfo {author} {\bibfnamefont {T.}~\bibnamefont {Still}},
  \bibinfo {author} {\bibfnamefont {Y.}~\bibnamefont {Xu}}, \bibinfo {author}
  {\bibfnamefont {Y.}~\bibnamefont {Zhang}}, \bibinfo {author} {\bibfnamefont
  {K.~N.}\ \bibnamefont {Nordstrom}}, \bibinfo {author} {\bibfnamefont {P.~E.}\
  \bibnamefont {Arratia}}, \bibinfo {author} {\bibfnamefont {R.~W.}\
  \bibnamefont {Carpick}}, \bibinfo {author} {\bibfnamefont {D.~J.}\
  \bibnamefont {Durian}}, \bibinfo {author} {\bibfnamefont {Z.}~\bibnamefont
  {Fakhraai}}, \bibinfo {author} {\bibfnamefont {D.~J.}\ \bibnamefont
  {Jerolmack}}, \bibinfo {author} {\bibfnamefont {D.}~\bibnamefont {Lee}},
  \bibinfo {author} {\bibfnamefont {J.}~\bibnamefont {Li}}, \bibinfo {author}
  {\bibfnamefont {R.}~\bibnamefont {Riggleman}}, \bibinfo {author}
  {\bibfnamefont {K.~T.}\ \bibnamefont {Turner}}, \bibinfo {author}
  {\bibfnamefont {A.~G.}\ \bibnamefont {Yodh}}, \bibinfo {author}
  {\bibfnamefont {D.~S.}\ \bibnamefont {Gianola}}, \ and\ \bibinfo {author}
  {\bibfnamefont {A.~J.}\ \bibnamefont {Liu}}} (\bibinfo {year} {2017}),\
  \href@noop {} {\bibfield  {journal} {\bibinfo  {journal} {Science}\ }\textbf
  {\bibinfo {volume} {358}}~(\bibinfo {number} {6366}),\ \bibinfo {pages}
  {1033}}\BibitemShut {NoStop}%
\bibitem [{\citenamefont {Cugliandolo}\ \emph {et~al.}(2004)\citenamefont
  {Cugliandolo}, \citenamefont {Lozano},\ and\ \citenamefont
  {Lozza}}]{Cugliandolo04}%
  \BibitemOpen
  \bibfield  {author} {\bibinfo {author} {\bibnamefont {Cugliandolo},
  \bibfnamefont {L.}}, \bibinfo {author} {\bibfnamefont {G.}~\bibnamefont
  {Lozano}}, \ and\ \bibinfo {author} {\bibfnamefont {H.}~\bibnamefont
  {Lozza}}} (\bibinfo {year} {2004}),\ \href@noop {} {\bibfield  {journal}
  {\bibinfo  {journal} {The European Physical Journal B-Condensed Matter and
  Complex Systems}\ }\textbf {\bibinfo {volume} {41}}~(\bibinfo {number} {1}),\
  \bibinfo {pages} {87}}\BibitemShut {NoStop}%
\bibitem [{\citenamefont {Deutsch}\ \emph {et~al.}(2004)\citenamefont
  {Deutsch}, \citenamefont {Dhar},\ and\ \citenamefont {Narayan}}]{Deutsch04}%
  \BibitemOpen
  \bibfield  {author} {\bibinfo {author} {\bibnamefont {Deutsch}, \bibfnamefont
  {J.~M.}}, \bibinfo {author} {\bibfnamefont {A.}~\bibnamefont {Dhar}}, \ and\
  \bibinfo {author} {\bibfnamefont {O.}~\bibnamefont {Narayan}}} (\bibinfo
  {year} {2004}),\ \href@noop {} {\bibfield  {journal} {\bibinfo  {journal}
  {Phys. Rev. Lett.}\ }\textbf {\bibinfo {volume} {92}},\ \bibinfo {pages}
  {227203}}\BibitemShut {NoStop}%
\bibitem [{\citenamefont {Deutsch}\ and\ \citenamefont
  {Narayan}(2003)}]{Deutsch03}%
  \BibitemOpen
  \bibfield  {author} {\bibinfo {author} {\bibnamefont {Deutsch}, \bibfnamefont
  {J.~M.}}, \ and\ \bibinfo {author} {\bibfnamefont {O.}~\bibnamefont
  {Narayan}}} (\bibinfo {year} {2003}),\ \href@noop {} {\bibfield  {journal}
  {\bibinfo  {journal} {Phys. Rev. Lett.}\ }\textbf {\bibinfo {volume}
  {91}}~(\bibinfo {number} {20}),\ \bibinfo {pages} {3}}\BibitemShut {NoStop}%
\bibitem [{\citenamefont {Diani}\ \emph {et~al.}(2009)\citenamefont {Diani},
  \citenamefont {Fayolle},\ and\ \citenamefont {Gilormini}}]{Diani09}%
  \BibitemOpen
  \bibfield  {author} {\bibinfo {author} {\bibnamefont {Diani}, \bibfnamefont
  {J.}}, \bibinfo {author} {\bibfnamefont {B.}~\bibnamefont {Fayolle}}, \ and\
  \bibinfo {author} {\bibfnamefont {P.}~\bibnamefont {Gilormini}}} (\bibinfo
  {year} {2009}),\ \href@noop {} {\bibfield  {journal} {\bibinfo  {journal}
  {European Polymer Journal}\ }\textbf {\bibinfo {volume} {45}}~(\bibinfo
  {number} {3}),\ \bibinfo {pages} {601}}\BibitemShut {NoStop}%
\bibitem [{\citenamefont {Dillavou}\ and\ \citenamefont
  {Rubinstein}(2018)}]{Dillavou18}%
  \BibitemOpen
  \bibfield  {author} {\bibinfo {author} {\bibnamefont {Dillavou},
  \bibfnamefont {S.}}, \ and\ \bibinfo {author} {\bibfnamefont {S.~M.}\
  \bibnamefont {Rubinstein}}} (\bibinfo {year} {2018}),\ \href@noop {}
  {\bibfield  {journal} {\bibinfo  {journal} {Phys. Rev. Lett.}\ }\textbf
  {\bibinfo {volume} {120}}~(\bibinfo {number} {22}),\ \bibinfo {pages}
  {224101}}\BibitemShut {NoStop}%
\bibitem [{\citenamefont {Eggers}(1997)}]{Eggers97}%
  \BibitemOpen
  \bibfield  {author} {\bibinfo {author} {\bibnamefont {Eggers}, \bibfnamefont
  {J.}}} (\bibinfo {year} {1997}),\ \href@noop {} {\bibfield  {journal}
  {\bibinfo  {journal} {Rev. Mod. Phys.}\ }\textbf {\bibinfo {volume} {69}},\
  \bibinfo {pages} {865}}\BibitemShut {NoStop}%
\bibitem [{\citenamefont {Eggers}\ and\ \citenamefont
  {Dupont}(1994)}]{Eggers94}%
  \BibitemOpen
  \bibfield  {author} {\bibinfo {author} {\bibnamefont {Eggers}, \bibfnamefont
  {J.}}, \ and\ \bibinfo {author} {\bibfnamefont {T.~F.}\ \bibnamefont
  {Dupont}}} (\bibinfo {year} {1994}),\ \href@noop {} {\bibfield  {journal}
  {\bibinfo  {journal} {Journal of Fluid Mechanics}\ }\textbf {\bibinfo
  {volume} {262}},\ \bibinfo {pages} {205}}\BibitemShut {NoStop}%
\bibitem [{\citenamefont {Emmett}\ and\ \citenamefont
  {Cines}(1947)}]{Emmett47}%
  \BibitemOpen
  \bibfield  {author} {\bibinfo {author} {\bibnamefont {Emmett}, \bibfnamefont
  {P.~H.}}, \ and\ \bibinfo {author} {\bibfnamefont {M.}~\bibnamefont {Cines}}}
  (\bibinfo {year} {1947}),\ \href@noop {} {\bibfield  {journal} {\bibinfo
  {journal} {J. Phys. Chem.}\ }\textbf {\bibinfo {volume} {51}}~(\bibinfo
  {number} {6}),\ \bibinfo {pages} {1248}}\BibitemShut {NoStop}%
\bibitem [{\citenamefont {Fiocco}\ \emph {et~al.}(2014)\citenamefont {Fiocco},
  \citenamefont {Foffi},\ and\ \citenamefont {Sastry}}]{Fiocco14}%
  \BibitemOpen
  \bibfield  {author} {\bibinfo {author} {\bibnamefont {Fiocco}, \bibfnamefont
  {D.}}, \bibinfo {author} {\bibfnamefont {G.}~\bibnamefont {Foffi}}, \ and\
  \bibinfo {author} {\bibfnamefont {S.}~\bibnamefont {Sastry}}} (\bibinfo
  {year} {2014}),\ \href@noop {} {\bibfield  {journal} {\bibinfo  {journal}
  {Phys. Rev. Lett.}\ }\textbf {\bibinfo {volume} {112}},\ \bibinfo {pages}
  {025702}}\BibitemShut {NoStop}%
\bibitem [{\citenamefont {Fiocco}\ \emph {et~al.}(2015)\citenamefont {Fiocco},
  \citenamefont {Foffi},\ and\ \citenamefont {Sastry}}]{Fiocco15}%
  \BibitemOpen
  \bibfield  {author} {\bibinfo {author} {\bibnamefont {Fiocco}, \bibfnamefont
  {D.}}, \bibinfo {author} {\bibfnamefont {G.}~\bibnamefont {Foffi}}, \ and\
  \bibinfo {author} {\bibfnamefont {S.}~\bibnamefont {Sastry}}} (\bibinfo
  {year} {2015}),\ \href@noop {} {\bibfield  {journal} {\bibinfo  {journal} {J.
  Phys: Cond. Matter}\ }\textbf {\bibinfo {volume} {27}}~(\bibinfo {number}
  {19}),\ \bibinfo {pages} {194130}}\BibitemShut {NoStop}%
\bibitem [{\citenamefont {Fleming}\ and\ \citenamefont
  {Schneemeyer}(1983)}]{Fleming83}%
  \BibitemOpen
  \bibfield  {author} {\bibinfo {author} {\bibnamefont {Fleming}, \bibfnamefont
  {R.}}, \ and\ \bibinfo {author} {\bibfnamefont {L.}~\bibnamefont
  {Schneemeyer}}} (\bibinfo {year} {1983}),\ \href@noop {} {\bibfield
  {journal} {\bibinfo  {journal} {Physical Review B}\ }\textbf {\bibinfo
  {volume} {28}}~(\bibinfo {number} {12}),\ \bibinfo {pages}
  {6996}}\BibitemShut {NoStop}%
\bibitem [{\citenamefont {Fleming}\ and\ \citenamefont
  {Schneemeyer}(1986)}]{Fleming86}%
  \BibitemOpen
  \bibfield  {author} {\bibinfo {author} {\bibnamefont {Fleming}, \bibfnamefont
  {R.~M.}}, \ and\ \bibinfo {author} {\bibfnamefont {L.~F.}\ \bibnamefont
  {Schneemeyer}}} (\bibinfo {year} {1986}),\ \href@noop {} {\bibfield
  {journal} {\bibinfo  {journal} {Phys. Rev. B}\ }\textbf {\bibinfo {volume}
  {33}},\ \bibinfo {pages} {2930}}\BibitemShut {NoStop}%
\bibitem [{\citenamefont {Fukao}\ and\ \citenamefont
  {Sakamoto}(2005)}]{Fukao05}%
  \BibitemOpen
  \bibfield  {author} {\bibinfo {author} {\bibnamefont {Fukao}, \bibfnamefont
  {K.}}, \ and\ \bibinfo {author} {\bibfnamefont {A.}~\bibnamefont {Sakamoto}}}
  (\bibinfo {year} {2005}),\ \href {\doibase 10.1103/PhysRevE.71.041803}
  {\bibfield  {journal} {\bibinfo  {journal} {Phys. Rev. E}\ }\textbf {\bibinfo
  {volume} {71}},\ \bibinfo {pages} {041803}}\BibitemShut {NoStop}%
\bibitem [{\citenamefont {Fusi}(2017)}]{Fusi2017}%
  \BibitemOpen
  \bibfield  {author} {\bibinfo {author} {\bibnamefont {Fusi}, \bibfnamefont
  {S.}}} (\bibinfo {year} {2017}),\ \href@noop {} {\bibfield  {journal}
  {\bibinfo  {journal} {arXiv.org}\ }}\Eprint {http://arxiv.org/abs/1706.04946}
  {1706.04946} \BibitemShut {NoStop}%
\bibitem [{\citenamefont {Fusi}\ and\ \citenamefont
  {Abbott}(2007)}]{FusiAbbott}%
  \BibitemOpen
  \bibfield  {author} {\bibinfo {author} {\bibnamefont {Fusi}, \bibfnamefont
  {S.}}, \ and\ \bibinfo {author} {\bibfnamefont {L.~F.}\ \bibnamefont
  {Abbott}}} (\bibinfo {year} {2007}),\ \href {\doibase
  https://doi.org/10.1038/nn1859} {\bibfield  {journal} {\bibinfo  {journal}
  {Nature Neuroscience}\ }\textbf {\bibinfo {volume} {10}},\
  https://doi.org/10.1038/nn1859}\BibitemShut {NoStop}%
\bibitem [{\citenamefont {Gadala-Maria}\ and\ \citenamefont
  {Acrivos}(1980)}]{Gadala-Maria80}%
  \BibitemOpen
  \bibfield  {author} {\bibinfo {author} {\bibnamefont {Gadala-Maria},
  \bibfnamefont {F.}}, \ and\ \bibinfo {author} {\bibfnamefont
  {A.}~\bibnamefont {Acrivos}}} (\bibinfo {year} {1980}),\ \href@noop {}
  {\bibfield  {journal} {\bibinfo  {journal} {Journal of Rheology}\ }\textbf
  {\bibinfo {volume} {24}}~(\bibinfo {number} {6}),\ \bibinfo {pages}
  {799}}\BibitemShut {NoStop}%
\bibitem [{\citenamefont {Gallardo}\ \emph {et~al.}(2010)\citenamefont
  {Gallardo}, \citenamefont {Manchado}, \citenamefont {Romero}, \citenamefont
  {Del~Cerro}, \citenamefont {Salje}, \citenamefont {Planes}, \citenamefont
  {Vives}, \citenamefont {Romero},\ and\ \citenamefont
  {Stipcich}}]{Gallardo10}%
  \BibitemOpen
  \bibfield  {author} {\bibinfo {author} {\bibnamefont {Gallardo},
  \bibfnamefont {M.~C.}}, \bibinfo {author} {\bibfnamefont {J.}~\bibnamefont
  {Manchado}}, \bibinfo {author} {\bibfnamefont {F.~J.}\ \bibnamefont
  {Romero}}, \bibinfo {author} {\bibfnamefont {J.}~\bibnamefont {Del~Cerro}},
  \bibinfo {author} {\bibfnamefont {E.~K.}\ \bibnamefont {Salje}}, \bibinfo
  {author} {\bibfnamefont {A.}~\bibnamefont {Planes}}, \bibinfo {author}
  {\bibfnamefont {E.}~\bibnamefont {Vives}}, \bibinfo {author} {\bibfnamefont
  {R.}~\bibnamefont {Romero}}, \ and\ \bibinfo {author} {\bibfnamefont
  {M.}~\bibnamefont {Stipcich}}} (\bibinfo {year} {2010}),\ \href@noop {}
  {\bibfield  {journal} {\bibinfo  {journal} {Physical Review B}\ }\textbf
  {\bibinfo {volume} {81}}~(\bibinfo {number} {17}),\ \bibinfo {pages}
  {174102}}\BibitemShut {NoStop}%
\bibitem [{\citenamefont {Gilbert}\ \emph {et~al.}(2015)\citenamefont
  {Gilbert}, \citenamefont {Chern}, \citenamefont {Fore}, \citenamefont {Lao},
  \citenamefont {Zhang}, \citenamefont {Nisoli},\ and\ \citenamefont
  {Schiffer}}]{Gilbert15}%
  \BibitemOpen
  \bibfield  {author} {\bibinfo {author} {\bibnamefont {Gilbert}, \bibfnamefont
  {I.}}, \bibinfo {author} {\bibfnamefont {G.-W.}\ \bibnamefont {Chern}},
  \bibinfo {author} {\bibfnamefont {B.}~\bibnamefont {Fore}}, \bibinfo {author}
  {\bibfnamefont {Y.}~\bibnamefont {Lao}}, \bibinfo {author} {\bibfnamefont
  {S.}~\bibnamefont {Zhang}}, \bibinfo {author} {\bibfnamefont
  {C.}~\bibnamefont {Nisoli}}, \ and\ \bibinfo {author} {\bibfnamefont
  {P.}~\bibnamefont {Schiffer}}} (\bibinfo {year} {2015}),\ \href@noop {}
  {\bibfield  {journal} {\bibinfo  {journal} {Phys. Rev. B}\ }\textbf {\bibinfo
  {volume} {92}}~(\bibinfo {number} {10}),\ \bibinfo {pages}
  {104417}}\BibitemShut {NoStop}%
\bibitem [{\citenamefont {Gill}(1981)}]{Gill81}%
  \BibitemOpen
  \bibfield  {author} {\bibinfo {author} {\bibnamefont {Gill}, \bibfnamefont
  {J.}}} (\bibinfo {year} {1981}),\ \href@noop {} {\bibfield  {journal}
  {\bibinfo  {journal} {Solid State Communications}\ }\textbf {\bibinfo
  {volume} {39}}~(\bibinfo {number} {11}),\ \bibinfo {pages}
  {1203}}\BibitemShut {NoStop}%
\bibitem [{\citenamefont {Golding}\ and\ \citenamefont
  {Graebner}(1976)}]{Golding76}%
  \BibitemOpen
  \bibfield  {author} {\bibinfo {author} {\bibnamefont {Golding}, \bibfnamefont
  {B.}}, \ and\ \bibinfo {author} {\bibfnamefont {J.~E.}\ \bibnamefont
  {Graebner}}} (\bibinfo {year} {1976}),\ \href@noop {} {\bibfield  {journal}
  {\bibinfo  {journal} {Phys. Rev. Lett.}\ }\textbf {\bibinfo {volume}
  {37}}~(\bibinfo {number} {13}),\ \bibinfo {pages} {852}}\BibitemShut
  {NoStop}%
\bibitem [{\citenamefont {Goldstein}\ \emph {et~al.}(1993)\citenamefont
  {Goldstein}, \citenamefont {Pesci},\ and\ \citenamefont
  {Shelley}}]{Goldstein93}%
  \BibitemOpen
  \bibfield  {author} {\bibinfo {author} {\bibnamefont {Goldstein},
  \bibfnamefont {R.~E.}}, \bibinfo {author} {\bibfnamefont {A.~I.}\
  \bibnamefont {Pesci}}, \ and\ \bibinfo {author} {\bibfnamefont {M.~J.}\
  \bibnamefont {Shelley}}} (\bibinfo {year} {1993}),\ \href@noop {} {\bibfield
  {journal} {\bibinfo  {journal} {Phys. Rev. Lett.}\ }\textbf {\bibinfo
  {volume} {70}}~(\bibinfo {number} {20}),\ \bibinfo {pages}
  {3043}}\BibitemShut {NoStop}%
\bibitem [{\citenamefont {Goodrich}\ \emph {et~al.}(2015)\citenamefont
  {Goodrich}, \citenamefont {Liu},\ and\ \citenamefont {Nagel}}]{Goodrich15}%
  \BibitemOpen
  \bibfield  {author} {\bibinfo {author} {\bibnamefont {Goodrich},
  \bibfnamefont {C.~P.}}, \bibinfo {author} {\bibfnamefont {A.~J.}\
  \bibnamefont {Liu}}, \ and\ \bibinfo {author} {\bibfnamefont {S.~R.}\
  \bibnamefont {Nagel}}} (\bibinfo {year} {2015}),\ \href@noop {} {\bibfield
  {journal} {\bibinfo  {journal} {Phys. Rev. Lett.}\ }\textbf {\bibinfo
  {volume} {114}}~(\bibinfo {number} {22}),\ \bibinfo {pages}
  {225501}}\BibitemShut {NoStop}%
\bibitem [{\citenamefont {Gould}(1965)}]{Gould65}%
  \BibitemOpen
  \bibfield  {author} {\bibinfo {author} {\bibnamefont {Gould}, \bibfnamefont
  {R.}}} (\bibinfo {year} {1965}),\ \href@noop {} {\bibfield  {journal}
  {\bibinfo  {journal} {Physics Letters}\ }\textbf {\bibinfo {volume}
  {19}}~(\bibinfo {number} {6}),\ \bibinfo {pages} {477}}\BibitemShut {NoStop}%
\bibitem [{\citenamefont {Hahn}(1950)}]{Hahn50}%
  \BibitemOpen
  \bibfield  {author} {\bibinfo {author} {\bibnamefont {Hahn}, \bibfnamefont
  {E.~L.}}} (\bibinfo {year} {1950}),\ \href@noop {} {\bibfield  {journal}
  {\bibinfo  {journal} {Physical Review}\ }\textbf {\bibinfo {volume}
  {80}}~(\bibinfo {number} {4}),\ \bibinfo {pages} {580}}\BibitemShut {NoStop}%
\bibitem [{\citenamefont {Haw}\ \emph {et~al.}(1998)\citenamefont {Haw},
  \citenamefont {Poon}, \citenamefont {Pusey}, \citenamefont {Hebraud},\ and\
  \citenamefont {Lequeux}}]{Haw98}%
  \BibitemOpen
  \bibfield  {author} {\bibinfo {author} {\bibnamefont {Haw}, \bibfnamefont
  {M.~D.}}, \bibinfo {author} {\bibfnamefont {W.~C.~K.}\ \bibnamefont {Poon}},
  \bibinfo {author} {\bibfnamefont {P.~N.}\ \bibnamefont {Pusey}}, \bibinfo
  {author} {\bibfnamefont {P.}~\bibnamefont {Hebraud}}, \ and\ \bibinfo
  {author} {\bibfnamefont {F.}~\bibnamefont {Lequeux}}} (\bibinfo {year}
  {1998}),\ \href@noop {} {\bibfield  {journal} {\bibinfo  {journal} {Phys.
  Rev. E}\ }\textbf {\bibinfo {volume} {58}}~(\bibinfo {number} {4}),\ \bibinfo
  {pages} {4673}}\BibitemShut {NoStop}%
\bibitem [{\citenamefont {H{\'e}braud}\ \emph {et~al.}(1997)\citenamefont
  {H{\'e}braud}, \citenamefont {Lequeux}, \citenamefont {Munch},\ and\
  \citenamefont {Pine}}]{Hebraud97}%
  \BibitemOpen
  \bibfield  {author} {\bibinfo {author} {\bibnamefont {H{\'e}braud},
  \bibfnamefont {P.}}, \bibinfo {author} {\bibfnamefont {F.}~\bibnamefont
  {Lequeux}}, \bibinfo {author} {\bibfnamefont {J.-P.}\ \bibnamefont {Munch}},
  \ and\ \bibinfo {author} {\bibfnamefont {D.~J.}\ \bibnamefont {Pine}}}
  (\bibinfo {year} {1997}),\ \href@noop {} {\bibfield  {journal} {\bibinfo
  {journal} {Phys. Rev. Lett.}\ }\textbf {\bibinfo {volume} {78}}~(\bibinfo
  {number} {24}),\ \bibinfo {pages} {4657}}\BibitemShut {NoStop}%
\bibitem [{\citenamefont {van Hecke}(2010)}]{VanHecke10}%
  \BibitemOpen
  \bibfield  {author} {\bibinfo {author} {\bibnamefont {van Hecke},
  \bibfnamefont {M.}}} (\bibinfo {year} {2010}),\ \href@noop {} {\bibfield
  {journal} {\bibinfo  {journal} {J. Phys: Cond. Matter}\ }\textbf {\bibinfo
  {volume} {22}}~(\bibinfo {number} {3}),\ \bibinfo {pages}
  {033101}}\BibitemShut {NoStop}%
\bibitem [{\citenamefont {Hertz}\ \emph {et~al.}(1991)\citenamefont {Hertz},
  \citenamefont {Krogh},\ and\ \citenamefont {Palmer}}]{Hertz:1991}%
  \BibitemOpen
  \bibfield  {author} {\bibinfo {author} {\bibnamefont {Hertz}, \bibfnamefont
  {J.}}, \bibinfo {author} {\bibfnamefont {A.}~\bibnamefont {Krogh}}, \ and\
  \bibinfo {author} {\bibfnamefont {R.~G.}\ \bibnamefont {Palmer}}} (\bibinfo
  {year} {1991}),\ \href@noop {} {\emph {\bibinfo {title} {Introduction to the
  Theory of Neural Computation}}}\ (\bibinfo  {publisher} {Addison-Wesley
  Longman Publishing Co., Inc.},\ \bibinfo {address} {Boston, MA,
  USA})\BibitemShut {NoStop}%
\bibitem [{\citenamefont {Hexner}\ \emph
  {et~al.}(2018{\natexlab{a}})\citenamefont {Hexner}, \citenamefont {Liu},\
  and\ \citenamefont {Nagel}}]{Hexner18PRE}%
  \BibitemOpen
  \bibfield  {author} {\bibinfo {author} {\bibnamefont {Hexner}, \bibfnamefont
  {D.}}, \bibinfo {author} {\bibfnamefont {A.~J.}\ \bibnamefont {Liu}}, \ and\
  \bibinfo {author} {\bibfnamefont {S.~R.}\ \bibnamefont {Nagel}}} (\bibinfo
  {year} {2018}{\natexlab{a}}),\ \href@noop {} {\bibfield  {journal} {\bibinfo
  {journal} {Physical Review E}\ }\textbf {\bibinfo {volume} {97}}~(\bibinfo
  {number} {6}),\ \bibinfo {pages} {063001}}\BibitemShut {NoStop}%
\bibitem [{\citenamefont {Hexner}\ \emph
  {et~al.}(2018{\natexlab{b}})\citenamefont {Hexner}, \citenamefont {Liu},\
  and\ \citenamefont {Nagel}}]{Hexner18}%
  \BibitemOpen
  \bibfield  {author} {\bibinfo {author} {\bibnamefont {Hexner}, \bibfnamefont
  {D.}}, \bibinfo {author} {\bibfnamefont {A.~J.}\ \bibnamefont {Liu}}, \ and\
  \bibinfo {author} {\bibfnamefont {S.~R.}\ \bibnamefont {Nagel}}} (\bibinfo
  {year} {2018}{\natexlab{b}}),\ \href@noop {} {\bibfield  {journal} {\bibinfo
  {journal} {Soft matter}\ }\textbf {\bibinfo {volume} {14}}~(\bibinfo {number}
  {2}),\ \bibinfo {pages} {312}}\BibitemShut {NoStop}%
\bibitem [{\citenamefont {Hill}\ and\ \citenamefont {Kaplan}(1965)}]{Hill65}%
  \BibitemOpen
  \bibfield  {author} {\bibinfo {author} {\bibnamefont {Hill}, \bibfnamefont
  {R.}}, \ and\ \bibinfo {author} {\bibfnamefont {D.}~\bibnamefont {Kaplan}}}
  (\bibinfo {year} {1965}),\ \href@noop {} {\bibfield  {journal} {\bibinfo
  {journal} {Phys. Rev. Lett.}\ }\textbf {\bibinfo {volume} {14}}~(\bibinfo
  {number} {26}),\ \bibinfo {pages} {1062}}\BibitemShut {NoStop}%
\bibitem [{\citenamefont {Hopfield}(1982)}]{Hopfield82}%
  \BibitemOpen
  \bibfield  {author} {\bibinfo {author} {\bibnamefont {Hopfield},
  \bibfnamefont {J.~J.}}} (\bibinfo {year} {1982}),\ \href@noop {} {\bibfield
  {journal} {\bibinfo  {journal} {Proceedings of the National Academy of
  Sciences}\ }\textbf {\bibinfo {volume} {79}}~(\bibinfo {number} {8}),\
  \bibinfo {pages} {2554}}\BibitemShut {NoStop}%
\bibitem [{\citenamefont {Hovorka}\ and\ \citenamefont
  {Friedman}(2008)}]{Hovorka08}%
  \BibitemOpen
  \bibfield  {author} {\bibinfo {author} {\bibnamefont {Hovorka}, \bibfnamefont
  {O.}}, \ and\ \bibinfo {author} {\bibfnamefont {G.}~\bibnamefont {Friedman}}}
  (\bibinfo {year} {2008}),\ \href@noop {} {\bibfield  {journal} {\bibinfo
  {journal} {Phys. Rev. Lett.}\ }\textbf {\bibinfo {volume} {100}}~(\bibinfo
  {number} {9}),\ \bibinfo {pages} {097201}}\BibitemShut {NoStop}%
\bibitem [{\citenamefont {Jaeger}\ \emph {et~al.}(1996)\citenamefont {Jaeger},
  \citenamefont {Nagel},\ and\ \citenamefont {Behringer}}]{Jaeger96}%
  \BibitemOpen
  \bibfield  {author} {\bibinfo {author} {\bibnamefont {Jaeger}, \bibfnamefont
  {H.~M.}}, \bibinfo {author} {\bibfnamefont {S.~R.}\ \bibnamefont {Nagel}}, \
  and\ \bibinfo {author} {\bibfnamefont {R.~P.}\ \bibnamefont {Behringer}}}
  (\bibinfo {year} {1996}),\ \href@noop {} {\bibfield  {journal} {\bibinfo
  {journal} {Rev. Mod. Phys.}\ }\textbf {\bibinfo {volume} {68}}~(\bibinfo
  {number} {4}),\ \bibinfo {pages} {1259}}\BibitemShut {NoStop}%
\bibitem [{\citenamefont {James}(2019)}]{James19}%
  \BibitemOpen
  \bibfield  {author} {\bibinfo {author} {\bibnamefont {James}, \bibfnamefont
  {R.~D.}}} (\bibinfo {year} {2019}),\ \href@noop {} {\bibfield  {journal}
  {\bibinfo  {journal} {Bulletin (New Series) of the American Mathematical
  Society}\ }\textbf {\bibinfo {volume} {26}}~(\bibinfo {number}
  {1})}\BibitemShut {NoStop}%
\bibitem [{\citenamefont {Jiles}(2016)}]{Jiles16}%
  \BibitemOpen
  \bibfield  {author} {\bibinfo {author} {\bibnamefont {Jiles}, \bibfnamefont
  {D.}}} (\bibinfo {year} {2016}),\ \href@noop {} {\emph {\bibinfo {title}
  {Introduction to magnetism and magnetic materials}}}\ (\bibinfo  {publisher}
  {CRC Press/Taylor \& Francis Group},\ \bibinfo {address} {Boca
  Raton})\BibitemShut {NoStop}%
\bibitem [{\citenamefont {Jim{\'e}nez}\ \emph {et~al.}(2005)\citenamefont
  {Jim{\'e}nez}, \citenamefont {Mart{\'\i}n-Mayor},\ and\ \citenamefont
  {Perez-Gaviro}}]{Jimenez05}%
  \BibitemOpen
  \bibfield  {author} {\bibinfo {author} {\bibnamefont {Jim{\'e}nez},
  \bibfnamefont {S.}}, \bibinfo {author} {\bibfnamefont {V.}~\bibnamefont
  {Mart{\'\i}n-Mayor}}, \ and\ \bibinfo {author} {\bibfnamefont
  {S.}~\bibnamefont {Perez-Gaviro}}} (\bibinfo {year} {2005}),\ \href@noop {}
  {\bibfield  {journal} {\bibinfo  {journal} {Physical Review B}\ }\textbf
  {\bibinfo {volume} {72}}~(\bibinfo {number} {5}),\ \bibinfo {pages}
  {054417}}\BibitemShut {NoStop}%
\bibitem [{\citenamefont {Jonason}\ \emph {et~al.}(2000)\citenamefont
  {Jonason}, \citenamefont {Nordblad}, \citenamefont {Vincent}, \citenamefont
  {Hammann},\ and\ \citenamefont {Bouchaud}}]{Jonason00}%
  \BibitemOpen
  \bibfield  {author} {\bibinfo {author} {\bibnamefont {Jonason}, \bibfnamefont
  {K.}}, \bibinfo {author} {\bibfnamefont {P.}~\bibnamefont {Nordblad}},
  \bibinfo {author} {\bibfnamefont {E.}~\bibnamefont {Vincent}}, \bibinfo
  {author} {\bibfnamefont {J.}~\bibnamefont {Hammann}}, \ and\ \bibinfo
  {author} {\bibfnamefont {J.-P.}\ \bibnamefont {Bouchaud}}} (\bibinfo {year}
  {2000}),\ \href@noop {} {\bibfield  {journal} {\bibinfo  {journal} {The
  European Physical Journal B-Condensed Matter and Complex Systems}\ }\textbf
  {\bibinfo {volume} {13}}~(\bibinfo {number} {1}),\ \bibinfo {pages}
  {99}}\BibitemShut {NoStop}%
\bibitem [{\citenamefont {Jonason}\ \emph {et~al.}(1998)\citenamefont
  {Jonason}, \citenamefont {Vincent}, \citenamefont {Hammann}, \citenamefont
  {Bouchaud},\ and\ \citenamefont {Nordblad}}]{Jonason98}%
  \BibitemOpen
  \bibfield  {author} {\bibinfo {author} {\bibnamefont {Jonason}, \bibfnamefont
  {K.}}, \bibinfo {author} {\bibfnamefont {E.}~\bibnamefont {Vincent}},
  \bibinfo {author} {\bibfnamefont {J.}~\bibnamefont {Hammann}}, \bibinfo
  {author} {\bibfnamefont {J.}~\bibnamefont {Bouchaud}}, \ and\ \bibinfo
  {author} {\bibfnamefont {P.}~\bibnamefont {Nordblad}}} (\bibinfo {year}
  {1998}),\ \href@noop {} {\bibfield  {journal} {\bibinfo  {journal} {Phys.
  Rev. Lett.}\ }\textbf {\bibinfo {volume} {81}}~(\bibinfo {number} {15}),\
  \bibinfo {pages} {3243}}\BibitemShut {NoStop}%
\bibitem [{\citenamefont {Kaiser}(1950)}]{Kaiser50}%
  \BibitemOpen
  \bibfield  {author} {\bibinfo {author} {\bibnamefont {Kaiser}, \bibfnamefont
  {J.}}} (\bibinfo {year} {1950}),\ \emph {\bibinfo {title} {An investigation
  into the occurrence of noises in tensile tests or a study of acoustic
  phenomena}},\ \href@noop {} {Ph.D. thesis}\ (\bibinfo  {school} {PhD thesis,
  Technical University, Munich, Germany})\BibitemShut {NoStop}%
\bibitem [{\citenamefont {Karmakar}\ \emph {et~al.}(2010)\citenamefont
  {Karmakar}, \citenamefont {Lerner},\ and\ \citenamefont
  {Procaccia}}]{Karmakar10}%
  \BibitemOpen
  \bibfield  {author} {\bibinfo {author} {\bibnamefont {Karmakar},
  \bibfnamefont {S.}}, \bibinfo {author} {\bibfnamefont {E.}~\bibnamefont
  {Lerner}}, \ and\ \bibinfo {author} {\bibfnamefont {I.}~\bibnamefont
  {Procaccia}}} (\bibinfo {year} {2010}),\ \href@noop {} {\bibfield  {journal}
  {\bibinfo  {journal} {Physical Review E}\ }\textbf {\bibinfo {volume}
  {82}}~(\bibinfo {number} {2}),\ \bibinfo {pages} {026104}}\BibitemShut
  {NoStop}%
\bibitem [{\citenamefont {Kegel}\ and\ \citenamefont {Gould}(1965)}]{Kegel65}%
  \BibitemOpen
  \bibfield  {author} {\bibinfo {author} {\bibnamefont {Kegel}, \bibfnamefont
  {W.}}, \ and\ \bibinfo {author} {\bibfnamefont {R.}~\bibnamefont {Gould}}}
  (\bibinfo {year} {1965}),\ \href@noop {} {\bibfield  {journal} {\bibinfo
  {journal} {Physics Letters}\ }\textbf {\bibinfo {volume} {19}}~(\bibinfo
  {number} {7}),\ \bibinfo {pages} {531}}\BibitemShut {NoStop}%
\bibitem [{\citenamefont {Keim}(2011)}]{Keim11b}%
  \BibitemOpen
  \bibfield  {author} {\bibinfo {author} {\bibnamefont {Keim}, \bibfnamefont
  {N.~C.}}} (\bibinfo {year} {2011}),\ \href@noop {} {\bibfield  {journal}
  {\bibinfo  {journal} {Phys. Rev. E}\ }\textbf {\bibinfo {volume}
  {83}}~(\bibinfo {number} {5}),\ \bibinfo {pages} {056325}}\BibitemShut
  {NoStop}%
\bibitem [{\citenamefont {Keim}\ and\ \citenamefont {Arratia}(2014)}]{Keim14}%
  \BibitemOpen
  \bibfield  {author} {\bibinfo {author} {\bibnamefont {Keim}, \bibfnamefont
  {N.~C.}}, \ and\ \bibinfo {author} {\bibfnamefont {P.~E.}\ \bibnamefont
  {Arratia}}} (\bibinfo {year} {2014}),\ \href@noop {} {\bibfield  {journal}
  {\bibinfo  {journal} {Phys. Rev. Lett.}\ }\textbf {\bibinfo {volume}
  {112}}~(\bibinfo {number} {2}),\ \bibinfo {pages} {028302}}\BibitemShut
  {NoStop}%
\bibitem [{\citenamefont {Keim}\ \emph {et~al.}(2018)\citenamefont {Keim},
  \citenamefont {Hass}, \citenamefont {Kroger},\ and\ \citenamefont
  {Wieker}}]{Keim18}%
  \BibitemOpen
  \bibfield  {author} {\bibinfo {author} {\bibnamefont {Keim}, \bibfnamefont
  {N.~C.}}, \bibinfo {author} {\bibfnamefont {J.}~\bibnamefont {Hass}},
  \bibinfo {author} {\bibfnamefont {B.}~\bibnamefont {Kroger}}, \ and\ \bibinfo
  {author} {\bibfnamefont {D.}~\bibnamefont {Wieker}}} (\bibinfo {year}
  {2018}),\ \href@noop {} {\bibinfo  {journal} {arXiv:1809.08505}\
  }\BibitemShut {NoStop}%
\bibitem [{\citenamefont {Keim}\ \emph {et~al.}(2006)\citenamefont {Keim},
  \citenamefont {M{\o}ller}, \citenamefont {Zhang},\ and\ \citenamefont
  {Nagel}}]{Keim06}%
  \BibitemOpen
\bibfield  {journal} {  }\bibfield  {author} {\bibinfo {author} {\bibnamefont
  {Keim}, \bibfnamefont {N.~C.}}, \bibinfo {author} {\bibfnamefont
  {P.}~\bibnamefont {M{\o}ller}}, \bibinfo {author} {\bibfnamefont {W.~W.}\
  \bibnamefont {Zhang}}, \ and\ \bibinfo {author} {\bibfnamefont {S.~R.}\
  \bibnamefont {Nagel}}} (\bibinfo {year} {2006}),\ \href@noop {} {\bibfield
  {journal} {\bibinfo  {journal} {Phys. Rev. Lett.}\ }\textbf {\bibinfo
  {volume} {97}}~(\bibinfo {number} {14}),\ \bibinfo {pages}
  {144503}}\BibitemShut {NoStop}%
\bibitem [{\citenamefont {Keim}\ and\ \citenamefont {Nagel}(2011)}]{Keim11}%
  \BibitemOpen
  \bibfield  {author} {\bibinfo {author} {\bibnamefont {Keim}, \bibfnamefont
  {N.~C.}}, \ and\ \bibinfo {author} {\bibfnamefont {S.~R.}\ \bibnamefont
  {Nagel}}} (\bibinfo {year} {2011}),\ \href@noop {} {\bibfield  {journal}
  {\bibinfo  {journal} {Phys. Rev. Lett.}\ }\textbf {\bibinfo {volume}
  {107}}~(\bibinfo {number} {1}),\ \bibinfo {pages} {010603}}\BibitemShut
  {NoStop}%
\bibitem [{\citenamefont {Keim}\ \emph {et~al.}(2013)\citenamefont {Keim},
  \citenamefont {Paulsen},\ and\ \citenamefont {Nagel}}]{Keim13b}%
  \BibitemOpen
  \bibfield  {author} {\bibinfo {author} {\bibnamefont {Keim}, \bibfnamefont
  {N.~C.}}, \bibinfo {author} {\bibfnamefont {J.~D.}\ \bibnamefont {Paulsen}},
  \ and\ \bibinfo {author} {\bibfnamefont {S.~R.}\ \bibnamefont {Nagel}}}
  (\bibinfo {year} {2013}),\ \href@noop {} {\bibfield  {journal} {\bibinfo
  {journal} {Phys. Rev. E}\ }\textbf {\bibinfo {volume} {88}}~(\bibinfo
  {number} {3}),\ \bibinfo {pages} {032306}}\BibitemShut {NoStop}%
\bibitem [{\citenamefont {Keller}\ and\ \citenamefont
  {Miksis}(1983)}]{Keller83}%
  \BibitemOpen
  \bibfield  {author} {\bibinfo {author} {\bibnamefont {Keller}, \bibfnamefont
  {J.~B.}}, \ and\ \bibinfo {author} {\bibfnamefont {M.~J.}\ \bibnamefont
  {Miksis}}} (\bibinfo {year} {1983}),\ \href@noop {} {\bibfield  {journal}
  {\bibinfo  {journal} {SIAM Journal on Applied Mathematics}\ }\textbf
  {\bibinfo {volume} {43}}~(\bibinfo {number} {2}),\ \bibinfo {pages}
  {268}}\BibitemShut {NoStop}%
\bibitem [{\citenamefont {Kim}\ and\ \citenamefont {Mason}(2017)}]{Kim17}%
  \BibitemOpen
  \bibfield  {author} {\bibinfo {author} {\bibnamefont {Kim}, \bibfnamefont
  {H.~S.}}, \ and\ \bibinfo {author} {\bibfnamefont {T.~G.}\ \bibnamefont
  {Mason}}} (\bibinfo {year} {2017}),\ \href@noop {} {\bibfield  {journal}
  {\bibinfo  {journal} {Advances in colloid and interface science}\ }\textbf
  {\bibinfo {volume} {247}},\ \bibinfo {pages} {397}}\BibitemShut {NoStop}%
\bibitem [{\citenamefont {Komori}\ \emph {et~al.}(2000)\citenamefont {Komori},
  \citenamefont {Yoshino},\ and\ \citenamefont {Takayama}}]{Komori00}%
  \BibitemOpen
  \bibfield  {author} {\bibinfo {author} {\bibnamefont {Komori}, \bibfnamefont
  {T.}}, \bibinfo {author} {\bibfnamefont {H.}~\bibnamefont {Yoshino}}, \ and\
  \bibinfo {author} {\bibfnamefont {H.}~\bibnamefont {Takayama}}} (\bibinfo
  {year} {2000}),\ \href@noop {} {\bibfield  {journal} {\bibinfo  {journal}
  {Journal of the Physical Society of Japan}\ }\textbf {\bibinfo {volume}
  {69}}~(\bibinfo {number} {4}),\ \bibinfo {pages} {1192}}\BibitemShut
  {NoStop}%
\bibitem [{\citenamefont {Korpel}\ and\ \citenamefont
  {Chatterjee}(1981)}]{korpel81}%
  \BibitemOpen
  \bibfield  {author} {\bibinfo {author} {\bibnamefont {Korpel}, \bibfnamefont
  {A.}}, \ and\ \bibinfo {author} {\bibfnamefont {M.}~\bibnamefont
  {Chatterjee}}} (\bibinfo {year} {1981}),\ \href@noop {} {\bibfield  {journal}
  {\bibinfo  {journal} {Proceedings of the IEEE}\ }\textbf {\bibinfo {volume}
  {69}}~(\bibinfo {number} {12}),\ \bibinfo {pages} {1539}}\BibitemShut
  {NoStop}%
\bibitem [{\citenamefont {Kovacs}(1963)}]{Kovacs63}%
  \BibitemOpen
  \bibfield  {author} {\bibinfo {author} {\bibnamefont {Kovacs}, \bibfnamefont
  {A.}}} (\bibinfo {year} {1963}),\ \href@noop {} {\bibfield  {journal}
  {\bibinfo  {journal} {Adv. Polym. Sci}\ }\textbf {\bibinfo {volume}
  {3}}~(\bibinfo {number} {3}),\ \bibinfo {pages} {394}}\BibitemShut {NoStop}%
\bibitem [{\citenamefont {Kovacs}\ \emph {et~al.}(1979)\citenamefont {Kovacs},
  \citenamefont {Aklonis}, \citenamefont {Hutchinson},\ and\ \citenamefont
  {Ramos}}]{Kovacs79}%
  \BibitemOpen
  \bibfield  {author} {\bibinfo {author} {\bibnamefont {Kovacs}, \bibfnamefont
  {A.~J.}}, \bibinfo {author} {\bibfnamefont {J.~J.}\ \bibnamefont {Aklonis}},
  \bibinfo {author} {\bibfnamefont {J.~M.}\ \bibnamefont {Hutchinson}}, \ and\
  \bibinfo {author} {\bibfnamefont {A.~R.}\ \bibnamefont {Ramos}}} (\bibinfo
  {year} {1979}),\ \href@noop {} {\bibfield  {journal} {\bibinfo  {journal}
  {Journal of Polymer Science: Polymer Physics Edition}\ }\textbf {\bibinfo
  {volume} {17}}~(\bibinfo {number} {7}),\ \bibinfo {pages} {1097}}\BibitemShut
  {NoStop}%
\bibitem [{\citenamefont {K{\"u}hner}\ \emph {et~al.}(2009)\citenamefont
  {K{\"u}hner}, \citenamefont {van Noort}, \citenamefont {Betts}, \citenamefont
  {Leo-Macias}, \citenamefont {Batisse}, \citenamefont {Rode}, \citenamefont
  {Yamada}, \citenamefont {Maier}, \citenamefont {Bader}, \citenamefont
  {Beltran-Alvarez}, \citenamefont {Casta{\~n}o-Diez}, \citenamefont {Chen},
  \citenamefont {Devos}, \citenamefont {G{\"u}ell}, \citenamefont {Norambuena},
  \citenamefont {Racke}, \citenamefont {Rybin}, \citenamefont {Schmidt},
  \citenamefont {Yus}, \citenamefont {Aebersold}, \citenamefont {Herrmann},
  \citenamefont {B{\"o}ttcher}, \citenamefont {Frangakis}, \citenamefont
  {Russell}, \citenamefont {Serrano}, \citenamefont {Bork},\ and\ \citenamefont
  {Gavin}}]{Kuhner09}%
  \BibitemOpen
  \bibfield  {author} {\bibinfo {author} {\bibnamefont {K{\"u}hner},
  \bibfnamefont {S.}}, \bibinfo {author} {\bibfnamefont {V.}~\bibnamefont {van
  Noort}}, \bibinfo {author} {\bibfnamefont {M.~J.}\ \bibnamefont {Betts}},
  \bibinfo {author} {\bibfnamefont {A.}~\bibnamefont {Leo-Macias}}, \bibinfo
  {author} {\bibfnamefont {C.}~\bibnamefont {Batisse}}, \bibinfo {author}
  {\bibfnamefont {M.}~\bibnamefont {Rode}}, \bibinfo {author} {\bibfnamefont
  {T.}~\bibnamefont {Yamada}}, \bibinfo {author} {\bibfnamefont
  {T.}~\bibnamefont {Maier}}, \bibinfo {author} {\bibfnamefont
  {S.}~\bibnamefont {Bader}}, \bibinfo {author} {\bibfnamefont
  {P.}~\bibnamefont {Beltran-Alvarez}}, \bibinfo {author} {\bibfnamefont
  {D.}~\bibnamefont {Casta{\~n}o-Diez}}, \bibinfo {author} {\bibfnamefont
  {W.-H.}\ \bibnamefont {Chen}}, \bibinfo {author} {\bibfnamefont
  {D.}~\bibnamefont {Devos}}, \bibinfo {author} {\bibfnamefont
  {M.}~\bibnamefont {G{\"u}ell}}, \bibinfo {author} {\bibfnamefont
  {T.}~\bibnamefont {Norambuena}}, \bibinfo {author} {\bibfnamefont
  {I.}~\bibnamefont {Racke}}, \bibinfo {author} {\bibfnamefont
  {V.}~\bibnamefont {Rybin}}, \bibinfo {author} {\bibfnamefont
  {A.}~\bibnamefont {Schmidt}}, \bibinfo {author} {\bibfnamefont
  {E.}~\bibnamefont {Yus}}, \bibinfo {author} {\bibfnamefont {R.}~\bibnamefont
  {Aebersold}}, \bibinfo {author} {\bibfnamefont {R.}~\bibnamefont {Herrmann}},
  \bibinfo {author} {\bibfnamefont {B.}~\bibnamefont {B{\"o}ttcher}}, \bibinfo
  {author} {\bibfnamefont {A.~S.}\ \bibnamefont {Frangakis}}, \bibinfo {author}
  {\bibfnamefont {R.~B.}\ \bibnamefont {Russell}}, \bibinfo {author}
  {\bibfnamefont {L.}~\bibnamefont {Serrano}}, \bibinfo {author} {\bibfnamefont
  {P.}~\bibnamefont {Bork}}, \ and\ \bibinfo {author} {\bibfnamefont {A.-C.}\
  \bibnamefont {Gavin}}} (\bibinfo {year} {2009}),\ \href {\doibase
  10.1126/science.1176343} {\bibfield  {journal} {\bibinfo  {journal}
  {Science}\ }\textbf {\bibinfo {volume} {326}}~(\bibinfo {number} {5957}),\
  \bibinfo {pages} {1235}}\BibitemShut {NoStop}%
\bibitem [{\citenamefont {Kurita}\ and\ \citenamefont
  {Fujii}(1979)}]{Kurita79}%
  \BibitemOpen
  \bibfield  {author} {\bibinfo {author} {\bibnamefont {Kurita}, \bibfnamefont
  {K.}}, \ and\ \bibinfo {author} {\bibfnamefont {N.}~\bibnamefont {Fujii}}}
  (\bibinfo {year} {1979}),\ \href@noop {} {\bibfield  {journal} {\bibinfo
  {journal} {Geophysical Research Letters}\ }\textbf {\bibinfo {volume}
  {6}}~(\bibinfo {number} {1}),\ \bibinfo {pages} {9}}\BibitemShut {NoStop}%
\bibitem [{\citenamefont {Kurnit}\ \emph {et~al.}(1964)\citenamefont {Kurnit},
  \citenamefont {Abella},\ and\ \citenamefont {Hartmann}}]{Kurnit64}%
  \BibitemOpen
  \bibfield  {author} {\bibinfo {author} {\bibnamefont {Kurnit}, \bibfnamefont
  {N.}}, \bibinfo {author} {\bibfnamefont {I.}~\bibnamefont {Abella}}, \ and\
  \bibinfo {author} {\bibfnamefont {S.}~\bibnamefont {Hartmann}}} (\bibinfo
  {year} {1964}),\ \href@noop {} {\bibfield  {journal} {\bibinfo  {journal}
  {Phys. Rev. Lett.}\ }\textbf {\bibinfo {volume} {13}}~(\bibinfo {number}
  {19}),\ \bibinfo {pages} {567}}\BibitemShut {NoStop}%
\bibitem [{\citenamefont {Lagoudas}(2008)}]{Lagoudas08}%
  \BibitemOpen
  \bibfield  {author} {\bibinfo {author} {\bibnamefont {Lagoudas},
  \bibfnamefont {D.~C.}}} (\bibinfo {year} {2008}),\ \href@noop {} {\emph
  {\bibinfo {title} {Shape memory alloys: modeling and engineering
  applications}}}\ (\bibinfo  {publisher} {Springer Science \& Business
  Media})\BibitemShut {NoStop}%
\bibitem [{\citenamefont {Lahini}\ \emph {et~al.}(2017)\citenamefont {Lahini},
  \citenamefont {Gottesman}, \citenamefont {Amir},\ and\ \citenamefont
  {Rubinstein}}]{Lahini17}%
  \BibitemOpen
  \bibfield  {author} {\bibinfo {author} {\bibnamefont {Lahini}, \bibfnamefont
  {Y.}}, \bibinfo {author} {\bibfnamefont {O.}~\bibnamefont {Gottesman}},
  \bibinfo {author} {\bibfnamefont {A.}~\bibnamefont {Amir}}, \ and\ \bibinfo
  {author} {\bibfnamefont {S.~M.}\ \bibnamefont {Rubinstein}}} (\bibinfo {year}
  {2017}),\ \href@noop {} {\bibfield  {journal} {\bibinfo  {journal} {Phys.
  Rev. Lett.}\ }\textbf {\bibinfo {volume} {118}}~(\bibinfo {number} {8}),\
  \bibinfo {pages} {085501}}\BibitemShut {NoStop}%
\bibitem [{\citenamefont {Lasanta}\ \emph {et~al.}(2018)\citenamefont
  {Lasanta}, \citenamefont {Reyes}, \citenamefont {Prados},\ and\ \citenamefont
  {Santos}}]{Lasanta18}%
  \BibitemOpen
  \bibfield  {author} {\bibinfo {author} {\bibnamefont {Lasanta}, \bibfnamefont
  {A.}}, \bibinfo {author} {\bibfnamefont {F.~V.}\ \bibnamefont {Reyes}},
  \bibinfo {author} {\bibfnamefont {A.}~\bibnamefont {Prados}}, \ and\ \bibinfo
  {author} {\bibfnamefont {A.}~\bibnamefont {Santos}}} (\bibinfo {year}
  {2018}),\ \href@noop {} {\bibinfo  {journal} {arXiv preprint
  arXiv:1808.08401}\ }\BibitemShut {NoStop}%
\bibitem [{\citenamefont {Laurson}\ and\ \citenamefont
  {Alava}(2012)}]{Laurson12}%
  \BibitemOpen
\bibfield  {journal} {  }\bibfield  {author} {\bibinfo {author} {\bibnamefont
  {Laurson}, \bibfnamefont {L.}}, \ and\ \bibinfo {author} {\bibfnamefont
  {M.~J.}\ \bibnamefont {Alava}}} (\bibinfo {year} {2012}),\ \href@noop {}
  {\bibfield  {journal} {\bibinfo  {journal} {Phys. Rev. Lett.}\ }\textbf
  {\bibinfo {volume} {109}}~(\bibinfo {number} {15}),\ \bibinfo {pages}
  {155504}}\BibitemShut {NoStop}%
\bibitem [{\citenamefont {Lavrentovich}\ \emph {et~al.}(2017)\citenamefont
  {Lavrentovich}, \citenamefont {Liu},\ and\ \citenamefont
  {Nagel}}]{Lavrentovich17}%
  \BibitemOpen
  \bibfield  {author} {\bibinfo {author} {\bibnamefont {Lavrentovich},
  \bibfnamefont {M.~O.}}, \bibinfo {author} {\bibfnamefont {A.~J.}\
  \bibnamefont {Liu}}, \ and\ \bibinfo {author} {\bibfnamefont {S.~R.}\
  \bibnamefont {Nagel}}} (\bibinfo {year} {2017}),\ \href@noop {} {\bibfield
  {journal} {\bibinfo  {journal} {Phys. Rev. E}\ }\textbf {\bibinfo {volume}
  {96}},\ \bibinfo {pages} {020101(R)}}\BibitemShut {NoStop}%
\bibitem [{\citenamefont {Lee}\ and\ \citenamefont {Furst}(2008)}]{Lee08}%
  \BibitemOpen
  \bibfield  {author} {\bibinfo {author} {\bibnamefont {Lee}, \bibfnamefont
  {M.~H.}}, \ and\ \bibinfo {author} {\bibfnamefont {E.~M.}\ \bibnamefont
  {Furst}}} (\bibinfo {year} {2008}),\ \href@noop {} {\bibfield  {journal}
  {\bibinfo  {journal} {Phys. Rev. E}\ }\textbf {\bibinfo {volume}
  {77}}~(\bibinfo {number} {4}),\ \bibinfo {pages} {041408}}\BibitemShut
  {NoStop}%
\bibitem [{\citenamefont {Lendlein}\ \emph {et~al.}(2005)\citenamefont
  {Lendlein}, \citenamefont {Jiang}, \citenamefont {J{\"u}nger},\ and\
  \citenamefont {Langer}}]{Lendlein05}%
  \BibitemOpen
  \bibfield  {author} {\bibinfo {author} {\bibnamefont {Lendlein},
  \bibfnamefont {A.}}, \bibinfo {author} {\bibfnamefont {H.}~\bibnamefont
  {Jiang}}, \bibinfo {author} {\bibfnamefont {O.}~\bibnamefont {J{\"u}nger}}, \
  and\ \bibinfo {author} {\bibfnamefont {R.}~\bibnamefont {Langer}}} (\bibinfo
  {year} {2005}),\ \href@noop {} {\bibfield  {journal} {\bibinfo  {journal}
  {Nature}\ }\textbf {\bibinfo {volume} {434}}~(\bibinfo {number} {7035}),\
  \bibinfo {pages} {879}}\BibitemShut {NoStop}%
\bibitem [{\citenamefont {Lendlein}\ and\ \citenamefont
  {Kelch}(2002)}]{Lendlein02}%
  \BibitemOpen
  \bibfield  {author} {\bibinfo {author} {\bibnamefont {Lendlein},
  \bibfnamefont {A.}}, \ and\ \bibinfo {author} {\bibfnamefont
  {S.}~\bibnamefont {Kelch}}} (\bibinfo {year} {2002}),\ \href@noop {}
  {\bibfield  {journal} {\bibinfo  {journal} {Angewandte Chemie International
  Edition}\ }\textbf {\bibinfo {volume} {41}}~(\bibinfo {number} {12}),\
  \bibinfo {pages} {2034}}\BibitemShut {NoStop}%
\bibitem [{\citenamefont {Lin}\ \emph {et~al.}(2016)\citenamefont {Lin},
  \citenamefont {Ness}, \citenamefont {Cates}, \citenamefont {Sun},\ and\
  \citenamefont {Cohen}}]{Lin16}%
  \BibitemOpen
  \bibfield  {author} {\bibinfo {author} {\bibnamefont {Lin}, \bibfnamefont
  {N.~Y.}}, \bibinfo {author} {\bibfnamefont {C.}~\bibnamefont {Ness}},
  \bibinfo {author} {\bibfnamefont {M.~E.}\ \bibnamefont {Cates}}, \bibinfo
  {author} {\bibfnamefont {J.}~\bibnamefont {Sun}}, \ and\ \bibinfo {author}
  {\bibfnamefont {I.}~\bibnamefont {Cohen}}} (\bibinfo {year} {2016}),\
  \href@noop {} {\bibfield  {journal} {\bibinfo  {journal} {Proceedings of the
  National Academy of Sciences}\ }\textbf {\bibinfo {volume} {113}}~(\bibinfo
  {number} {39}),\ \bibinfo {pages} {10774}}\BibitemShut {NoStop}%
\bibitem [{\citenamefont {Liu}\ and\ \citenamefont {Nagel}(2010)}]{Liu10}%
  \BibitemOpen
  \bibfield  {author} {\bibinfo {author} {\bibnamefont {Liu}, \bibfnamefont
  {A.~J.}}, \ and\ \bibinfo {author} {\bibfnamefont {S.~R.}\ \bibnamefont
  {Nagel}}} (\bibinfo {year} {2010}),\ \href@noop {} {\bibfield  {journal}
  {\bibinfo  {journal} {Annual Reviews of Condensed Matter Physics}\ }\textbf
  {\bibinfo {volume} {1}},\ \bibinfo {pages} {347}}\BibitemShut {NoStop}%
\bibitem [{\citenamefont {Lundberg}\ \emph {et~al.}(2008)\citenamefont
  {Lundberg}, \citenamefont {Krishan}, \citenamefont {Xu}, \citenamefont
  {O'Hern},\ and\ \citenamefont {Dennin}}]{Lundberg08}%
  \BibitemOpen
  \bibfield  {author} {\bibinfo {author} {\bibnamefont {Lundberg},
  \bibfnamefont {M.}}, \bibinfo {author} {\bibfnamefont {K.}~\bibnamefont
  {Krishan}}, \bibinfo {author} {\bibfnamefont {N.}~\bibnamefont {Xu}},
  \bibinfo {author} {\bibfnamefont {C.~S.}\ \bibnamefont {O'Hern}}, \ and\
  \bibinfo {author} {\bibfnamefont {M.}~\bibnamefont {Dennin}}} (\bibinfo
  {year} {2008}),\ \href@noop {} {\bibfield  {journal} {\bibinfo  {journal}
  {Phys. Rev. E}\ }\textbf {\bibinfo {volume} {77}}~(\bibinfo {number} {4}),\
  \bibinfo {pages} {041505}}\BibitemShut {NoStop}%
\bibitem [{\citenamefont {Maiorano}\ \emph {et~al.}(2005)\citenamefont
  {Maiorano}, \citenamefont {Marinari},\ and\ \citenamefont
  {Ricci-Tersenghi}}]{Maiorano05}%
  \BibitemOpen
  \bibfield  {author} {\bibinfo {author} {\bibnamefont {Maiorano},
  \bibfnamefont {A.}}, \bibinfo {author} {\bibfnamefont {E.}~\bibnamefont
  {Marinari}}, \ and\ \bibinfo {author} {\bibfnamefont {F.}~\bibnamefont
  {Ricci-Tersenghi}}} (\bibinfo {year} {2005}),\ \href@noop {} {\bibfield
  {journal} {\bibinfo  {journal} {Physical Review B}\ }\textbf {\bibinfo
  {volume} {72}}~(\bibinfo {number} {10}),\ \bibinfo {pages}
  {104411}}\BibitemShut {NoStop}%
\bibitem [{\citenamefont {Majumdar}\ \emph {et~al.}(2018)\citenamefont
  {Majumdar}, \citenamefont {Foucard}, \citenamefont {Levine},\ and\
  \citenamefont {Gardel}}]{Majumdar:2018kc}%
  \BibitemOpen
  \bibfield  {author} {\bibinfo {author} {\bibnamefont {Majumdar},
  \bibfnamefont {S.}}, \bibinfo {author} {\bibfnamefont {L.~C.}\ \bibnamefont
  {Foucard}}, \bibinfo {author} {\bibfnamefont {A.~J.}\ \bibnamefont {Levine}},
  \ and\ \bibinfo {author} {\bibfnamefont {M.~L.}\ \bibnamefont {Gardel}}}
  (\bibinfo {year} {2018}),\ \href@noop {} {\bibfield  {journal} {\bibinfo
  {journal} {Soft Matter}\ }\textbf {\bibinfo {volume} {14}}~(\bibinfo {number}
  {11}),\ \bibinfo {pages} {2052}}\BibitemShut {NoStop}%
\bibitem [{\citenamefont {Mangan}\ \emph {et~al.}(2008)\citenamefont {Mangan},
  \citenamefont {Reichhardt},\ and\ \citenamefont {Reichhardt}}]{Mangan08}%
  \BibitemOpen
  \bibfield  {author} {\bibinfo {author} {\bibnamefont {Mangan}, \bibfnamefont
  {N.}}, \bibinfo {author} {\bibfnamefont {C.}~\bibnamefont {Reichhardt}}, \
  and\ \bibinfo {author} {\bibfnamefont {C.~J.~O.}\ \bibnamefont {Reichhardt}}}
  (\bibinfo {year} {2008}),\ \href@noop {} {\bibfield  {journal} {\bibinfo
  {journal} {Phys. Rev. Lett.}\ }\textbf {\bibinfo {volume} {100}}~(\bibinfo
  {number} {18}),\ \bibinfo {pages} {187002}}\BibitemShut {NoStop}%
\bibitem [{\citenamefont {Mano}(2008)}]{Mano08}%
  \BibitemOpen
  \bibfield  {author} {\bibinfo {author} {\bibnamefont {Mano}, \bibfnamefont
  {J.~F.}}} (\bibinfo {year} {2008}),\ \href@noop {} {\bibfield  {journal}
  {\bibinfo  {journal} {Advanced Engineering Materials}\ }\textbf {\bibinfo
  {volume} {10}}~(\bibinfo {number} {6}),\ \bibinfo {pages} {515}}\BibitemShut
  {NoStop}%
\bibitem [{\citenamefont {Matan}\ \emph {et~al.}(2002)\citenamefont {Matan},
  \citenamefont {Williams}, \citenamefont {Witten},\ and\ \citenamefont
  {Nagel}}]{Matan02}%
  \BibitemOpen
  \bibfield  {author} {\bibinfo {author} {\bibnamefont {Matan}, \bibfnamefont
  {K.}}, \bibinfo {author} {\bibfnamefont {R.~B.}\ \bibnamefont {Williams}},
  \bibinfo {author} {\bibfnamefont {T.~A.}\ \bibnamefont {Witten}}, \ and\
  \bibinfo {author} {\bibfnamefont {S.~R.}\ \bibnamefont {Nagel}}} (\bibinfo
  {year} {2002}),\ \href@noop {} {\bibfield  {journal} {\bibinfo  {journal}
  {Phys. Rev. Lett.}\ }\textbf {\bibinfo {volume} {88}}~(\bibinfo {number}
  {7}),\ \bibinfo {pages} {076101}}\BibitemShut {NoStop}%
\bibitem [{\citenamefont {Mather}\ \emph {et~al.}(2009)\citenamefont {Mather},
  \citenamefont {Luo},\ and\ \citenamefont {Rousseau}}]{Mather09}%
  \BibitemOpen
  \bibfield  {author} {\bibinfo {author} {\bibnamefont {Mather}, \bibfnamefont
  {P.~T.}}, \bibinfo {author} {\bibfnamefont {X.}~\bibnamefont {Luo}}, \ and\
  \bibinfo {author} {\bibfnamefont {I.~A.}\ \bibnamefont {Rousseau}}} (\bibinfo
  {year} {2009}),\ \href@noop {} {\bibfield  {journal} {\bibinfo  {journal}
  {Annual Review of Materials Research}\ }\textbf {\bibinfo {volume} {39}},\
  \bibinfo {pages} {445}}\BibitemShut {NoStop}%
\bibitem [{\citenamefont {Mehta}\ \emph {et~al.}(2018)\citenamefont {Mehta},
  \citenamefont {Bukov}, \citenamefont {Wang}, \citenamefont {Day},
  \citenamefont {Richardson}, \citenamefont {Fisher},\ and\ \citenamefont
  {Schwab}}]{Pankaj18}%
  \BibitemOpen
  \bibfield  {author} {\bibinfo {author} {\bibnamefont {Mehta}, \bibfnamefont
  {P.}}, \bibinfo {author} {\bibfnamefont {M.}~\bibnamefont {Bukov}}, \bibinfo
  {author} {\bibfnamefont {C.-H.}\ \bibnamefont {Wang}}, \bibinfo {author}
  {\bibfnamefont {A.~G.}\ \bibnamefont {Day}}, \bibinfo {author} {\bibfnamefont
  {C.}~\bibnamefont {Richardson}}, \bibinfo {author} {\bibfnamefont {C.~K.}\
  \bibnamefont {Fisher}}, \ and\ \bibinfo {author} {\bibfnamefont {D.~J.}\
  \bibnamefont {Schwab}}} (\bibinfo {year} {2018}),\ \href@noop {} {\bibinfo
  {journal} {arXiv:1803.08823}\ }\BibitemShut {NoStop}%
\bibitem [{\citenamefont {Middleton}(1992)}]{Middleton92}%
  \BibitemOpen
\bibfield  {journal} {  }\bibfield  {author} {\bibinfo {author} {\bibnamefont
  {Middleton}, \bibfnamefont {A.~A.}}} (\bibinfo {year} {1992}),\ \href@noop {}
  {\bibfield  {journal} {\bibinfo  {journal} {Phys. Rev. Lett.}\ }\textbf
  {\bibinfo {volume} {68}},\ \bibinfo {pages} {670}}\BibitemShut {NoStop}%
\bibitem [{\citenamefont {Mossa}\ and\ \citenamefont
  {Sciortino}(2004)}]{Mossa2004d}%
  \BibitemOpen
  \bibfield  {author} {\bibinfo {author} {\bibnamefont {Mossa}, \bibfnamefont
  {S.}}, \ and\ \bibinfo {author} {\bibfnamefont {F.}~\bibnamefont
  {Sciortino}}} (\bibinfo {year} {2004}),\ \href {\doibase
  10.1103/PhysRevLett.92.045504} {\bibfield  {journal} {\bibinfo  {journal}
  {Phys. Rev. Lett.}\ }\textbf {\bibinfo {volume} {92}}~(\bibinfo {number}
  {4}),\ \bibinfo {pages} {4}}\BibitemShut {NoStop}%
\bibitem [{\citenamefont {Mueggenburg}(2005)}]{Mueggenburg05}%
  \BibitemOpen
  \bibfield  {author} {\bibinfo {author} {\bibnamefont {Mueggenburg},
  \bibfnamefont {N.}}} (\bibinfo {year} {2005}),\ \href@noop {} {\bibfield
  {journal} {\bibinfo  {journal} {Phys. Rev. E}\ }\textbf {\bibinfo {volume}
  {71}}~(\bibinfo {number} {3}),\ \bibinfo {pages} {031301}}\BibitemShut
  {NoStop}%
\bibitem [{\citenamefont {Mukherji}\ \emph {et~al.}(2018)\citenamefont
  {Mukherji}, \citenamefont {Kandula}, \citenamefont {Sood},\ and\
  \citenamefont {Ganapathy}}]{Sood18}%
  \BibitemOpen
  \bibfield  {author} {\bibinfo {author} {\bibnamefont {Mukherji},
  \bibfnamefont {S.}}, \bibinfo {author} {\bibfnamefont {N.}~\bibnamefont
  {Kandula}}, \bibinfo {author} {\bibfnamefont {A.~K.}\ \bibnamefont {Sood}}, \
  and\ \bibinfo {author} {\bibfnamefont {R.}~\bibnamefont {Ganapathy}}}
  (\bibinfo {year} {2018}),\ \href@noop {} {}\bibinfo {howpublished}
  {arXiv:1808.07701}\BibitemShut {NoStop}%
\bibitem [{\citenamefont {Mullins}(1948)}]{Mullins48}%
  \BibitemOpen
  \bibfield  {author} {\bibinfo {author} {\bibnamefont {Mullins}, \bibfnamefont
  {L.}}} (\bibinfo {year} {1948}),\ \href@noop {} {\bibfield  {journal}
  {\bibinfo  {journal} {Rubber Chemistry and Technology}\ }\textbf {\bibinfo
  {volume} {21}}~(\bibinfo {number} {2}),\ \bibinfo {pages} {281}}\BibitemShut
  {NoStop}%
\bibitem [{\citenamefont {Mungan}\ and\ \citenamefont
  {Terzi}(2018)}]{Mungan18}%
  \BibitemOpen
  \bibfield  {author} {\bibinfo {author} {\bibnamefont {Mungan}, \bibfnamefont
  {M.}}, \ and\ \bibinfo {author} {\bibfnamefont {M.~M.}\ \bibnamefont
  {Terzi}}} (\bibinfo {year} {2018}),\ \href@noop {} {}\bibinfo {howpublished}
  {arXiv:1802.03096}\BibitemShut {NoStop}%
\bibitem [{\citenamefont {Mungan}\ and\ \citenamefont
  {Witten}(2019)}]{Mungan19}%
  \BibitemOpen
  \bibfield  {author} {\bibinfo {author} {\bibnamefont {Mungan}, \bibfnamefont
  {M.}}, \ and\ \bibinfo {author} {\bibfnamefont {T.~A.}\ \bibnamefont
  {Witten}}} (\bibinfo {year} {2019}),\ \href@noop {} {\bibinfo  {journal}
  {arXiv preprint arXiv:1902.08088}\ }\BibitemShut {NoStop}%
\bibitem [{\citenamefont {Murugan}\ \emph {et~al.}(2015)\citenamefont
  {Murugan}, \citenamefont {Zeravcic}, \citenamefont {Brenner},\ and\
  \citenamefont {Leibler}}]{Murugan15}%
  \BibitemOpen
\bibfield  {journal} {  }\bibfield  {author} {\bibinfo {author} {\bibnamefont
  {Murugan}, \bibfnamefont {A.}}, \bibinfo {author} {\bibfnamefont
  {Z.}~\bibnamefont {Zeravcic}}, \bibinfo {author} {\bibfnamefont {M.~P.}\
  \bibnamefont {Brenner}}, \ and\ \bibinfo {author} {\bibfnamefont
  {S.}~\bibnamefont {Leibler}}} (\bibinfo {year} {2015}),\ \href {\doibase
  10.1073/pnas.1413941112} {\bibfield  {journal} {\bibinfo  {journal}
  {Proceedings of the National Academy of Sciences}\ }\textbf {\bibinfo
  {volume} {112}}~(\bibinfo {number} {1}),\ \bibinfo {pages} {54}}\BibitemShut
  {NoStop}%
\bibitem [{\citenamefont {Nagamanasa}\ \emph {et~al.}(2014)\citenamefont
  {Nagamanasa}, \citenamefont {Gokhale}, \citenamefont {Sood},\ and\
  \citenamefont {Ganapathy}}]{Nagamanasa14}%
  \BibitemOpen
  \bibfield  {author} {\bibinfo {author} {\bibnamefont {Nagamanasa},
  \bibfnamefont {K.~H.}}, \bibinfo {author} {\bibfnamefont {S.}~\bibnamefont
  {Gokhale}}, \bibinfo {author} {\bibfnamefont {A.~K.}\ \bibnamefont {Sood}}, \
  and\ \bibinfo {author} {\bibfnamefont {R.}~\bibnamefont {Ganapathy}}}
  (\bibinfo {year} {2014}),\ \href@noop {} {\bibfield  {journal} {\bibinfo
  {journal} {Phys. Rev. E}\ }\textbf {\bibinfo {volume} {89}}~(\bibinfo
  {number} {6}),\ \bibinfo {pages} {062308}}\BibitemShut {NoStop}%
\bibitem [{\citenamefont {Nagel}\ \emph {et~al.}(1983)\citenamefont {Nagel},
  \citenamefont {Grest},\ and\ \citenamefont {Rahman}}]{Nagel83}%
  \BibitemOpen
  \bibfield  {author} {\bibinfo {author} {\bibnamefont {Nagel}, \bibfnamefont
  {S.~R.}}, \bibinfo {author} {\bibfnamefont {G.~S.}\ \bibnamefont {Grest}}, \
  and\ \bibinfo {author} {\bibfnamefont {A.}~\bibnamefont {Rahman}}} (\bibinfo
  {year} {1983}),\ \href@noop {} {\bibfield  {journal} {\bibinfo  {journal}
  {Physics Today}\ }\textbf {\bibinfo {volume} {36}},\ \bibinfo {pages}
  {24}}\BibitemShut {NoStop}%
\bibitem [{\citenamefont {Ness}\ \emph {et~al.}(2018)\citenamefont {Ness},
  \citenamefont {Mari},\ and\ \citenamefont {Cates}}]{Ness2018}%
  \BibitemOpen
  \bibfield  {author} {\bibinfo {author} {\bibnamefont {Ness}, \bibfnamefont
  {C.}}, \bibinfo {author} {\bibfnamefont {R.}~\bibnamefont {Mari}}, \ and\
  \bibinfo {author} {\bibfnamefont {M.~E.}\ \bibnamefont {Cates}}} (\bibinfo
  {year} {2018}),\ \href {\doibase 10.1126/sciadv.aar3296} {\bibfield
  {journal} {\bibinfo  {journal} {Sci. Adv.}\ }\textbf {\bibinfo {volume}
  {4}}~(\bibinfo {number} {3}),\ \bibinfo {pages} {4}}\BibitemShut {NoStop}%
\bibitem [{\citenamefont {Ong}\ \emph {et~al.}(2017)\citenamefont {Ong},
  \citenamefont {Hanikel}, \citenamefont {Yaghi}, \citenamefont {Strauss},
  \citenamefont {Bron}, \citenamefont {Lai-Kee-Him}, \citenamefont {Schueder},
  \citenamefont {Wang}, \citenamefont {Wang}, \citenamefont {Kishi},
  \citenamefont {Myhrvold}, \citenamefont {Zhu}, \citenamefont {Jungmann},
  \citenamefont {Bellot}, \citenamefont {Ke},\ and\ \citenamefont
  {Yin}}]{Ong2017}%
  \BibitemOpen
  \bibfield  {author} {\bibinfo {author} {\bibnamefont {Ong}, \bibfnamefont
  {L.~L.}}, \bibinfo {author} {\bibfnamefont {N.}~\bibnamefont {Hanikel}},
  \bibinfo {author} {\bibfnamefont {C.}~\bibnamefont {Yaghi}, \bibfnamefont
  {Omar K.and~Grun}}, \bibinfo {author} {\bibfnamefont {M.~T.}\ \bibnamefont
  {Strauss}}, \bibinfo {author} {\bibfnamefont {P.}~\bibnamefont {Bron}},
  \bibinfo {author} {\bibfnamefont {J.}~\bibnamefont {Lai-Kee-Him}}, \bibinfo
  {author} {\bibfnamefont {F.}~\bibnamefont {Schueder}}, \bibinfo {author}
  {\bibfnamefont {B.}~\bibnamefont {Wang}}, \bibinfo {author} {\bibfnamefont
  {P.}~\bibnamefont {Wang}}, \bibinfo {author} {\bibfnamefont {J.~Y.}\
  \bibnamefont {Kishi}}, \bibinfo {author} {\bibfnamefont {C.}~\bibnamefont
  {Myhrvold}}, \bibinfo {author} {\bibfnamefont {A.}~\bibnamefont {Zhu}},
  \bibinfo {author} {\bibfnamefont {R.}~\bibnamefont {Jungmann}}, \bibinfo
  {author} {\bibfnamefont {G.}~\bibnamefont {Bellot}}, \bibinfo {author}
  {\bibfnamefont {Y.}~\bibnamefont {Ke}}, \ and\ \bibinfo {author}
  {\bibfnamefont {P.}~\bibnamefont {Yin}}} (\bibinfo {year} {2017}),\
  \href@noop {} {\bibfield  {journal} {\bibinfo  {journal} {Nature}\ }\textbf
  {\bibinfo {volume} {552}}}\BibitemShut {NoStop}%
\bibitem [{\citenamefont {Ort{\'\i}n}(1991)}]{Ortin91}%
  \BibitemOpen
  \bibfield  {author} {\bibinfo {author} {\bibnamefont {Ort{\'\i}n},
  \bibfnamefont {J.}}} (\bibinfo {year} {1991}),\ \href@noop {} {\bibfield
  {journal} {\bibinfo  {journal} {J. Appl. Phys}\ }\textbf {\bibinfo {volume}
  {71}}~(\bibinfo {number} {3}),\ \bibinfo {pages} {1454}}\BibitemShut
  {NoStop}%
\bibitem [{\citenamefont {Osterholm}\ \emph {et~al.}(2012)\citenamefont
  {Osterholm}, \citenamefont {Kelley}, \citenamefont {Sommer},\ and\
  \citenamefont {Belongia}}]{Osterholm12}%
  \BibitemOpen
  \bibfield  {author} {\bibinfo {author} {\bibnamefont {Osterholm},
  \bibfnamefont {M.~T.}}, \bibinfo {author} {\bibfnamefont {N.~S.}\
  \bibnamefont {Kelley}}, \bibinfo {author} {\bibfnamefont {A.}~\bibnamefont
  {Sommer}}, \ and\ \bibinfo {author} {\bibfnamefont {E.~A.}\ \bibnamefont
  {Belongia}}} (\bibinfo {year} {2012}),\ \href@noop {} {\bibfield  {journal}
  {\bibinfo  {journal} {The Lancet infectious diseases}\ }\textbf {\bibinfo
  {volume} {12}}~(\bibinfo {number} {1}),\ \bibinfo {pages} {36}}\BibitemShut
  {NoStop}%
\bibitem [{\citenamefont {Packard}\ \emph {et~al.}(2010)\citenamefont
  {Packard}, \citenamefont {Homer}, \citenamefont {Al-Aqeeli},\ and\
  \citenamefont {Schuh}}]{Packard10}%
  \BibitemOpen
  \bibfield  {author} {\bibinfo {author} {\bibnamefont {Packard}, \bibfnamefont
  {C.~E.}}, \bibinfo {author} {\bibfnamefont {E.~R.}\ \bibnamefont {Homer}},
  \bibinfo {author} {\bibfnamefont {N.}~\bibnamefont {Al-Aqeeli}}, \ and\
  \bibinfo {author} {\bibfnamefont {C.~A.}\ \bibnamefont {Schuh}}} (\bibinfo
  {year} {2010}),\ \href@noop {} {\bibfield  {journal} {\bibinfo  {journal}
  {Philos. Mag.}\ }\textbf {\bibinfo {volume} {90}}~(\bibinfo {number} {10}),\
  \bibinfo {pages} {1373}}\BibitemShut {NoStop}%
\bibitem [{\citenamefont {Panagopoulos}\ \emph {et~al.}(2006)\citenamefont
  {Panagopoulos}, \citenamefont {Majoros}, \citenamefont {Nishizaki},\ and\
  \citenamefont {Iwasaki}}]{Panagopoulos06}%
  \BibitemOpen
  \bibfield  {author} {\bibinfo {author} {\bibnamefont {Panagopoulos},
  \bibfnamefont {C.}}, \bibinfo {author} {\bibfnamefont {M.}~\bibnamefont
  {Majoros}}, \bibinfo {author} {\bibfnamefont {T.}~\bibnamefont {Nishizaki}},
  \ and\ \bibinfo {author} {\bibfnamefont {H.}~\bibnamefont {Iwasaki}}}
  (\bibinfo {year} {2006}),\ \href@noop {} {\bibfield  {journal} {\bibinfo
  {journal} {Phys. Rev. Lett.}\ }\textbf {\bibinfo {volume} {96}}~(\bibinfo
  {number} {4}),\ \bibinfo {pages} {047002}}\BibitemShut {NoStop}%
\bibitem [{\citenamefont {Parisi}(1986)}]{Parisi86}%
  \BibitemOpen
  \bibfield  {author} {\bibinfo {author} {\bibnamefont {Parisi}, \bibfnamefont
  {G.}}} (\bibinfo {year} {1986}),\ \href
  {http://stacks.iop.org/0305-4470/19/i=10/a=011} {\bibfield  {journal}
  {\bibinfo  {journal} {Journal of Physics A: Mathematical and General}\
  }\textbf {\bibinfo {volume} {19}}~(\bibinfo {number} {10}),\ \bibinfo {pages}
  {L617}}\BibitemShut {NoStop}%
\bibitem [{\citenamefont {Pashine}\ \emph {et~al.}(2019)\citenamefont
  {Pashine}, \citenamefont {Hexner}, \citenamefont {Liu},\ and\ \citenamefont
  {Nagel}}]{Pashine2019}%
  \BibitemOpen
  \bibfield  {author} {\bibinfo {author} {\bibnamefont {Pashine}, \bibfnamefont
  {N.}}, \bibinfo {author} {\bibfnamefont {D.}~\bibnamefont {Hexner}}, \bibinfo
  {author} {\bibfnamefont {A.~J.}\ \bibnamefont {Liu}}, \ and\ \bibinfo
  {author} {\bibfnamefont {S.~R.}\ \bibnamefont {Nagel}}} (\bibinfo {year}
  {2019}),\ \href@noop {} {\bibinfo  {journal} {arXiv preprint
  arXiv:1903.05776}\ }\BibitemShut {NoStop}%
\bibitem [{\citenamefont {Paulsen}\ and\ \citenamefont
  {Keim}(2018)}]{Paulsen18}%
  \BibitemOpen
\bibfield  {journal} {  }\bibfield  {author} {\bibinfo {author} {\bibnamefont
  {Paulsen}, \bibfnamefont {J.~D.}}, \ and\ \bibinfo {author} {\bibfnamefont
  {N.~C.}\ \bibnamefont {Keim}}} (\bibinfo {year} {2018}),\ \href@noop {}
  {\bibinfo  {journal} {arXiv:1809.09715}\ }\BibitemShut {NoStop}%
\bibitem [{\citenamefont {Paulsen}\ \emph {et~al.}(2014)\citenamefont
  {Paulsen}, \citenamefont {Keim},\ and\ \citenamefont {Nagel}}]{Paulsen14}%
  \BibitemOpen
\bibfield  {journal} {  }\bibfield  {author} {\bibinfo {author} {\bibnamefont
  {Paulsen}, \bibfnamefont {J.~D.}}, \bibinfo {author} {\bibfnamefont {N.~C.}\
  \bibnamefont {Keim}}, \ and\ \bibinfo {author} {\bibfnamefont {S.~R.}\
  \bibnamefont {Nagel}}} (\bibinfo {year} {2014}),\ \href@noop {} {\bibfield
  {journal} {\bibinfo  {journal} {Phys. Rev. Lett.}\ }\textbf {\bibinfo
  {volume} {113}}~(\bibinfo {number} {6}),\ \bibinfo {pages}
  {068301}}\BibitemShut {NoStop}%
\bibitem [{\citenamefont {P{\'e}rez-Reche}\ \emph {et~al.}(2016)\citenamefont
  {P{\'e}rez-Reche}, \citenamefont {Triguero}, \citenamefont {Zanzotto},\ and\
  \citenamefont {Truskinovsky}}]{Perez-Reche16}%
  \BibitemOpen
  \bibfield  {author} {\bibinfo {author} {\bibnamefont {P{\'e}rez-Reche},
  \bibfnamefont {F.-J.}}, \bibinfo {author} {\bibfnamefont {C.}~\bibnamefont
  {Triguero}}, \bibinfo {author} {\bibfnamefont {G.}~\bibnamefont {Zanzotto}},
  \ and\ \bibinfo {author} {\bibfnamefont {L.}~\bibnamefont {Truskinovsky}}}
  (\bibinfo {year} {2016}),\ \href@noop {} {\bibfield  {journal} {\bibinfo
  {journal} {Physical Review B}\ }\textbf {\bibinfo {volume} {94}}~(\bibinfo
  {number} {14}),\ \bibinfo {pages} {144102}}\BibitemShut {NoStop}%
\bibitem [{\citenamefont {P{\'e}rez-Reche}\ \emph {et~al.}(2007)\citenamefont
  {P{\'e}rez-Reche}, \citenamefont {Truskinovsky},\ and\ \citenamefont
  {Zanzotto}}]{Perez-Reche07}%
  \BibitemOpen
  \bibfield  {author} {\bibinfo {author} {\bibnamefont {P{\'e}rez-Reche},
  \bibfnamefont {F.-J.}}, \bibinfo {author} {\bibfnamefont {L.}~\bibnamefont
  {Truskinovsky}}, \ and\ \bibinfo {author} {\bibfnamefont {G.}~\bibnamefont
  {Zanzotto}}} (\bibinfo {year} {2007}),\ \href@noop {} {\bibfield  {journal}
  {\bibinfo  {journal} {Phys. Rev. Lett.}\ }\textbf {\bibinfo {volume}
  {99}}~(\bibinfo {number} {7}),\ \bibinfo {pages} {075501}}\BibitemShut
  {NoStop}%
\bibitem [{\citenamefont {Perkovi{\'c}}\ and\ \citenamefont
  {Sethna}(1997)}]{Perkovic97}%
  \BibitemOpen
  \bibfield  {author} {\bibinfo {author} {\bibnamefont {Perkovi{\'c}},
  \bibfnamefont {O.}}, \ and\ \bibinfo {author} {\bibfnamefont {J.~P.}\
  \bibnamefont {Sethna}}} (\bibinfo {year} {1997}),\ \href@noop {} {\bibfield
  {journal} {\bibinfo  {journal} {J. Appl. Phys}\ }\textbf {\bibinfo {volume}
  {81}},\ \bibinfo {pages} {1590}}\BibitemShut {NoStop}%
\bibitem [{\citenamefont {Petekidis}\ \emph {et~al.}(2002)\citenamefont
  {Petekidis}, \citenamefont {Moussa{\"\i}d},\ and\ \citenamefont
  {Pusey}}]{Petekidis02}%
  \BibitemOpen
  \bibfield  {author} {\bibinfo {author} {\bibnamefont {Petekidis},
  \bibfnamefont {G.}}, \bibinfo {author} {\bibfnamefont {A.}~\bibnamefont
  {Moussa{\"\i}d}}, \ and\ \bibinfo {author} {\bibfnamefont {P.~N.}\
  \bibnamefont {Pusey}}} (\bibinfo {year} {2002}),\ \href@noop {} {\bibfield
  {journal} {\bibinfo  {journal} {Phys. Rev. E}\ }\textbf {\bibinfo {volume}
  {66}}~(\bibinfo {number} {5}),\ \bibinfo {pages} {051402}}\BibitemShut
  {NoStop}%
\bibitem [{\citenamefont {Pham}\ \emph {et~al.}(2015)\citenamefont {Pham},
  \citenamefont {Metzger},\ and\ \citenamefont {Butler}}]{Pham15}%
  \BibitemOpen
  \bibfield  {author} {\bibinfo {author} {\bibnamefont {Pham}, \bibfnamefont
  {P.}}, \bibinfo {author} {\bibfnamefont {B.}~\bibnamefont {Metzger}}, \ and\
  \bibinfo {author} {\bibfnamefont {J.~E.}\ \bibnamefont {Butler}}} (\bibinfo
  {year} {2015}),\ \href@noop {} {\bibfield  {journal} {\bibinfo  {journal}
  {Phys. Fluids}\ }\textbf {\bibinfo {volume} {27}}~(\bibinfo {number} {5}),\
  \bibinfo {pages} {051701}}\BibitemShut {NoStop}%
\bibitem [{\citenamefont {Picco}\ \emph {et~al.}(2001)\citenamefont {Picco},
  \citenamefont {Ricci-Tersenghi},\ and\ \citenamefont {Ritort}}]{Picco01}%
  \BibitemOpen
  \bibfield  {author} {\bibinfo {author} {\bibnamefont {Picco}, \bibfnamefont
  {M.}}, \bibinfo {author} {\bibfnamefont {F.}~\bibnamefont {Ricci-Tersenghi}},
  \ and\ \bibinfo {author} {\bibfnamefont {F.}~\bibnamefont {Ritort}}}
  (\bibinfo {year} {2001}),\ \href@noop {} {\bibfield  {journal} {\bibinfo
  {journal} {Physical Review B}\ }\textbf {\bibinfo {volume} {63}}~(\bibinfo
  {number} {17}),\ \bibinfo {pages} {174412}}\BibitemShut {NoStop}%
\bibitem [{\citenamefont {Pine}\ \emph {et~al.}(2005)\citenamefont {Pine},
  \citenamefont {Gollub}, \citenamefont {Brady},\ and\ \citenamefont
  {Leshansky}}]{Pine05}%
  \BibitemOpen
  \bibfield  {author} {\bibinfo {author} {\bibnamefont {Pine}, \bibfnamefont
  {D.~J.}}, \bibinfo {author} {\bibfnamefont {J.~P.}\ \bibnamefont {Gollub}},
  \bibinfo {author} {\bibfnamefont {J.~F.}\ \bibnamefont {Brady}}, \ and\
  \bibinfo {author} {\bibfnamefont {A.~M.}\ \bibnamefont {Leshansky}}}
  (\bibinfo {year} {2005}),\ \href@noop {} {\bibfield  {journal} {\bibinfo
  {journal} {Nature}\ }\textbf {\bibinfo {volume} {438}}~(\bibinfo {number}
  {7070}),\ \bibinfo {pages} {997}}\BibitemShut {NoStop}%
\bibitem [{\citenamefont {Popova}\ \emph {et~al.}(2007)\citenamefont {Popova},
  \citenamefont {Vorobieff}, \citenamefont {Ingber},\ and\ \citenamefont
  {Graham}}]{Popova07}%
  \BibitemOpen
  \bibfield  {author} {\bibinfo {author} {\bibnamefont {Popova}, \bibfnamefont
  {M.}}, \bibinfo {author} {\bibfnamefont {P.}~\bibnamefont {Vorobieff}},
  \bibinfo {author} {\bibfnamefont {M.~S.}\ \bibnamefont {Ingber}}, \ and\
  \bibinfo {author} {\bibfnamefont {A.~L.}\ \bibnamefont {Graham}}} (\bibinfo
  {year} {2007}),\ \href@noop {} {\bibfield  {journal} {\bibinfo  {journal}
  {Physical Review E}\ }\textbf {\bibinfo {volume} {75}}~(\bibinfo {number}
  {6}),\ \bibinfo {pages} {066309}}\BibitemShut {NoStop}%
\bibitem [{\citenamefont {Povinelli}\ \emph {et~al.}(1999)\citenamefont
  {Povinelli}, \citenamefont {Coppersmith}, \citenamefont {Kadanoff},
  \citenamefont {Nagel},\ and\ \citenamefont {Venkataramani}}]{Povinelli99}%
  \BibitemOpen
  \bibfield  {author} {\bibinfo {author} {\bibnamefont {Povinelli},
  \bibfnamefont {M.~L.}}, \bibinfo {author} {\bibfnamefont {S.~N.}\
  \bibnamefont {Coppersmith}}, \bibinfo {author} {\bibfnamefont {L.~P.}\
  \bibnamefont {Kadanoff}}, \bibinfo {author} {\bibfnamefont {S.~R.}\
  \bibnamefont {Nagel}}, \ and\ \bibinfo {author} {\bibfnamefont {S.~C.}\
  \bibnamefont {Venkataramani}}} (\bibinfo {year} {1999}),\ \href@noop {}
  {\bibfield  {journal} {\bibinfo  {journal} {Phys. Rev. E}\ }\textbf {\bibinfo
  {volume} {59}}~(\bibinfo {number} {5}),\ \bibinfo {pages} {4970}}\BibitemShut
  {NoStop}%
\bibitem [{\citenamefont {Preisach}(1935)}]{Preisach35}%
  \BibitemOpen
  \bibfield  {author} {\bibinfo {author} {\bibnamefont {Preisach},
  \bibfnamefont {F.}}} (\bibinfo {year} {1935}),\ \href@noop {} {\bibfield
  {journal} {\bibinfo  {journal} {Z. Physik}\ }\textbf {\bibinfo {volume}
  {94}}~(\bibinfo {number} {5-6}),\ \bibinfo {pages} {277}}\BibitemShut
  {NoStop}%
\bibitem [{\citenamefont {Regev}\ \emph {et~al.}(2013)\citenamefont {Regev},
  \citenamefont {Lookman},\ and\ \citenamefont {Reichhardt}}]{Regev13}%
  \BibitemOpen
  \bibfield  {author} {\bibinfo {author} {\bibnamefont {Regev}, \bibfnamefont
  {I.}}, \bibinfo {author} {\bibfnamefont {T.}~\bibnamefont {Lookman}}, \ and\
  \bibinfo {author} {\bibfnamefont {C.}~\bibnamefont {Reichhardt}}} (\bibinfo
  {year} {2013}),\ \href@noop {} {\bibfield  {journal} {\bibinfo  {journal}
  {Phys. Rev. E}\ }\textbf {\bibinfo {volume} {88}}~(\bibinfo {number} {6}),\
  \bibinfo {pages} {062401}}\BibitemShut {NoStop}%
\bibitem [{\citenamefont {Ren}\ \emph {et~al.}(2013)\citenamefont {Ren},
  \citenamefont {Dijksman},\ and\ \citenamefont {Behringer}}]{Ren13}%
  \BibitemOpen
  \bibfield  {author} {\bibinfo {author} {\bibnamefont {Ren}, \bibfnamefont
  {J.}}, \bibinfo {author} {\bibfnamefont {J.~A.}\ \bibnamefont {Dijksman}}, \
  and\ \bibinfo {author} {\bibfnamefont {R.~P.}\ \bibnamefont {Behringer}}}
  (\bibinfo {year} {2013}),\ \href@noop {} {\bibfield  {journal} {\bibinfo
  {journal} {Phys. Rev. Lett.}\ }\textbf {\bibinfo {volume} {110}}~(\bibinfo
  {number} {1}),\ \bibinfo {pages} {018302}}\BibitemShut {NoStop}%
\bibitem [{\citenamefont {Rocks}\ \emph {et~al.}(2017)\citenamefont {Rocks},
  \citenamefont {Pashine}, \citenamefont {Bischofberger}, \citenamefont
  {Goodrich}, \citenamefont {Liu},\ and\ \citenamefont {Nagel}}]{Rocks17}%
  \BibitemOpen
  \bibfield  {author} {\bibinfo {author} {\bibnamefont {Rocks}, \bibfnamefont
  {J.~W.}}, \bibinfo {author} {\bibfnamefont {N.}~\bibnamefont {Pashine}},
  \bibinfo {author} {\bibfnamefont {I.}~\bibnamefont {Bischofberger}}, \bibinfo
  {author} {\bibfnamefont {C.~P.}\ \bibnamefont {Goodrich}}, \bibinfo {author}
  {\bibfnamefont {A.~J.}\ \bibnamefont {Liu}}, \ and\ \bibinfo {author}
  {\bibfnamefont {S.~R.}\ \bibnamefont {Nagel}}} (\bibinfo {year} {2017}),\
  \href@noop {} {\bibfield  {journal} {\bibinfo  {journal} {Proceedings of the
  National Academy of Sciences}\ }\textbf {\bibinfo {volume} {114}}~(\bibinfo
  {number} {10}),\ \bibinfo {pages} {2520}}\BibitemShut {NoStop}%
\bibitem [{\citenamefont {Rocks}\ \emph {et~al.}(2019)\citenamefont {Rocks},
  \citenamefont {Ronellenfitsch}, \citenamefont {Liu}, \citenamefont {Nagel},\
  and\ \citenamefont {Katifori}}]{Rocks19}%
  \BibitemOpen
  \bibfield  {author} {\bibinfo {author} {\bibnamefont {Rocks}, \bibfnamefont
  {J.~W.}}, \bibinfo {author} {\bibfnamefont {H.}~\bibnamefont
  {Ronellenfitsch}}, \bibinfo {author} {\bibfnamefont {A.~J.}\ \bibnamefont
  {Liu}}, \bibinfo {author} {\bibfnamefont {S.~R.}\ \bibnamefont {Nagel}}, \
  and\ \bibinfo {author} {\bibfnamefont {E.}~\bibnamefont {Katifori}}}
  (\bibinfo {year} {2019}),\ \href@noop {} {\bibinfo  {journal} {Proceedings of
  the National Academy of Sciences}\ ,\ \bibinfo {pages}
  {201806790}}\BibitemShut {NoStop}%
\bibitem [{\citenamefont {Rogers}\ \emph {et~al.}(2016)\citenamefont {Rogers},
  \citenamefont {Shih},\ and\ \citenamefont {Manoharan}}]{Rogers16}%
  \BibitemOpen
\bibfield  {journal} {  }\bibfield  {author} {\bibinfo {author} {\bibnamefont
  {Rogers}, \bibfnamefont {W.~B.}}, \bibinfo {author} {\bibfnamefont {W.~M.}\
  \bibnamefont {Shih}}, \ and\ \bibinfo {author} {\bibfnamefont {V.~N.}\
  \bibnamefont {Manoharan}}} (\bibinfo {year} {2016}),\ \href
  {http://dx.doi.org/10.1038/natrevmats.2016.8} {\bibfield  {journal} {\bibinfo
   {journal} {Nature Reviews Materials}\ }\textbf {\bibinfo {volume} {1}},\
  \bibinfo {pages} {16008}}\BibitemShut {NoStop}%
\bibitem [{\citenamefont {Royer}\ and\ \citenamefont
  {Chaikin}(2015)}]{Royer15}%
  \BibitemOpen
  \bibfield  {author} {\bibinfo {author} {\bibnamefont {Royer}, \bibfnamefont
  {J.~R.}}, \ and\ \bibinfo {author} {\bibfnamefont {P.~M.}\ \bibnamefont
  {Chaikin}}} (\bibinfo {year} {2015}),\ \href@noop {} {\bibfield  {journal}
  {\bibinfo  {journal} {Proc. Natl. Acad. Sci.}\ }\textbf {\bibinfo {volume}
  {112}}~(\bibinfo {number} {1}),\ \bibinfo {pages} {49}}\BibitemShut {NoStop}%
\bibitem [{\citenamefont {Schmidt}\ \emph {et~al.}(2009)\citenamefont
  {Schmidt}, \citenamefont {Keim}, \citenamefont {Zhang},\ and\ \citenamefont
  {Nagel}}]{Schmidt09}%
  \BibitemOpen
  \bibfield  {author} {\bibinfo {author} {\bibnamefont {Schmidt}, \bibfnamefont
  {L.~E.}}, \bibinfo {author} {\bibfnamefont {N.~C.}\ \bibnamefont {Keim}},
  \bibinfo {author} {\bibfnamefont {W.~W.}\ \bibnamefont {Zhang}}, \ and\
  \bibinfo {author} {\bibfnamefont {S.~R.}\ \bibnamefont {Nagel}}} (\bibinfo
  {year} {2009}),\ \href@noop {} {\bibfield  {journal} {\bibinfo  {journal}
  {Nat. Phys.}\ }\textbf {\bibinfo {volume} {5}}~(\bibinfo {number} {5}),\
  \bibinfo {pages} {343}}\BibitemShut {NoStop}%
\bibitem [{\citenamefont {Schreck}\ \emph {et~al.}(2013)\citenamefont
  {Schreck}, \citenamefont {Hoy}, \citenamefont {Shattuck},\ and\ \citenamefont
  {O'Hern}}]{Schreck13}%
  \BibitemOpen
  \bibfield  {author} {\bibinfo {author} {\bibnamefont {Schreck}, \bibfnamefont
  {C.~F.}}, \bibinfo {author} {\bibfnamefont {R.~S.}\ \bibnamefont {Hoy}},
  \bibinfo {author} {\bibfnamefont {M.~D.}\ \bibnamefont {Shattuck}}, \ and\
  \bibinfo {author} {\bibfnamefont {C.~S.}\ \bibnamefont {O'Hern}}} (\bibinfo
  {year} {2013}),\ \href@noop {} {\bibfield  {journal} {\bibinfo  {journal}
  {Phys. Rev. E}\ }\textbf {\bibinfo {volume} {88}},\ \bibinfo {pages}
  {052205}}\BibitemShut {NoStop}%
\bibitem [{\citenamefont {Sethna}\ \emph {et~al.}(2017)\citenamefont {Sethna},
  \citenamefont {Bierbaum}, \citenamefont {Dahmen}, \citenamefont {Goodrich},
  \citenamefont {Greer}, \citenamefont {Hayden}, \citenamefont {Kent-Dobias},
  \citenamefont {Lee}, \citenamefont {Liarte}, \citenamefont {Ni},
  \citenamefont {Quinn}, \citenamefont {Raju}, \citenamefont {Rocklin},
  \citenamefont {Shekhawat},\ and\ \citenamefont {Zapperi}}]{Sethna17}%
  \BibitemOpen
  \bibfield  {author} {\bibinfo {author} {\bibnamefont {Sethna}, \bibfnamefont
  {J.~P.}}, \bibinfo {author} {\bibfnamefont {M.~K.}\ \bibnamefont {Bierbaum}},
  \bibinfo {author} {\bibfnamefont {K.~A.}\ \bibnamefont {Dahmen}}, \bibinfo
  {author} {\bibfnamefont {C.~P.}\ \bibnamefont {Goodrich}}, \bibinfo {author}
  {\bibfnamefont {J.~R.}\ \bibnamefont {Greer}}, \bibinfo {author}
  {\bibfnamefont {L.~X.}\ \bibnamefont {Hayden}}, \bibinfo {author}
  {\bibfnamefont {J.~P.}\ \bibnamefont {Kent-Dobias}}, \bibinfo {author}
  {\bibfnamefont {E.~D.}\ \bibnamefont {Lee}}, \bibinfo {author} {\bibfnamefont
  {D.~B.}\ \bibnamefont {Liarte}}, \bibinfo {author} {\bibfnamefont
  {X.}~\bibnamefont {Ni}}, \bibinfo {author} {\bibfnamefont {K.~N.}\
  \bibnamefont {Quinn}}, \bibinfo {author} {\bibfnamefont {A.}~\bibnamefont
  {Raju}}, \bibinfo {author} {\bibfnamefont {D.~Z.}\ \bibnamefont {Rocklin}},
  \bibinfo {author} {\bibfnamefont {A.}~\bibnamefont {Shekhawat}}, \ and\
  \bibinfo {author} {\bibfnamefont {S.}~\bibnamefont {Zapperi}}} (\bibinfo
  {year} {2017}),\ \href@noop {} {\bibfield  {journal} {\bibinfo  {journal}
  {Annu. Rev. Mater. Res.}\ }\textbf {\bibinfo {volume} {47}},\ \bibinfo
  {pages} {217}}\BibitemShut {NoStop}%
\bibitem [{\citenamefont {Sethna}\ \emph {et~al.}(1993)\citenamefont {Sethna},
  \citenamefont {Dahmen}, \citenamefont {Kartha}, \citenamefont {Krumhansl},
  \citenamefont {Roberts},\ and\ \citenamefont {Shore}}]{Sethna93}%
  \BibitemOpen
  \bibfield  {author} {\bibinfo {author} {\bibnamefont {Sethna}, \bibfnamefont
  {J.~P.}}, \bibinfo {author} {\bibfnamefont {K.}~\bibnamefont {Dahmen}},
  \bibinfo {author} {\bibfnamefont {S.}~\bibnamefont {Kartha}}, \bibinfo
  {author} {\bibfnamefont {J.~A.}\ \bibnamefont {Krumhansl}}, \bibinfo {author}
  {\bibfnamefont {B.~W.}\ \bibnamefont {Roberts}}, \ and\ \bibinfo {author}
  {\bibfnamefont {J.~D.}\ \bibnamefont {Shore}}} (\bibinfo {year} {1993}),\
  \href@noop {} {\bibfield  {journal} {\bibinfo  {journal} {Phys. Rev. Lett.}\
  }\textbf {\bibinfo {volume} {70}},\ \bibinfo {pages} {3347}}\BibitemShut
  {NoStop}%
\bibitem [{\citenamefont {Sethna}\ \emph {et~al.}(2001)\citenamefont {Sethna},
  \citenamefont {Dahmen},\ and\ \citenamefont {Myers}}]{Sethna01}%
  \BibitemOpen
  \bibfield  {author} {\bibinfo {author} {\bibnamefont {Sethna}, \bibfnamefont
  {J.~P.}}, \bibinfo {author} {\bibfnamefont {K.~A.}\ \bibnamefont {Dahmen}}, \
  and\ \bibinfo {author} {\bibfnamefont {C.~R.}\ \bibnamefont {Myers}}}
  (\bibinfo {year} {2001}),\ \href@noop {} {\bibfield  {journal} {\bibinfo
  {journal} {Nature}\ }\textbf {\bibinfo {volume} {410}}~(\bibinfo {number}
  {6825}),\ \bibinfo {pages} {242}}\BibitemShut {NoStop}%
\bibitem [{\citenamefont {Shi}\ \emph {et~al.}(1994)\citenamefont {Shi},
  \citenamefont {Brenner},\ and\ \citenamefont {Nagel}}]{Shi94}%
  \BibitemOpen
  \bibfield  {author} {\bibinfo {author} {\bibnamefont {Shi}, \bibfnamefont
  {X.}}, \bibinfo {author} {\bibfnamefont {M.~P.}\ \bibnamefont {Brenner}}, \
  and\ \bibinfo {author} {\bibfnamefont {S.~R.}\ \bibnamefont {Nagel}}}
  (\bibinfo {year} {1994}),\ \href@noop {} {\bibfield  {journal} {\bibinfo
  {journal} {Science}\ }\textbf {\bibinfo {volume} {265}}~(\bibinfo {number}
  {5169}),\ \bibinfo {pages} {219}}\BibitemShut {NoStop}%
\bibitem [{\citenamefont {Sircar}\ and\ \citenamefont {Wang}(2010)}]{Sircar10}%
  \BibitemOpen
  \bibfield  {author} {\bibinfo {author} {\bibnamefont {Sircar}, \bibfnamefont
  {S.}}, \ and\ \bibinfo {author} {\bibfnamefont {Q.}~\bibnamefont {Wang}}}
  (\bibinfo {year} {2010}),\ \href@noop {} {\bibfield  {journal} {\bibinfo
  {journal} {Rheol Acta}\ }\textbf {\bibinfo {volume} {49}}~(\bibinfo {number}
  {7}),\ \bibinfo {pages} {699}}\BibitemShut {NoStop}%
\bibitem [{\citenamefont {Slotterback}\ \emph {et~al.}(2012)\citenamefont
  {Slotterback}, \citenamefont {Mailman}, \citenamefont {Ronaszegi},
  \citenamefont {van Hecke}, \citenamefont {Girvan},\ and\ \citenamefont
  {Losert}}]{Slotterback12}%
  \BibitemOpen
  \bibfield  {author} {\bibinfo {author} {\bibnamefont {Slotterback},
  \bibfnamefont {S.}}, \bibinfo {author} {\bibfnamefont {M.}~\bibnamefont
  {Mailman}}, \bibinfo {author} {\bibfnamefont {K.}~\bibnamefont {Ronaszegi}},
  \bibinfo {author} {\bibfnamefont {M.}~\bibnamefont {van Hecke}}, \bibinfo
  {author} {\bibfnamefont {M.}~\bibnamefont {Girvan}}, \ and\ \bibinfo {author}
  {\bibfnamefont {W.}~\bibnamefont {Losert}}} (\bibinfo {year} {2012}),\
  \href@noop {} {\bibfield  {journal} {\bibinfo  {journal} {Phys. Rev. E}\
  }\textbf {\bibinfo {volume} {85}}~(\bibinfo {number} {2}),\ \bibinfo {pages}
  {021309}}\BibitemShut {NoStop}%
\bibitem [{\citenamefont {Sokolowski}\ and\ \citenamefont
  {Tan}(2007)}]{Sokolowski07}%
  \BibitemOpen
  \bibfield  {author} {\bibinfo {author} {\bibnamefont {Sokolowski},
  \bibfnamefont {W.~M.}}, \ and\ \bibinfo {author} {\bibfnamefont {S.~C.}\
  \bibnamefont {Tan}}} (\bibinfo {year} {2007}),\ \href@noop {} {\bibfield
  {journal} {\bibinfo  {journal} {Journal of spacecraft and rockets}\ }\textbf
  {\bibinfo {volume} {44}}~(\bibinfo {number} {4}),\ \bibinfo {pages}
  {750}}\BibitemShut {NoStop}%
\bibitem [{\citenamefont {Song}\ \emph {et~al.}(2013)\citenamefont {Song},
  \citenamefont {Chen}, \citenamefont {Dabade}, \citenamefont {Shield},\ and\
  \citenamefont {James}}]{Song13}%
  \BibitemOpen
  \bibfield  {author} {\bibinfo {author} {\bibnamefont {Song}, \bibfnamefont
  {Y.}}, \bibinfo {author} {\bibfnamefont {X.}~\bibnamefont {Chen}}, \bibinfo
  {author} {\bibfnamefont {V.}~\bibnamefont {Dabade}}, \bibinfo {author}
  {\bibfnamefont {T.~W.}\ \bibnamefont {Shield}}, \ and\ \bibinfo {author}
  {\bibfnamefont {R.~D.}\ \bibnamefont {James}}} (\bibinfo {year} {2013}),\
  \href@noop {} {\bibfield  {journal} {\bibinfo  {journal} {Nature}\ }\textbf
  {\bibinfo {volume} {502}}~(\bibinfo {number} {7469}),\ \bibinfo {pages}
  {85}}\BibitemShut {NoStop}%
\bibitem [{\citenamefont {Takayama}\ and\ \citenamefont
  {Hukushima}(2002)}]{Takayama02}%
  \BibitemOpen
  \bibfield  {author} {\bibinfo {author} {\bibnamefont {Takayama},
  \bibfnamefont {H.}}, \ and\ \bibinfo {author} {\bibfnamefont
  {K.}~\bibnamefont {Hukushima}}} (\bibinfo {year} {2002}),\ \href@noop {}
  {\bibfield  {journal} {\bibinfo  {journal} {Journal of the Physical Society
  of Japan}\ }\textbf {\bibinfo {volume} {71}}~(\bibinfo {number} {12}),\
  \bibinfo {pages} {3003}}\BibitemShut {NoStop}%
\bibitem [{\citenamefont {Tang}\ \emph {et~al.}(1987)\citenamefont {Tang},
  \citenamefont {Wiesenfeld}, \citenamefont {Bak}, \citenamefont
  {Coppersmith},\ and\ \citenamefont {Littlewood}}]{Tang87}%
  \BibitemOpen
  \bibfield  {author} {\bibinfo {author} {\bibnamefont {Tang}, \bibfnamefont
  {C.}}, \bibinfo {author} {\bibfnamefont {K.}~\bibnamefont {Wiesenfeld}},
  \bibinfo {author} {\bibfnamefont {P.}~\bibnamefont {Bak}}, \bibinfo {author}
  {\bibfnamefont {S.}~\bibnamefont {Coppersmith}}, \ and\ \bibinfo {author}
  {\bibfnamefont {P.}~\bibnamefont {Littlewood}}} (\bibinfo {year} {1987}),\
  \href@noop {} {\bibfield  {journal} {\bibinfo  {journal} {Phys. Rev. Lett.}\
  }\textbf {\bibinfo {volume} {58}}~(\bibinfo {number} {12}),\ \bibinfo {pages}
  {1161}}\BibitemShut {NoStop}%
\bibitem [{\citenamefont {Taylor}(1985)}]{Taylor85}%
  \BibitemOpen
  \bibfield  {author} {\bibinfo {author} {\bibnamefont {Taylor}, \bibfnamefont
  {G.~I.}}} (\bibinfo {year} {1985}),\ \href@noop {} {\emph {\bibinfo {title}
  {Low Reynolds number flows}}},\ National Committee for Fluid Mechanics Films\
  (\bibinfo  {publisher} {Released by Encyclopaedia Britannica Educational
  Corporation})\BibitemShut {NoStop}%
\bibitem [{\citenamefont {Thomas}\ \emph {et~al.}(2008)\citenamefont {Thomas},
  \citenamefont {White},\ and\ \citenamefont {Middleton}}]{Thomas08}%
  \BibitemOpen
  \bibfield  {author} {\bibinfo {author} {\bibnamefont {Thomas}, \bibfnamefont
  {C.~K.}}, \bibinfo {author} {\bibfnamefont {O.~L.}\ \bibnamefont {White}}, \
  and\ \bibinfo {author} {\bibfnamefont {A.~A.}\ \bibnamefont {Middleton}}}
  (\bibinfo {year} {2008}),\ \href@noop {} {\bibfield  {journal} {\bibinfo
  {journal} {Physical Review B}\ }\textbf {\bibinfo {volume} {77}}~(\bibinfo
  {number} {9}),\ \bibinfo {pages} {092415}}\BibitemShut {NoStop}%
\bibitem [{\citenamefont {Thorne}(1996)}]{Thorne96}%
  \BibitemOpen
  \bibfield  {author} {\bibinfo {author} {\bibnamefont {Thorne}, \bibfnamefont
  {R.~E.}}} (\bibinfo {year} {1996}),\ \href@noop {} {\bibfield  {journal}
  {\bibinfo  {journal} {Phys. Today}\ }\textbf {\bibinfo {volume}
  {59}}~(\bibinfo {number} {5}),\ \bibinfo {pages} {42}}\BibitemShut {NoStop}%
\bibitem [{\citenamefont {Toiya}\ \emph {et~al.}(2004)\citenamefont {Toiya},
  \citenamefont {Stambaugh},\ and\ \citenamefont {Losert}}]{Toiya04}%
  \BibitemOpen
  \bibfield  {author} {\bibinfo {author} {\bibnamefont {Toiya}, \bibfnamefont
  {M.}}, \bibinfo {author} {\bibfnamefont {J.}~\bibnamefont {Stambaugh}}, \
  and\ \bibinfo {author} {\bibfnamefont {W.}~\bibnamefont {Losert}}} (\bibinfo
  {year} {2004}),\ \href@noop {} {\bibfield  {journal} {\bibinfo  {journal}
  {Phys. Rev. Lett.}\ }\textbf {\bibinfo {volume} {93}}~(\bibinfo {number}
  {8}),\ \bibinfo {pages} {088001}}\BibitemShut {NoStop}%
\bibitem [{\citenamefont {Vincent}(2007)}]{Vincent2006b}%
  \BibitemOpen
  \bibfield  {author} {\bibinfo {author} {\bibnamefont {Vincent}, \bibfnamefont
  {E.}}} (\bibinfo {year} {2007}),\ in\ \href@noop {} {\emph {\bibinfo
  {booktitle} {Ageing and the glass transition}}}\ (\bibinfo  {publisher}
  {Springer})\ pp.\ \bibinfo {pages} {7--60}\BibitemShut {NoStop}%
\bibitem [{\citenamefont {Volkert}\ and\ \citenamefont
  {Spaepen}(1989)}]{Volkert89}%
  \BibitemOpen
  \bibfield  {author} {\bibinfo {author} {\bibnamefont {Volkert}, \bibfnamefont
  {C.}}, \ and\ \bibinfo {author} {\bibfnamefont {F.}~\bibnamefont {Spaepen}}}
  (\bibinfo {year} {1989}),\ \href@noop {} {\bibfield  {journal} {\bibinfo
  {journal} {Acta Metallurgica}\ }\textbf {\bibinfo {volume} {37}}~(\bibinfo
  {number} {5}),\ \bibinfo {pages} {1355}}\BibitemShut {NoStop}%
\bibitem [{\citenamefont {Xie}\ and\ \citenamefont {Xiao}(2008)}]{Xie08}%
  \BibitemOpen
  \bibfield  {author} {\bibinfo {author} {\bibnamefont {Xie}, \bibfnamefont
  {T.}}, \ and\ \bibinfo {author} {\bibfnamefont {X.}~\bibnamefont {Xiao}}}
  (\bibinfo {year} {2008}),\ \href@noop {} {\bibfield  {journal} {\bibinfo
  {journal} {Chemistry of Materials}\ }\textbf {\bibinfo {volume}
  {20}}~(\bibinfo {number} {9}),\ \bibinfo {pages} {2866}}\BibitemShut
  {NoStop}%
\bibitem [{\citenamefont {Yakacki}\ \emph {et~al.}(2007)\citenamefont
  {Yakacki}, \citenamefont {Shandas}, \citenamefont {Lanning}, \citenamefont
  {Rech}, \citenamefont {Eckstein},\ and\ \citenamefont {Gall}}]{Yakacki07}%
  \BibitemOpen
  \bibfield  {author} {\bibinfo {author} {\bibnamefont {Yakacki}, \bibfnamefont
  {C.~M.}}, \bibinfo {author} {\bibfnamefont {R.}~\bibnamefont {Shandas}},
  \bibinfo {author} {\bibfnamefont {C.}~\bibnamefont {Lanning}}, \bibinfo
  {author} {\bibfnamefont {B.}~\bibnamefont {Rech}}, \bibinfo {author}
  {\bibfnamefont {A.}~\bibnamefont {Eckstein}}, \ and\ \bibinfo {author}
  {\bibfnamefont {K.}~\bibnamefont {Gall}}} (\bibinfo {year} {2007}),\
  \href@noop {} {\bibfield  {journal} {\bibinfo  {journal} {Biomaterials}\
  }\textbf {\bibinfo {volume} {28}}~(\bibinfo {number} {14}),\ \bibinfo {pages}
  {2255}}\BibitemShut {NoStop}%
\bibitem [{\citenamefont {Yan}\ \emph {et~al.}(2017)\citenamefont {Yan},
  \citenamefont {Ravasio}, \citenamefont {Brito},\ and\ \citenamefont
  {Wyart}}]{Yan17}%
  \BibitemOpen
  \bibfield  {author} {\bibinfo {author} {\bibnamefont {Yan}, \bibfnamefont
  {L.}}, \bibinfo {author} {\bibfnamefont {R.}~\bibnamefont {Ravasio}},
  \bibinfo {author} {\bibfnamefont {C.}~\bibnamefont {Brito}}, \ and\ \bibinfo
  {author} {\bibfnamefont {M.}~\bibnamefont {Wyart}}} (\bibinfo {year}
  {2017}),\ \href@noop {} {\bibfield  {journal} {\bibinfo  {journal}
  {Proceedings of the National Academy of Sciences}\ }\textbf {\bibinfo
  {volume} {114}}~(\bibinfo {number} {10}),\ \bibinfo {pages}
  {2526}}\BibitemShut {NoStop}%
\bibitem [{\citenamefont {Yang}\ and\ \citenamefont
  {Middleton}(2017)}]{Yang17}%
  \BibitemOpen
  \bibfield  {author} {\bibinfo {author} {\bibnamefont {Yang}, \bibfnamefont
  {J.}}, \ and\ \bibinfo {author} {\bibfnamefont {A.~A.}\ \bibnamefont
  {Middleton}}} (\bibinfo {year} {2017}),\ \href@noop {} {\bibfield  {journal}
  {\bibinfo  {journal} {Physical Review B}\ }\textbf {\bibinfo {volume}
  {96}}~(\bibinfo {number} {21}),\ \bibinfo {pages} {214208}}\BibitemShut
  {NoStop}%
\bibitem [{\citenamefont {Yardimci}\ and\ \citenamefont
  {Leheny}(2003)}]{Yardimci03}%
  \BibitemOpen
  \bibfield  {author} {\bibinfo {author} {\bibnamefont {Yardimci},
  \bibfnamefont {H.}}, \ and\ \bibinfo {author} {\bibfnamefont
  {R.}~\bibnamefont {Leheny}}} (\bibinfo {year} {2003}),\ \href@noop {}
  {\bibfield  {journal} {\bibinfo  {journal} {EPL (Europhysics Letters)}\
  }\textbf {\bibinfo {volume} {62}}~(\bibinfo {number} {2}),\ \bibinfo {pages}
  {203}}\BibitemShut {NoStop}%
\bibitem [{\citenamefont {Zeravcic}\ \emph {et~al.}(2017)\citenamefont
  {Zeravcic}, \citenamefont {Manoharan},\ and\ \citenamefont
  {Brenner}}]{Zeravcic17}%
  \BibitemOpen
  \bibfield  {author} {\bibinfo {author} {\bibnamefont {Zeravcic},
  \bibfnamefont {Z.}}, \bibinfo {author} {\bibfnamefont {V.~N.}\ \bibnamefont
  {Manoharan}}, \ and\ \bibinfo {author} {\bibfnamefont {M.~P.}\ \bibnamefont
  {Brenner}}} (\bibinfo {year} {2017}),\ \href@noop {} {\bibfield  {journal}
  {\bibinfo  {journal} {Reviews of Modern Physics}\ }\textbf {\bibinfo {volume}
  {89}}~(\bibinfo {number} {3}),\ \bibinfo {pages} {031001}}\BibitemShut
  {NoStop}%
\bibitem [{\citenamefont {Zhang}\ \emph {et~al.}(2018)\citenamefont {Zhang},
  \citenamefont {He}, \citenamefont {Zhuo}, \citenamefont {Sha}, \citenamefont
  {Brujic}, \citenamefont {Seeman},\ and\ \citenamefont {Chaikin}}]{Zhang2018}%
  \BibitemOpen
  \bibfield  {author} {\bibinfo {author} {\bibnamefont {Zhang}, \bibfnamefont
  {Y.}}, \bibinfo {author} {\bibfnamefont {X.}~\bibnamefont {He}}, \bibinfo
  {author} {\bibfnamefont {R.}~\bibnamefont {Zhuo}}, \bibinfo {author}
  {\bibfnamefont {R.}~\bibnamefont {Sha}}, \bibinfo {author} {\bibfnamefont
  {J.}~\bibnamefont {Brujic}}, \bibinfo {author} {\bibfnamefont {N.~C.}\
  \bibnamefont {Seeman}}, \ and\ \bibinfo {author} {\bibfnamefont {P.~M.}\
  \bibnamefont {Chaikin}}} (\bibinfo {year} {2018}),\ \href@noop {} {\bibfield
  {journal} {\bibinfo  {journal} {Proceedings of the National Academy of
  Sciences}\ }\textbf {\bibinfo {volume} {115}}~(\bibinfo {number} {37}),\
  \bibinfo {pages} {9086}}\BibitemShut {NoStop}%
\bibitem [{\citenamefont {Zhong}\ \emph {et~al.}(2017)\citenamefont {Zhong},
  \citenamefont {Schwab},\ and\ \citenamefont {Murugan}}]{Zhong17}%
  \BibitemOpen
  \bibfield  {author} {\bibinfo {author} {\bibnamefont {Zhong}, \bibfnamefont
  {W.}}, \bibinfo {author} {\bibfnamefont {D.~J.}\ \bibnamefont {Schwab}}, \
  and\ \bibinfo {author} {\bibfnamefont {A.}~\bibnamefont {Murugan}}} (\bibinfo
  {year} {2017}),\ \href@noop {} {\bibfield  {journal} {\bibinfo  {journal}
  {Journal of Statistical Physics}\ }\textbf {\bibinfo {volume}
  {167}}~(\bibinfo {number} {3-4}),\ \bibinfo {pages} {806}}\BibitemShut
  {NoStop}%
\bibitem [{\citenamefont {Zou}\ and\ \citenamefont {Nagel}(2010)}]{Zou10}%
  \BibitemOpen
  \bibfield  {author} {\bibinfo {author} {\bibnamefont {Zou}, \bibfnamefont
  {L.-N.}}, \ and\ \bibinfo {author} {\bibfnamefont {S.~R.}\ \bibnamefont
  {Nagel}}} (\bibinfo {year} {2010}),\ \href@noop {} {\bibfield  {journal}
  {\bibinfo  {journal} {Phys. Rev. Lett.}\ }\textbf {\bibinfo {volume}
  {104}}~(\bibinfo {number} {25}),\ \bibinfo {pages} {257201}}\BibitemShut
  {NoStop}%
\end{thebibliography}%

\end{document}